\documentclass{emulateapj}

\usepackage{epstopdf}

\usepackage{graphicx}
\usepackage{pstricks}
\usepackage{lscape}
\usepackage[T1]{fontenc}
\usepackage{ae,aecompl}
\usepackage{remreset}
\makeatletter\@removefromreset{footnote}{chapter}\makeatother
\RequirePackage{natbib}

\begin{document}
\title{On the Origin of Cool Core Galaxy Clusters: Comparing X-ray Observations with Numerical Simulations}
\shorttitle{On the Origin of Cool Core Galaxy Clusters}
\author{Jason~W.~Henning\altaffilmark{1}, Brennan Gantner\altaffilmark{1}, Jack~O.~Burns\altaffilmark{1}, Eric~J.~Hallman\altaffilmark{1,2}}
\shortauthors{Henning et al.}
\altaffiltext{1}{Center for Astrophysics and Space Astronomy, Department of Astrophysical \& Planetary Science, University of Colorado, Boulder, 389 UCB, CO 80309-0389, USA}
\altaffiltext{2}{National Science Foundation Astronomy and Astrophysics Postdoctoral Fellow}
\email{jason.henning@colorado.edu}


\begin{abstract}
To better constrain models of cool core galaxy cluster formation, we have used X-ray observations taken from the \textit{Chandra} and \textit{ROSAT} archives to examine the properties of cool core and non-cool core clusters, especially beyond the cluster cores.  Using an optimized reduction process, we produced X-ray images, surface brightness profiles, and hardness ratio maps of 30 nearby rich Abell clusters (17 cool cores and 13 non-cool cores).  We show that the use of double $\beta$-models with cool core surface brightness profiles and single $\beta$-models for non-cool core profiles yield statistically significant differences in the slopes (i.e., $\beta$ values) of the outer surface brightness profiles, but similar cluster core radii, for the two types of clusters.  Hardness ratio profiles as well as spectroscopically-fit temperatures suggest that non-cool core clusters are warmer than cool core clusters of comparable mass beyond the cluster cores.  We compared the properties of these clusters with the results from analogously reduced simulations of 88 numerical clusters created by the AMR Enzo code.  The simulated surface brightness profiles have steeper $\beta$-model fits in the outer cluster regions for both cool cores and non-cool cores, suggesting additional ICM heating is required compared to observed cluster ICMs.  Temperature and surface brightness profiles reveal that the simulated clusters are over-cooled in their cores.  As in the observations, however, simulated hardness ratio and temperature profiles indicate that non-cool core clusters are warmer than cool core clusters of comparable mass far beyond the cluster cores.  The general similarities between observations and simulations support a model described in Paper I \citep{paper_one} suggesting that non-cool core clusters suffered early major mergers destroying nascent cool cores.  Differences between simulations and observations will be used to motivate new approaches to feedback in subsequent numerical models.
\end{abstract}


\addtocounter{footnote}{2}

\section{Introduction}
Due to their high overdensity relative to the cosmic mean and their subsequent relative rarity, rich galaxy clusters have previously been treated as physically isolated objects \citep{voit_2005}.  The combination of large mass and relative isolation make them an attractive tool for precisely measuring cosmological parameters such as $\Omega_{m}$, $\Omega_{b}$ and the dark energy equation of state \citep[e.g.][]{lin_2003, sanderson_2003, laroque_2006, allen_2008,vikhlinin_2009}.  Assuming that the collapse of the cluster has captured all the mass within its gravitational sphere of influence, the relative densities of dark matter and baryons should be the same as in the wider universe \citep{White_1993, Eke_1998, Frenk_1999}.  However, accurate determination of the baryon fraction in the cluster, and thus for the universe as a whole, requires accurate measurements of both the baryon and dark matter mass of the system.  Of the two types of galaxy clusters, cool core (hereafter CC) and non-cool core (hereafter NCC), CC clusters are more often chosen for these measurements because they are assumed to be dynamically relaxed and, thus, may be accurately fit with a density profile that provides the gas mass to dark matter ratio \citep{allen_2004, allen_2008,vikhlinin_2009}.

Cool core clusters demonstrate sharply peaked X-ray emission at their centers due to condensed regions of cooler gas that are brighter than the surrounding material (see e.g. \citet{Fabian_1994,donahue_and_voit_2004} for reviews of cool core clusters).  In recent flux-limited samples, approximately half of the clusters contain cool cores \citep{chen_2007}.  While flux-limited samples are likely to increase the number of CC clusters found through the selection bias of a brighter core, CC clusters still clearly represent a significant fraction of galaxy clusters.  The lack of observed strong cooling flows in these cores \citep{Tamura_2001, Peterson_2001, Peterson_2003} has been presented in tandem with several recent AGN observations \citep{Fabian_2006, McNamara_Nulsen_2007, Wise_2007} suggesting some form of AGN feedback inhibits runaway cooling \citep{Heinz_2006, Mathews_2006, Nusser_2006,  Sijacki_Springel_2006, Binney_2007, Cattaneo_Teyssier_2007, Ciotti_Ostriker_2007}.  The lack of runaway cooling can be interpreted as potential evidence that CC clusters have relaxed and achieved a relatively steady state of cooling and accretion making them more attractive for cosmological studies. 

One may find the baryon mass of galaxy clusters utilizing measured X-ray surface brightness radial profiles by fitting them with a $\beta$-model, 
\begin{equation}
S_x \left( r \right) = S_{x0} \left( 1 + \left( \frac{r}{r_{core}} \right) ^2 \right) ^{\frac{1}{2} - 3 \beta},
\end{equation}
where $S_{x0}$ is the normalization factor, $r_{core}$ is the size of the cluster `core' and $\beta$ is the slope of the power law.  This parametric fit of $S_x$ has been successful for some galaxy clusters \citep{Jones_Forman_1984, Jones_Forman_1999, vik_forman_jones_1999}, however, the fit implies no physical processes or assumptions about the cluster.  The $\beta$-model was originally developed for an isothermal system by \citet{king_1966}, and though clusters are known to be non-isothermal \citep[e.g.][]{Markevitch_1998, vik_2005, vik_doublebeta_2006}, the $\beta$-model often yields reasonable empirical fits to observed profiles \citep{cff_1976}.  Nevertheless, if a cluster is assumed to be isothermal, then the X-ray surface brightness can directly translate into a function of gas density,
\begin{equation}
n _e \left( r \right) = n _{e0} \left( 1 + \left( \frac{r}{r_{core}} \right) ^2 \right) ^{\frac{-3 \beta}{2}}.
\end{equation}
Additionally, detailed \textit{Chandra} and \textit{XMM} observations show the dark matter potential is consistent with the NFW profile parameterization of \citet{nfw_1997} \citep{mushotzky_2004}.  The total cluster mass and relative quantities of baryonic and dark matter constrain the values of $\Omega_{m}$ and $\Omega_{b}$ \citep{wang_steinhardt_1998, haiman_2001, allen_2008,vikhlinin_2009}.  While $\beta$-models have been shown to fit NCC clusters well, they are typically inadequate for CC clusters, owing to the excess surface brightness from their denser, cooler cores \citep[e.g.][]{vik_doublebeta_2006}.  To combat this, the over-dense CC cores are often excised from the $\beta$-model fit, or a second $\beta$-model is included as well to fit the core \citep{pratt_2002, point_2004, vik_doublebeta_2006, santos_2008}.

In the light of recent work \citep[e.g.][]{paper_one, jeltema_2008}, two questions beg to be asked: What causes the X-ray properties of CC and NCC clusters to be so different?  Are CC clusters as simple as previously assumed?  The results of Paper I \citep[][hereafter B08]{paper_one} predict some observed differences between CC and NCC clusters may result from different merger histories.  Cosmological N-body/hydrodynamic adaptive mesh refinement simulations with cooling and feedback indicate that NCC clusters suffer major mergers earlier in their development than CC clusters, disrupting nascent cool cores.  We defined a ``major merger" in B08 as a cluster encounter during which $\sim 50\%$ of the cluster mass is accreted over roughly 1 Gyr, a merger large enough to potentially destroy any existing cool cores.  CC clusters, on the other hand, undergo only minor mergers so these cool cores are less susceptible to disruption.  In the intermediate cluster region just beyond the cluster cool cores ($~0.05-0.3\,r_{200}$ where $r_{200}$ is approximately the virial radius) the simulations find $\sim 40\%$ more cool gas in CC clusters than NCC clusters.  This excess cool gas is also indicated by an offset between CC and NCC hardness ratio profiles.  Additionally, single $\beta$-model fits to surface brightness profiles in this intermediate region indicate that CC cluster flux is substantially overestimated there.  This overestimation leads to a significant bias in estimated cluster mass \citep{hallman_2006}, and suggests that CC clusters as cosmological probes must be used with caution.  Motivated by these predictions, this paper looks for evidence of differing merger histories in observed clusters, further evidence that may cast doubt on the assumption that CC clusters are simple, relaxed objects.  We also examine differences between observations and our simulations in order to improve our cluster models.  Using a sample of cluster observations from the \textit{Chandra} and \textit{ROSAT} archives, we look for evidence of differing merger histories between CC and NCC clusters in the hardness ratio profiles and spectroscopically fit temperatures of the observed clusters.  We also compare observed surface brightness profiles and calculated $\beta$-model fits to the simulations of \citet{paper_one}.   

In $\S$ \ref{obs_data_sample} we describe the observational cluster sample and the individual \textit{Chandra} and \textit{ROSAT} data used.  We discuss in $\S$ \ref{data_analysis} the data reduction pipeline for both \textit{Chandra} and \textit{ROSAT} data, detailing differences between the standard pipeline and our process.  We present observational results including surface brightness profiles, $\beta$-model fits, hardness ratio profiles, spectroscopically fit temperatures, and comparisons with published temperature maps from different X-ray observatories for various clusters in $\S$ \ref{obs_results}.  In $\S$ \ref{sim_comparisons}, we compare the simulations described in B08 with our observational results to determine how our simulations may be improved in the future, and also find that the observed clusters indicate similar signs of different merger histories as found in our simulations.  Finally, conclusions and a summary of the paper are in $\S$ \ref{summary}.

\section{Observational Data Sample}\label{obs_data_sample}

Using both the \textit{Chandra} and \textit{ROSAT} archives, we chose 30 rich clusters to analyze (17 CC and 13 NCC).  \citet{edge_1990} defines a sample of 55 clusters with fluxes greater than 1.7 x10$^{-11}$ ergs cm$^{-2}$ s$^{-1}$ as measured with the \textit{Einstein} or EXOSAT satellites, a reasonably complete list of clusters for these fluxes and above.  We used a subset of the \citet{edge_1990} list as our data sample;  the sample consists of Abell clusters with at least one observation of 3 ksec or greater from each of the \textit{Chandra}, \textit{ROSAT} and \textit{XMM} archives.  In the final sample, detailed in Table \ref{clusters_table}, total \textit{Chandra} exposures are all $\geq10$ ksec and total \textit{ROSAT} exposures are all $\geq3$ ksec.  While the \textit{XMM} data are not used in this paper, we plan to combine it with the \textit{ROSAT} and \textit{Chandra} results in a future publication.  Clusters are included regardless of morphology or substructure present in an effort to limit selection bias.  Given the diverse population of X-ray observations in the archives, our sample is not statistically complete, however we believe it is representative of CC and NCC clusters as described below.  This is important if any significant conclusions about the properties and merger histories of general CC and NCC clusters are to be made.  The clusters we analyzed are listed in Table \ref{clusters_table} with their X-ray properties as well as \textit{Chandra} and \textit{ROSAT} observation details.  

Determinations of NCC or CC were taken from the \citet{ohara_2006} list that includes all of our clusters.  However, we differed from their list at one point.  While \citet{ohara_2006} lists Abell 3562 as a CC cluster, we changed its designation to a NCC cluster.  \citet{chen_2007} lists mass accretion into the core of the cluster as 0.0 M$_{\odot}$/year and a cooling time of $1.3\times10^{10}$ years, both typical values for NCC clusters.  The results we obtained from the data analysis also place it solidly into the NCC category.  

Redshifts were drawn from \citet{ohara_2006}, while $M_{200}$ and $r_{200}$ were calculated from $M_{500}$ and the temperature, respectively.  Note that subscripts on $M$ and $r$ denote cluster mass and radius where the cluster overdensities with respect to the critical density $\rho_{cr}$ equal these subscripts.  Further note that the virial radius of the cluster is at $r_{178}$, which is $\sim r_{200}$.  For $M_{200}$, we used the values of $M_{500}$ taken from \citet{chen_2007} and the scaling relation $M_{200} = 1.377 M_{500}$ from \citet{voit_2005}.  For the values of $r_{200}$, we calculated $r_{500}$ from the \citet{chen_2007} temperatures using the relation $r_{500} = 0.447 h^{-1}_{70} T^{0.527}_x$ Mpc \citep{ohara_2006, fino_2001}.  The temperatures from \citet{chen_2007} were used in particular as they are corrected for the biasing effects of cluster cool cores in average cluster temperature estimates.  In reality, these $T_{Chen}$ temperatures are $T_h$, the hot-component cluster temperatures from \citet{Ikebe_2002}.  We thus believe that the temperatures given by \citet{chen_2007} are more physically relevant estimates of overall cluster temperatures.  Values of $r_{500}$ were then scaled to $r_{200}$ via $r_{200} = 1.51 r_{500}$, which is based on an NFW dark matter density profile \citep{nfw_1997}.  The cosmological values assumed throughout this paper for the observations are $H_0 = 71$ km s$^{-1}$ Mpc$^{-1}$, $\Omega_M = 0.27$ and a ``flat'' universe where $\Omega = 1$\footnote{Note that the cosmological values used in \citet{chen_2007} are $H_0 = 50$km s$^{-1}$ Mpc$^{-1}$, $\Omega_M = 1$, and $\Omega_{\Lambda} = 0$.  The masses quoted in Table \ref{clusters_table} and used throughout this work are corrected to our cosmology.}.

\begin{figure}[h]
\plotone{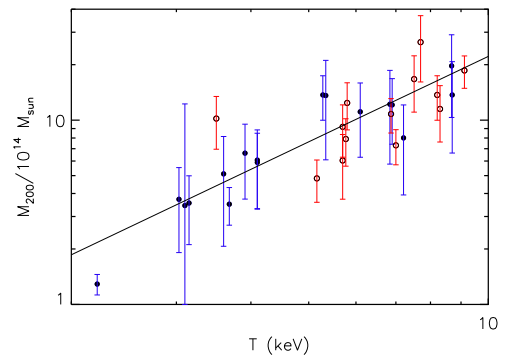}
\caption{Plot of cluster mass versus temperature for the sample in Table \ref{clusters_table}.  Red points correspond to NCC clusters and blue points correspond to CC clusters.  The mass-temperature relation from \citet{chen_2007} is over-plotted, showing that our cluster sample is similar to a more statistically complete cluster set.}
\label{fig_temp_mass}
\end{figure}

\begin{figure}[h]
\epsscale{0.75}
\plotone{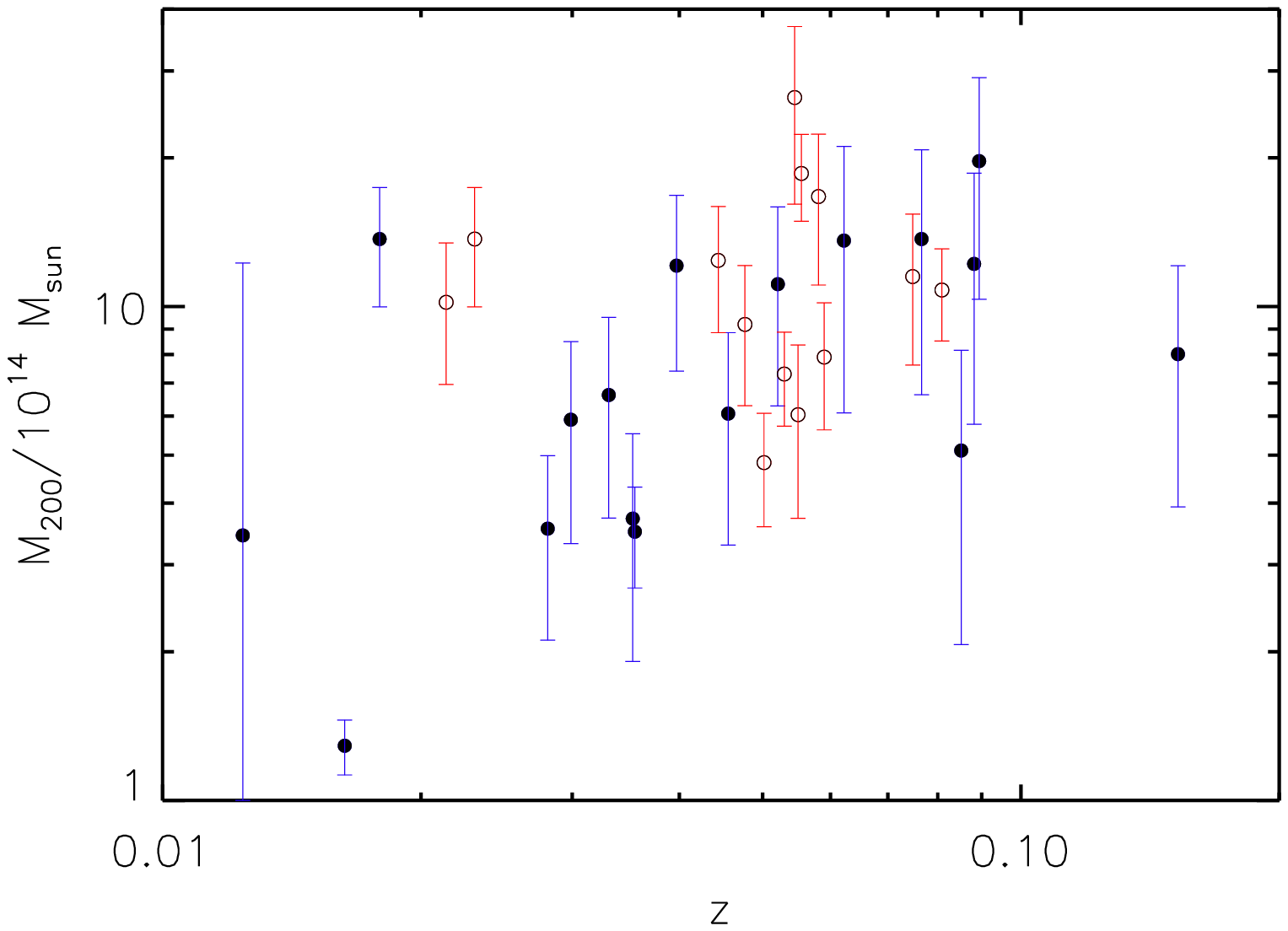}
\plotone{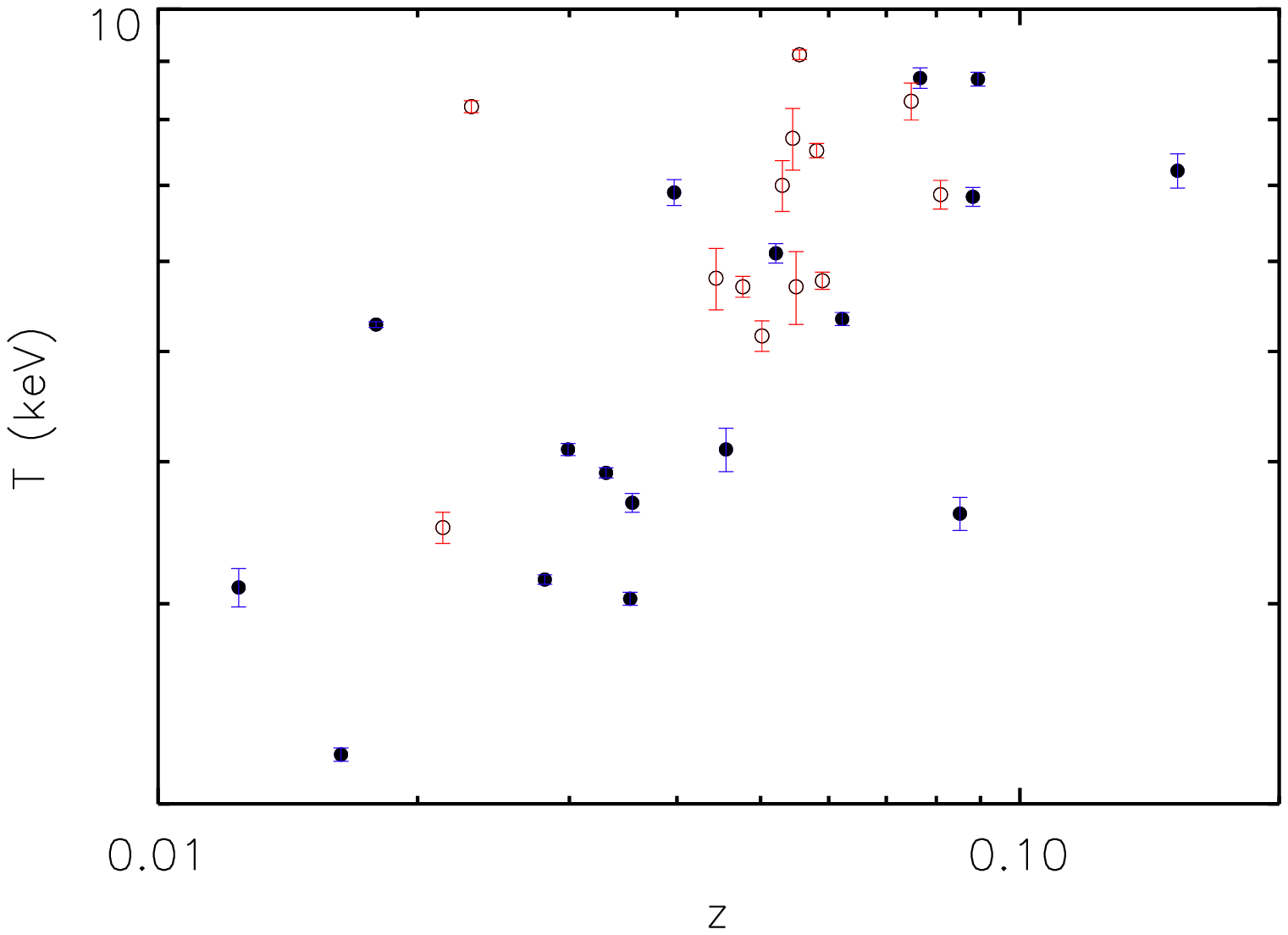}
\caption{Plots of cluster temperature and mass versus redshift for the sample in Table \ref{clusters_table}.  Red points correspond to NCC clusters and blue points correspond to CC clusters.  A weak correlation in mass and temperature with redshift is detectable, but this is an expected consequence of the cosmological volume element.}
\label{fig_temp_mass_z}
\end{figure}

Figures \ref{fig_temp_mass} and \ref{fig_temp_mass_z} demonstrate the range of clusters in our sample and hopefully further demonstrate that our sample is representative of CC and NCC clusters, if not a statistically complete set.  If clusters are scale-free, one should find a correlation between temperature and mass \citep{Kaiser_1986}, which is clearly visible for our sample in Figure \ref{fig_temp_mass}.  An empirical mass-temperature relation for a large sample of CC and NCC clusters was found by \citet{chen_2007} to be
\begin{equation}\label{eq_chen_M_T_relation}
\log _{10}\left( \frac{M_{500}}{5\times10^{14} \mbox{M}_{\odot}} \right) = -0.112 + 1.54 \log_{10}\left( \frac{T}{4 \mbox{keV}}\right).
\end{equation}
This relation is over-plotted in Figure \ref{fig_temp_mass} and shows that our clusters contain quite representative masses and temperatures from the broader sample.

As noted in B08 and seen in the data from \citet{ohara_2006} and \citet{chen_2007}, CC clusters are somewhat segregated from NCC clusters in temperature and mass as we see in Figures \ref{fig_temp_mass} and \ref{fig_temp_mass_z}.  In both observed and simulated samples, NCC clusters tend to have higher overall temperatures and mass.  The CC clusters, on the other hand, are observed to fill the whole range of temperatures and mass in the sample but have somewhat lower mass and are cooler on average than NCC clusters.  The fact that NCC clusters are only found with relatively high mass and temperatures corroborates our suggestions that CC and NCC clusters experience different merger histories as discussed in B08.  The segregation in cluster types is fully explained if NCC clusters experience major mergers during their formation, inhibiting their ability to retain cool cores and warming their ICM in the merger process, while CC clusters grow via only relatively minor mergers later in their history once their cool cores have been established.

With the exception of Abell 2204, the sample consists of nearby rich clusters with redshifts between $0.01 < z < 0.09$.  In Figure \ref{fig_temp_mass_z}, one notices that the (related) cluster distributions in temperature and mass versus redshift are reasonably random.  A slight correlation, though, is visibly present in that the NCC clusters are clumped at higher redshift and temperature.  This correlation is not necessarily a selection bias, but an expected consequence of the cosmological volume element.  At higher redshifts, the larger cosmological volume and larger numbers of clusters yields more of the rarer high mass clusters (NCC clusters).

Though statistically incomplete, our \textit{Chandra}/\textit{ROSAT} X-ray subsample appears broadly representative of the X-ray properties of the larger statistical sample of clusters from \citet{edge_1990} and expectations from models.  The temperature-mass correlation in our subsample follows the empirical relation for general CC and NCC clusters found by \citet{chen_2007}.  The clusters are also distributed as expected in redshift space.  Finally, the slight correlation of mass and temperature with redshift observed in our subsample is in accordance with effects from the cosmological volume element.  We feel our subsample is representative enough of a larger, more statistically complete cluster set to draw \textit{general} conclusions from comparisons with simulated clusters in B08.

\section{X-ray Data Analysis}\label{data_analysis}

\subsection{\textit{Chandra} Data Reduction}

While reducing the \textit{Chandra} data, we carefully followed the well-tested procedures outlined in the CIAO ACIS data pipeline threads\footnote{http://cxc.harvard.edu/ciao/guides/acis\_data.html} and the Markevitch cookbook for ACIS background datasets\footnote{http://cxc.harvard.edu/contrib/maxim/acisbg/COOKBOOK}.  In order to ensure that all of the data used were reduced accurately with the most recent calibration files in a consistent manner, we started observational reductions with the raw Level 1 telemetry files and reset all status bits associated with each event.  Using CIAO (version 3.3.0), we calculated new ``bad pixel'' maps with the \textbf{acis\_run\_hotpix} tool for each individual observation.  Unlike the standard data pipeline, however, we did not discard the columns next to bad pixels, consistent with recent advice \citep{mark_2003}.  Using the CIAO tool \textbf{acis\_process\_events}, we corrected each observation for differing gains across the CCDs, time-dependent gain, and CTIs (charge transfer inefficiencies) to recreate a Level 1 file using the latest calibration tools and files.  When an observation included chip S4, we utilized the tool \textbf{destreak} to remove systematically hot pixel columns that occasionally appear in that chip.  Events with unacceptable ASCA-based grades \citep{asca_grades} and point sources were discarded.  We also used the \textbf{acis\_process\_events} tool to apply a randomization to both the position of the event within the given pixel that detected the photon as well as the recorded energy within the PHA (Pulse Height Amplitude) bins.  This randomization removes stratification of the data at fine position and energy bins.  

To remove unresolved flares in the same manner as is used for the ACIS background datasets, event data for front- and back-illuminated chips were first separated.  Good Time Intervals (GTI) were determined using a non-CIAO tool called \textbf{lc\_clean}, again recommended and written by \citet{mark_2003}.  This resulted in excised periods of time where the incident flux differed from the observational average by more than 20\%.  The cleaned event files were then merged, resulting in the new Level 2 event files for each cluster observation.

\subsubsection{Background Subtraction}

Due to the extended spatial structure of the clusters and the limited field of view of \textit{Chandra}, using an exterior portion of each image as a representative background is inadequate.  Instead, we used the ACIS background files discussed in the Markevitch cookbook.  We followed the procedures laid out in the CIAO threads\footnote{http://cxc.harvard.edu/ciao/threads/acisbackground/} in order to appropriately match the backgrounds to each cluster observation.  Most of our cluster observations span multiple chips, so the background files, which are separated by chip, were then merged while maintaining separate files for front- and back-illuminated chip sets.  If the processing of the background and event files used different gain files, then the backgrounds were re-calibrated with the CIAO tool \textbf{acis\_process\_events} to match the gain files used on the observation event files.  Ideally, the calibration files used for CTIs  and time-dependent gains would also match between our event files and the ACIS background files, though it is not possible to update these gains.  According to the CIAO threads, however, the induced error is expected to be minimal for extended sources like the clusters in our observations.  Lastly, we re-projected the merged backgrounds into the coordinates of the observations using the tool \textbf{reproject\_events} and the aspect solution files from the observations.

To determine the exposure duration of the backgrounds in order to normalize them to each observation, the image GTI duration was multiplied by the ratio of the 9-12 keV flux between the image and background files.  In this band \textit{Chandra}'s effective area is essentially zero, thus nearly all the counts are particle background.  The observations and backgrounds were then split into two energy bands (0.5-2.0 keV and 2.0-8.0 keV) and each band was thereafter treated as a separate observation in order to construct hardness ratio maps later in the analysis.  Multiple observations of the same cluster were co-added into a single image and backgrounds were subtracted with the tool \textbf{dmimgcalc}.  These background-subtracted mosaic images constitute the final ``counts'' images and were used for the hardness ratio analysis described below.

\subsubsection{Exposure Maps}

Each \textit{Chandra} observation was taken over an extended period of time (tens of kiloseconds) and the pointing during that time both drifts and is dithered to prevent pixel differences from biasing the image.  While pointing is corrected within the Level 1 file, aspect files from the pointing history were calculated to help determine the effective area convolved across the CCD surface.  For this correction, we combined the spatial region of the cluster (minus the cool core ($<0.05r_{200}$) when appropriate) and extracted the observational spectrum.  This was then fit with a combination of galactic $N_H$ absorption \citep{morrison_1983} and a Raymond-Smith plasma model \citep{raymond_smith_1977} within the \textbf{Sherpa} program to recreate the expected spectrum reaching the front of the telescope.  This spectrum, the instrument responses (CIAO's ARF and RMF files) and observational GTIs were then folded together as detailed in the standard CIAO pipeline to create an exposure map for each observation, containing effective area, quantum efficiency and pointing.  We then divided the exposure maps from the counts images to give a final product in units of surface brightness (photons s$^{-1}$cm$^{-2}$arcsec$^{-2}$).
  
These exposure maps were used only for surface brightness profiles (0.5-2 keV) and not in the hardness ratio images (0.5-2 keV and 2-8 keV) as we discuss in Section \ref{obs_results}.  Because of the wide energy bands and spatial coverage, there is no exposure map that can accurately represent the effective area convolved with the source spectrum, especially over the 2-8 keV range.  The lower energy range of 0.5-2 keV has a relatively flat effective area function and the spatial change across the field of view utilized is on the order of ~1\%.  Thus, while the exposure maps were used to calculate fluxes in this band for surface brightnesses, we followed a different approach for the images used in the hardness ratios.

\begin{figure}[!h]
\epsscale{0.9}
\plotone{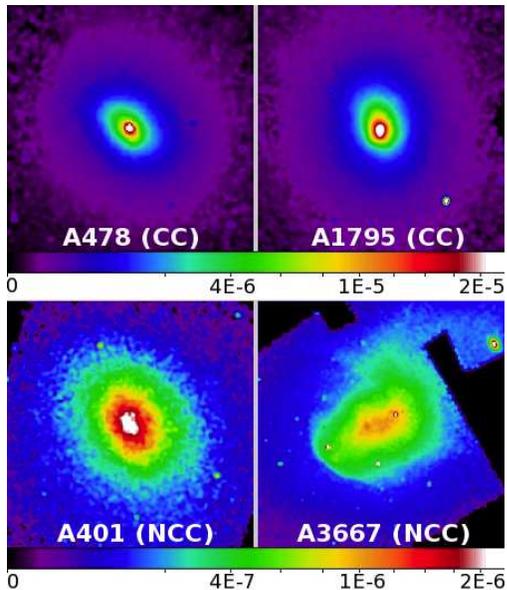}
\caption{Four example \textit{Chandra} flux images of CC and NCC clusters over the soft band of 0.5-2.0 keV.  Each is the result of multiple individual observations that have been combined to form a final data product and then smoothed with a 3 pixel ($\sim12$ arcsec) FWHM gaussian.  The FWHM for A478, A1795, A401, and A3667, respectively,  are the following: 19.6 kpc, 14.2 kpc, 16.9 kpc, 12.2 kpc.  Each image is cut to a radial metric field-of-view of $0.3\,r_{200}$.  All four images are square-root scaled and show $0-2\times10^{-5}$ (CC) and $0-2\times10^{-6}$ (NCC) photons s$^{-1}$cm$^{-2}$arcsec$^{-2}$, respectively.}
\label{Chandra_flux_collage}
\end{figure}

Figure \ref{Chandra_flux_collage} shows four examples of reduced \textit{Chandra} flux images, two CC and two NCC clusters.  The final mosaicked images have square-root scaling and cover the soft band 0.5-2.0 keV.  The images were also smoothed with a three pixel FWHM gaussian.  The flux images were binned by a factor of eight so that one pixel corresponds to 12 arcsec.  The CC clusters are very distinct from the NCC clusters, with significantly lower flux in the outer cluster extent in contrast with a  sharp flux rise in their cores.  The NCC surface brightnesses exhibit a much shallower drop-off moving outward from the cluster centers.  The images for A478, A1795, and A3667 are mosaics of three individual pointings, while the image of A401 is made from two.

\subsubsection{CIAO Calibration Updates}

Since the original work in this analysis was completed with CIAO 3.3.0, a newer version of CIAO and the CALDB (both versions 4.1.1) have been released.  The new calibration files take into account a contaminant originally missed during ground calibration tests that helps explain both discrepancies in high-temperature clusters between \textit{Chandra} and \textit{XMM-Newton} observations and between the \textit{Chandra} ACIS-I and ACIS-S instruments themselves\footnote{http://cxc.harvard.edu/caldb/downloads/Release\_notes/\linebreak CALDB\_v4.1.1.html}.  To make sure the use of the new calibration files would not significantly alter our findings, we re-analyzed six clusters with temperature $\gtrsim6$ keV (A85, A2255, A478, A401, A2319, A3266).  We found changes of less than $5\%$ in derived hardness ratios (see Section \ref{hr_maps}) for all clusters other than A478, which itself saw only $5-10\%$ deviations.  Derived temperatures for each of the clusters in our temperature analysis below (A85, A2255, A478, and A401; see Section \ref{bulk_temperatures}) all dropped slightly compared to the results using CIAO 3.3.0.  Because the changes in hardness ratios were so small, the results discussed below in Section \ref{hr_maps} derive from the analysis with CIAO 3.3.0 and its corresponding CALDB.  The findings in Section \ref{bulk_temperatures} for clusters A85, A2255, A478, and A401, however, arise from the use of CIAO and CALDB 4.1.1 to take advantage of the new calibrations.

\subsection{\textit{ROSAT} Data Reduction}

While \textit{Chandra} is the superior instrument in terms of effective area, calibrations, and breadth of energy response, the \textit{ROSAT} PSPC has a field of view approximately four times larger.  It is important to note, however, that \textit{ROSAT} observations in our cluster subsample generally have much shorter exposure times than the \textit{Chandra} observations of the same cluster.  Data were reduced using the process and tools as described by \citet{snowden_1994}, and all the \textit{ROSAT} pointed observations in our subsample used the PSPC detector.  \textit{ROSAT} data are not available at an uncorrected level equivalent to \textit{Chandra} data.  Thus, the event files provided were fully corrected for most inconsistencies and we did not recalibrate the events other than the screening discussed below.  Using the same tools and limits as the \textit{Chandra} pipeline, the time delineated flux was screened for flare events (time periods with fluxes more than 20\% from the observational mean), which were removed from the observation.  Events also determined to be ``spurious trailing'' events (recorded events that represent the residual charge from an actual prior photon) and point sources were excised.  The events were binned by a factor of 30 to form 15 arcsec x 15 arcsec pixels with the goal of increasing the S/N, and finally cut to an energy range of 0.42-2.01 keV.

\subsubsection{Exposure Maps}

Unlike \textit{Chandra}, the \textit{ROSAT} data reduction pipeline does not consider the spectrum of the source when creating an exposure map.  While this method is inherently inaccurate, the changes in both the spectrum and \textit{ROSAT}'s effective area across the 0.42-2.01 keV energy band are small enough to make the result acceptable.  The pointing of the telescope during the exposure and the GTI time were combined with the effective area and quantum efficiency calibrations.  These effects are contained in a previously calibrated ``detector map'' provided by the Snowden package.  Since the closest energy band of the detector maps is 0.42-2.01 keV, the \textit{ROSAT} and \textit{Chandra} images do not exactly match in energy.  Counts images were finally combined into a mosaic and corrected by a similar mosaic of exposure maps to create a final image.

\subsubsection{Background}

Unlike the \textit{Chandra} images, \textit{ROSAT} has a sufficient field of view to use the outer regions of an image as a background source.  A flux average of the annulus at the exterior of an image was subtracted from the rest of the image to create the final background-subtracted product.  This annulus varied in radius between images due to detector edge blurring and noise.  This was typically 20 pixels in width and ended a few pixels short of the largest radii deemed free of these edge effects.

\section{Observational Results}\label{obs_results}

Numerical cluster simulations from B08 suggest observable differences in the X-ray properties between CC and NCC clusters, owing to different evolutionary pathways.  In particular, our numerical models predict differences in surface brightness and temperature/hardness ratio profiles in the region beyond the cluster cores.  To test these predictions, we studied surface brightness, hardness ratio, and temperature profiles derived from our reduced \textit{Chandra} and \textit{ROSAT} observations, which are described in detail below.  In \S \ref{sim_comparisons}, we compare analogous simulated profiles to these observed results to draw conclusions not only about the merger histories of observed CC and NCC clusters, but also about what physics and feedback parameters need further development in next-generation simulations.

\subsection{Radial Surface Brightness Profiles}\label{surface_brightness_profs}

\begin{figure}[hbp]
\epsscale{1.1}
\plottwo{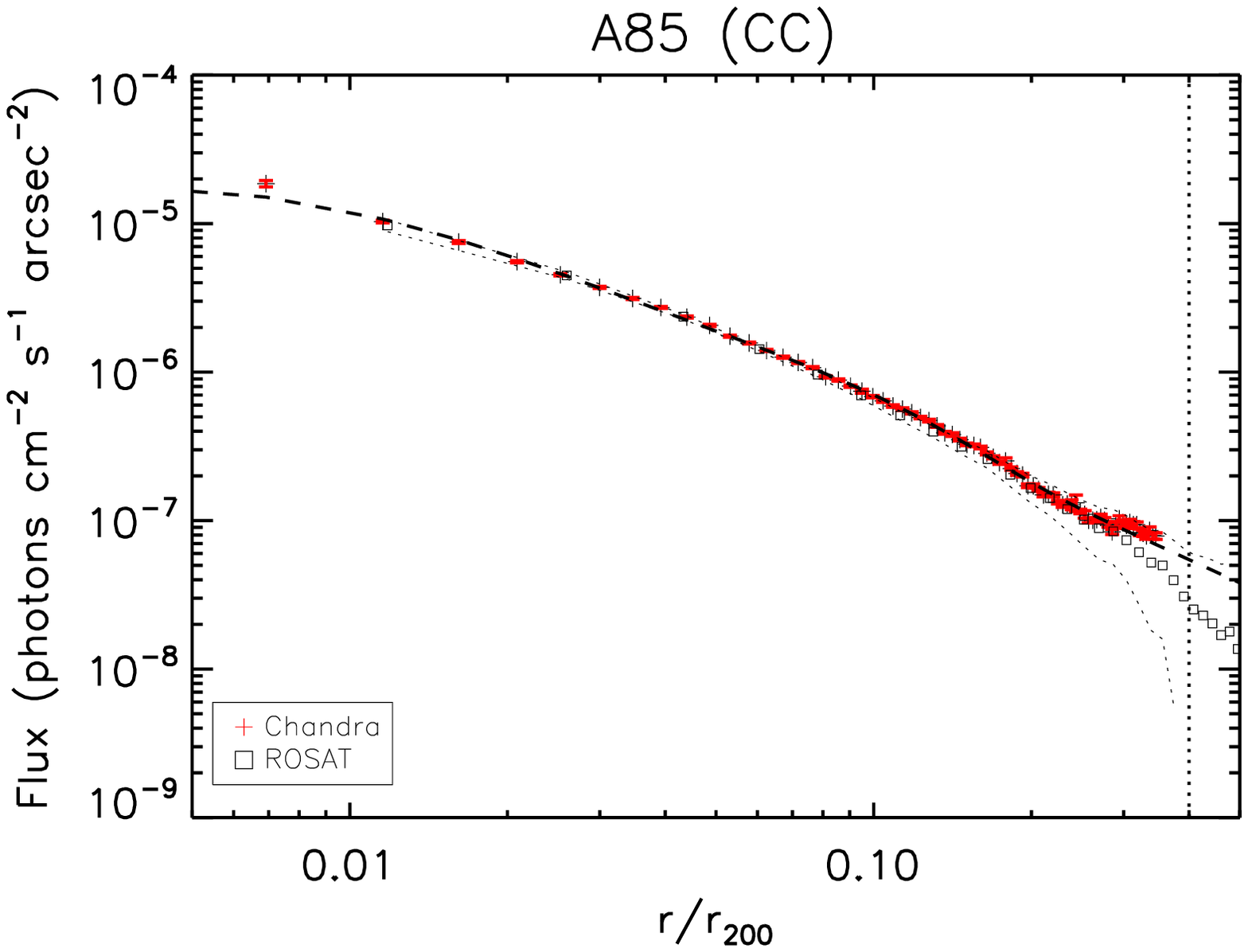}{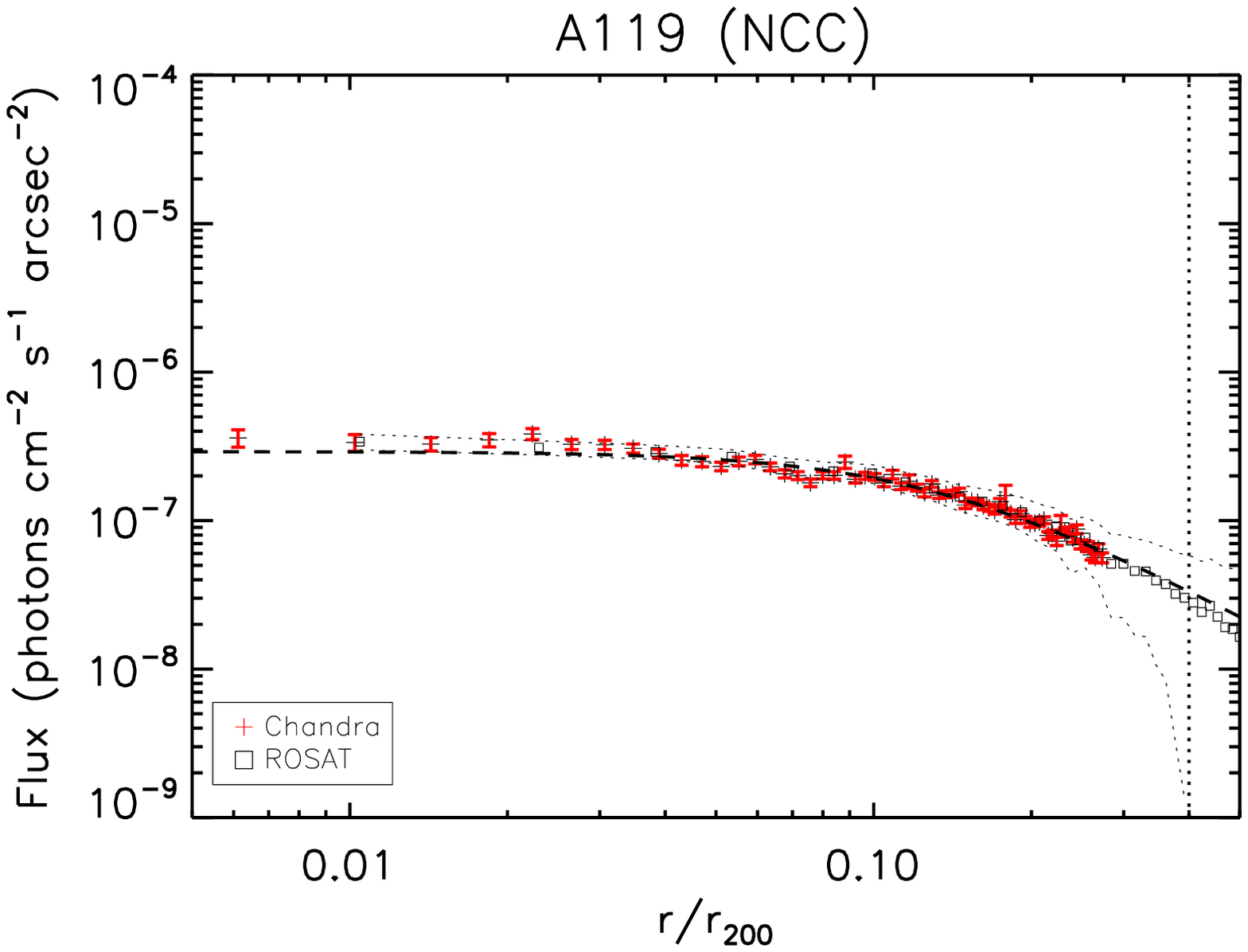}
\plottwo{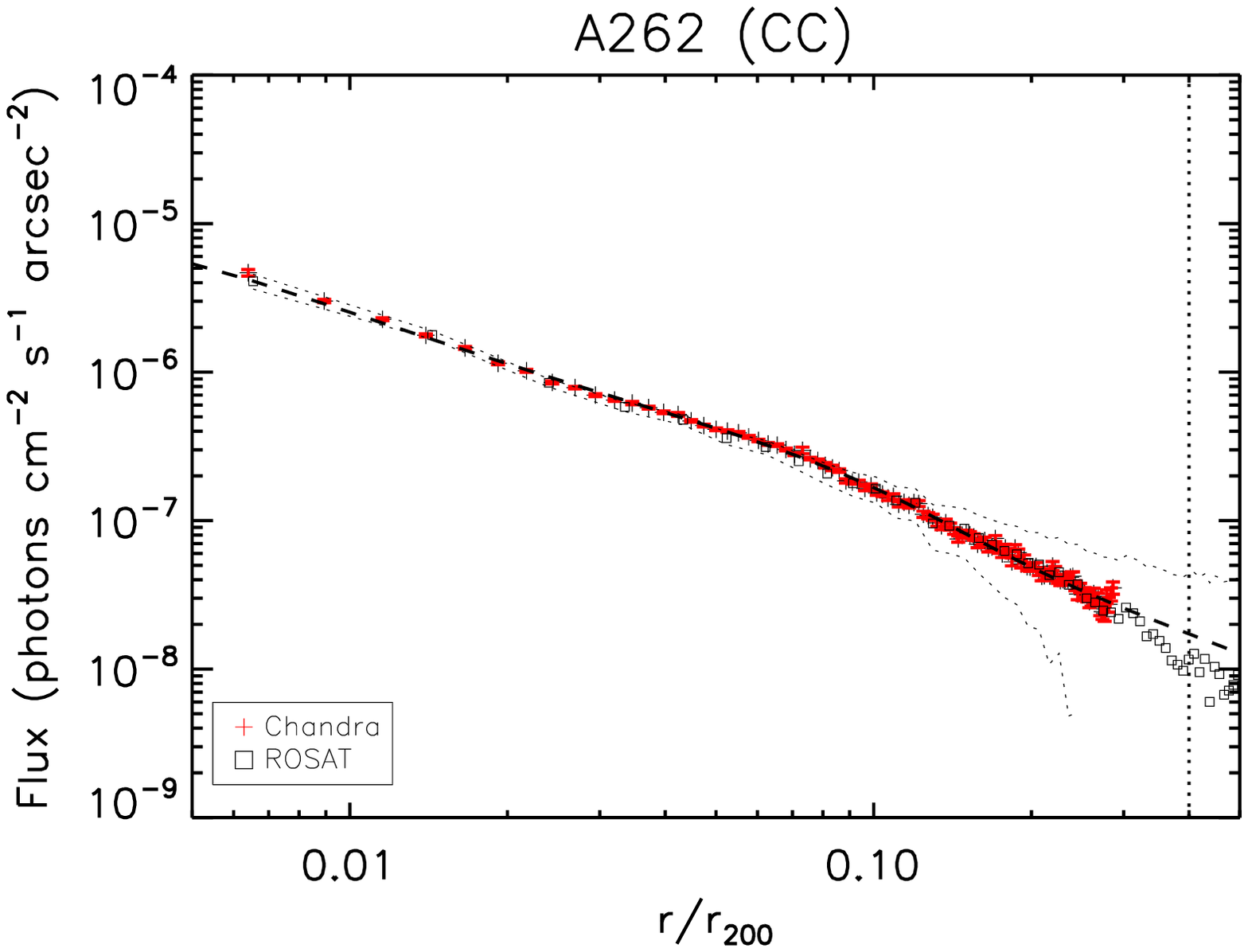}{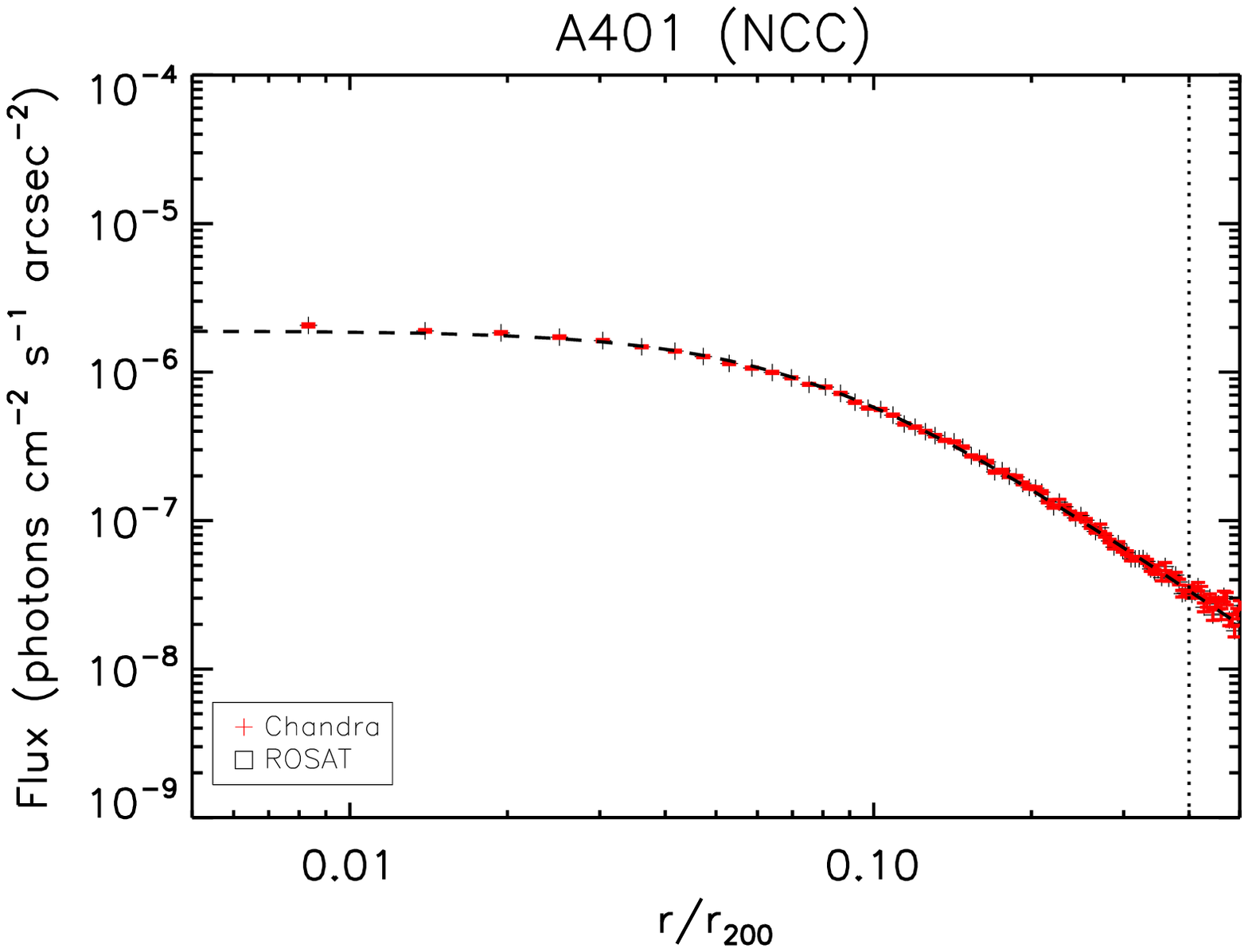}
\plottwo{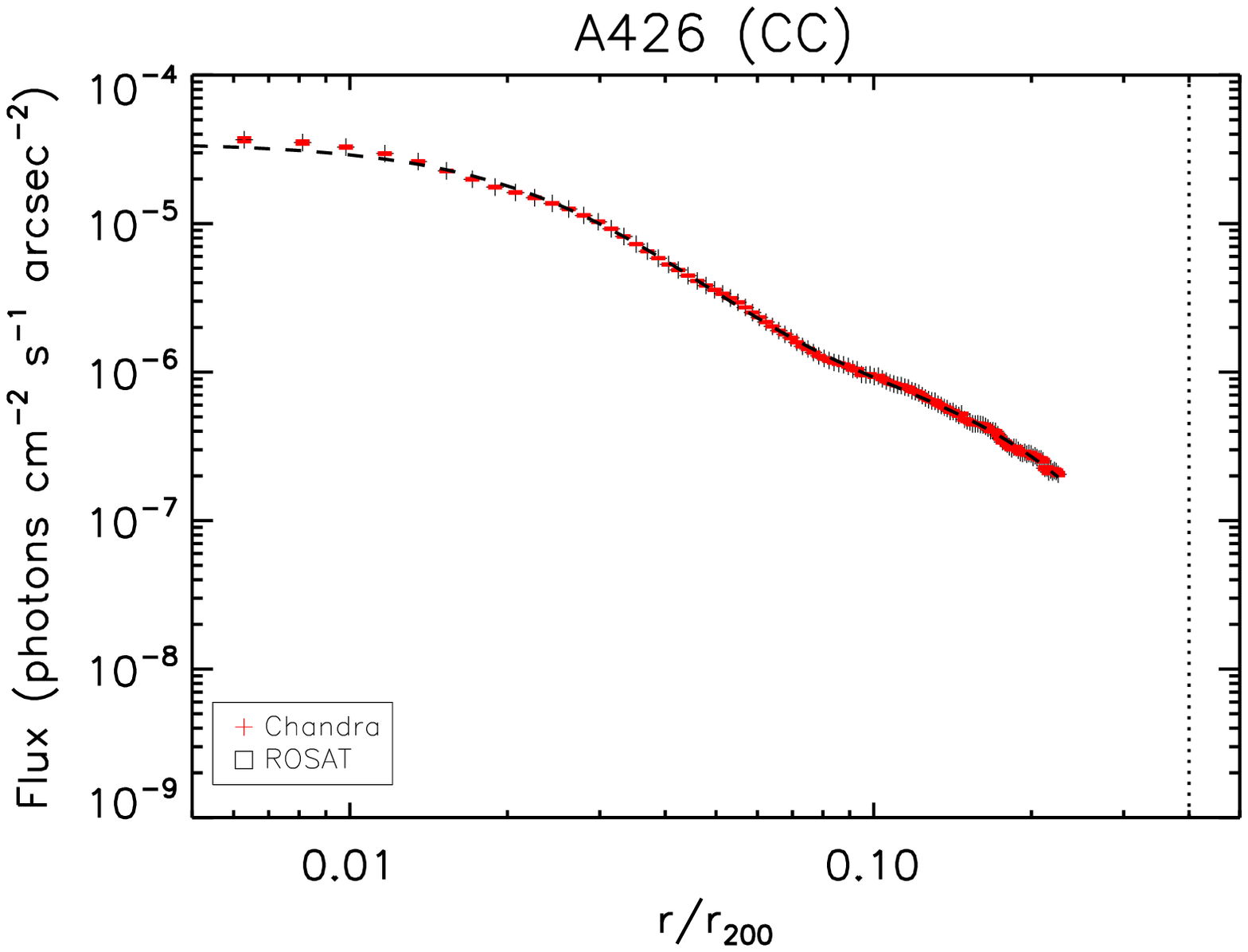}{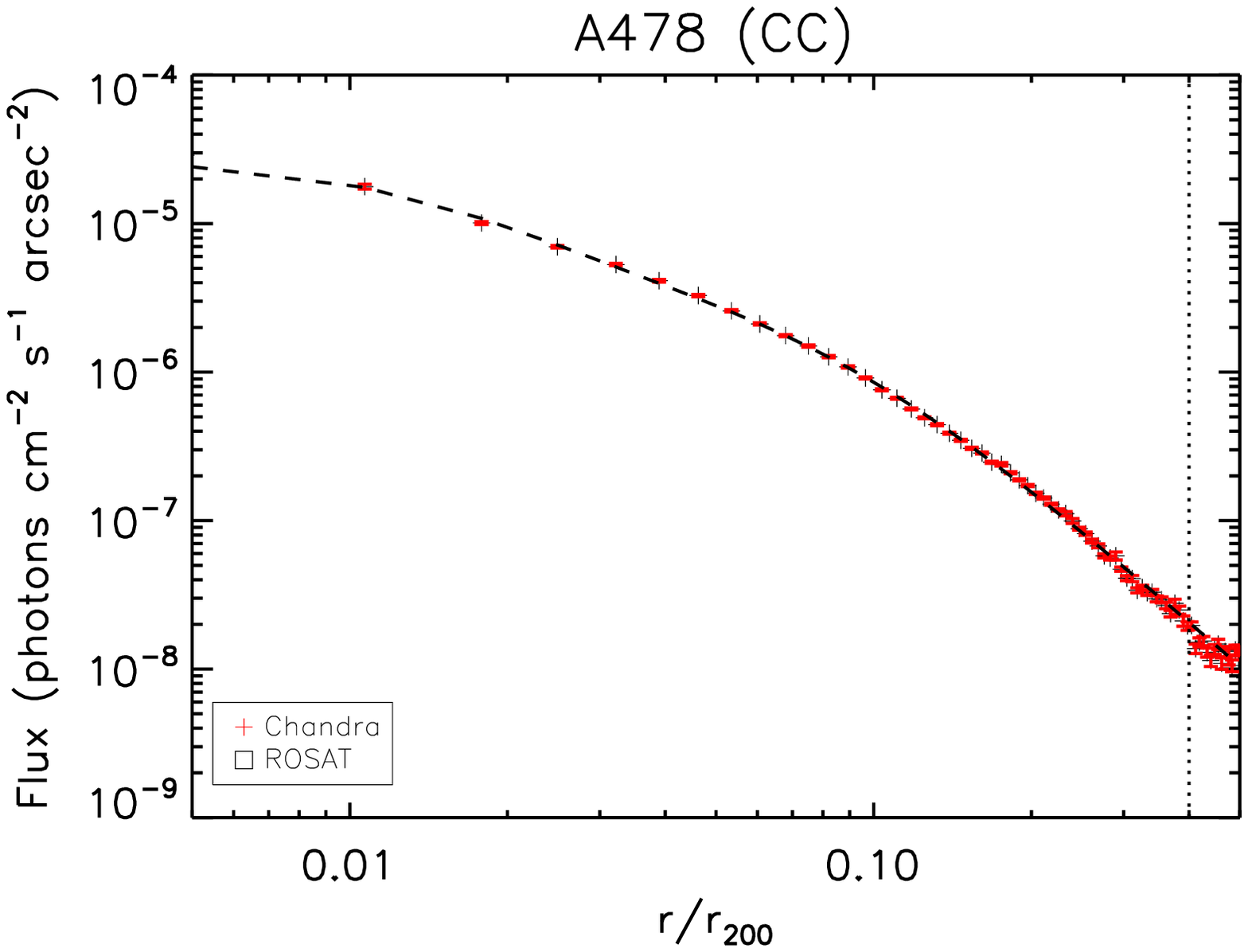}
\caption{Observational surface brightness profiles with both \textit{Chandra} (red) and \textit{ROSAT} (black) data.  Error bars are the standard deviation of each data bin from their respective telescope.  The \textit{ROSAT} error bars are represented by the black dotted lines for clarity purposes and disappear when the measurement minus the error becomes negative.  Black dashed lines show the $\beta$-model fits from 0 to $0.4\,r_{200}$ for each cluster.  A double $\beta$-model was used for CC clusters to account for the presence of the cores, while single $\beta$-model fits were used for NCC clusters.  Vertical dashed lines at $0.4\,r_{200}$ show the extent of data used for the $\beta$-model fits.   The double $\beta$-model fits for CC clusters agree well with the observed data, except for A2204 where there is reason to believe the \textit{Chandra} calibration data is suspect \citep[see][]{Reiprich_2008}.}
\label{obs_surbri}
\end{figure}

We created flux images (photons s$^{-1}$cm$^{-2}$arcsec$^{-2}$) for both \textit{Chandra} (0.5-2.0 keV) and \textit{ROSAT} (0.42-2.01 keV) observations from the reduced counts images by performing exposure map corrections, as detailed above.  The point sources, as well as obvious instrument artifacts such as CCD boundaries, were masked out in each image before taking averages of radial annuli.  While real galaxy clusters are not perfectly symmetric azimuthally, the differences are usually minimal even for clusters that have obvious structure and an annular average is a reasonable representation of the state of the cluster at that radius \citep[e.g.][]{vik_forman_jones_1999}.  The size of radial bins were adaptively binned for each cluster separately to maximize both the number of bins and S/N.  Figure \ref{obs_surbri} shows the combined \textit{Chandra} and \textit{ROSAT} surface brightness profiles for all clusters in our subsample. 

Several traits are common across most of the observations as shown in Figure \ref{obs_surbri}.  First, the \textit{ROSAT} data were included in this work because it was hoped that with its larger field of view the surface brightness profiles could be extended to larger radii.  This was largely untrue.  The errors of the \textit{ROSAT} data are often large enough to preclude a confident fit even where the \textit{Chandra} data have low errors.  While some nearby clusters do gain from the \textit{ROSAT} inclusion, on the whole it does not add significantly to the results.  As a consequence, \textit{ROSAT} data were excluded from cluster profiles for which \textit{Chandra} data extended to roughly $0.4\,r_{200}$.  Second, the reliable \textit{ROSAT} data and the \textit{Chandra} data match well across every cluster but two, A426 and A2063, so \textit{ROSAT} data are excluded for these clusters as well.  It is unknown why the data for A2063 do not agree as well as in the other clusters.  \textit{Chandra} data for A426 show obvious substructure that does not appear in the \textit{ROSAT} data, leading to deviations in the radial profile averages that account for the poorer match.  In the end, \textit{ROSAT} data were used for A85, A119, A262, A1060, A1367, A1656, A2199, A2319, and A4038.  Finally, one may notice that $0.3-0.5\,r_{200}$ is a reasonable median of the outer radius at which clusters can be reliably fit with a $\beta$-model based on the standard deviations of values at each radial bin.  NCC clusters are fit with a single $\beta$-model, while CC clusters are fit with a double $\beta$-model \citep[e.g.][]{santos_2008}.  The $\beta$-model fits are done using the \textbf{mpfit} routine, which uses a Levenberg-Marquardt least-squares algorithm for minimization \citep{markwardt_2009,more_1977}.  The use of a double $\beta$-model is to account for the influence of the cool cores, which may contaminate the surface brightness profiles beyond the cores themselves, potentially biasing a single $\beta$-model fit that only excludes the fiducial cores.  

\begin{figure}[h]
\epsscale{1.0}
\plotone{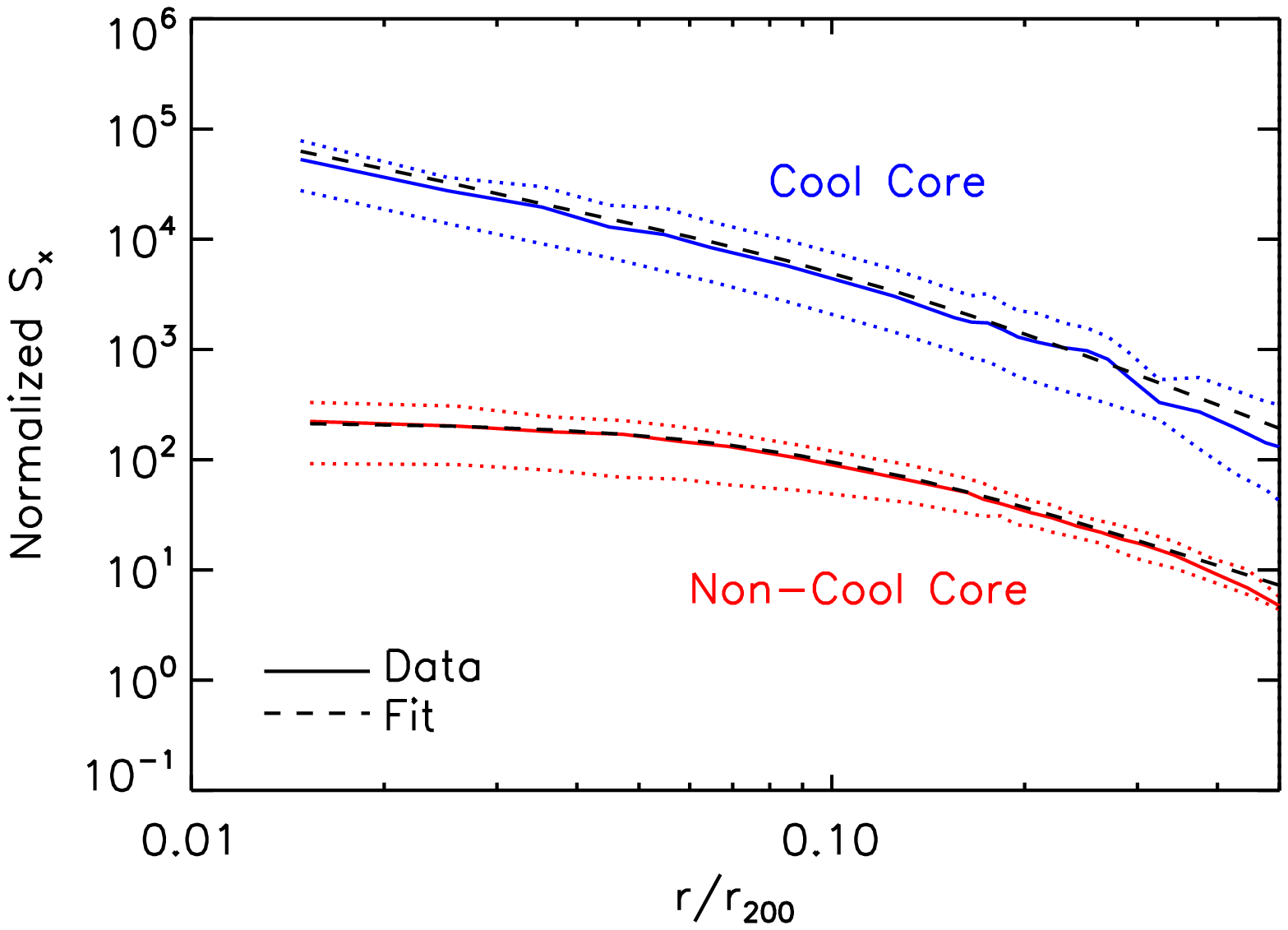}
\caption{Average surface brightness of 15 CC (blue) and 13 NCC (red) clusters observed with \textit{Chandra} and \textit{ROSAT}.  (The $\beta$-model fits for A426 and A478 are far outliers and are not included here.)  Each cluster is scaled by $M_{200}^{3/2}$ to account for luminosity differences due to mass.  Dotted lines delineate the 90\% regions of each data bin.  The NCC clusters are arbitrarily divided by 10 to separate the curves for clarity.  Dashed lines are the averaged individual beta model fits out to $0.4\,r_{200}$.  A double $\beta$-model was used for CC clusters, while a single $\beta$-model was used for NCC clusters.}
\label{obs_surbri_avg}
\end{figure}

In Figure \ref{obs_surbri_avg}, the averaged surface brightness profiles of 30 clusters separated by cluster type are shown with the average of $\beta$-model fits (dashed lines).   The averaging was done to increase S/N in the outer profiles.  To create the averaged profiles, each cluster was first scaled by M$_{200}^{3/2}$ to account for luminosity differences due to the differing cluster masses \citep{fino_2001}.  The rescaled profiles were radially binned between fixed fractions of $r_{200}$ to standardize the data, and these binned cluster fluxes were then averaged to produce the data in Figure \ref{obs_surbri_avg}.  Uncertainties are shown as 90\% regions, demarcated by dotted lines, within which 90\% of our data can be found.  Each cluster was fit individually with a $\beta$-model between 0 and $0.4\,r_{200}$.  The results of the individual cluster fits were then averaged to obtain an average $r_{core}$ and $\beta$ values in the case of NCC clusters or two $r_{core}$ and two $\beta$ values in the case of CC clusters.  The resulting average $\beta$-model fits are overplotted in Figure \ref{obs_surbri_avg}.  The size of deviations from the mean increase past $\sim0.2\,r_{200}$ in CC clusters, demonstrating that it is difficult to assess the validity of a $\beta$-model fit to the outer region of these clusters.  As is evident in Figure \ref{obs_surbri}, the surface brightness profiles for many CC clusters do not go out as far as those for NCC clusters, which is contributing to the larger uncertainties in the outer CC profile.  This is likely to be a selection effect as NCC clusters are in general farther away and so cover a smaller angular area, which results in better sampled surface brightness profiles farther from the center of the cluster.  However, this does not exclude other potential reasons for larger deviations in the outer average CC profile.  Perhaps some physical effect has a role.  Are the CC clusters truncated compared to NCC clusters?  Are NCC clusters being ``puffed out" by massive mergers and heating?  Further investigation is needed to explain where the larger deviations from the mean outer CC surface brightness profiles may be originating. 

\begin{figure}[h]
\epsscale{0.75}
\plotone{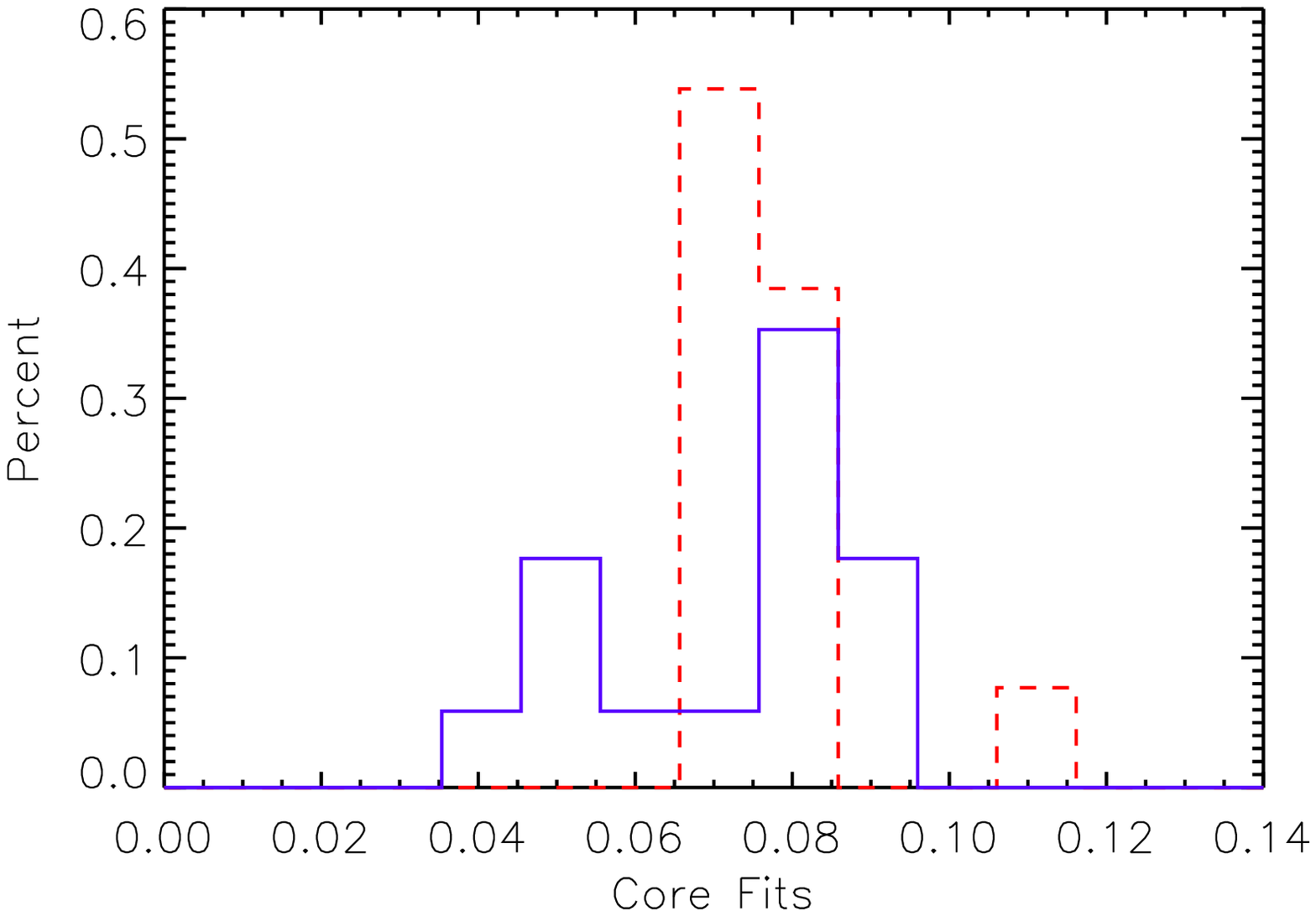}
\plotone{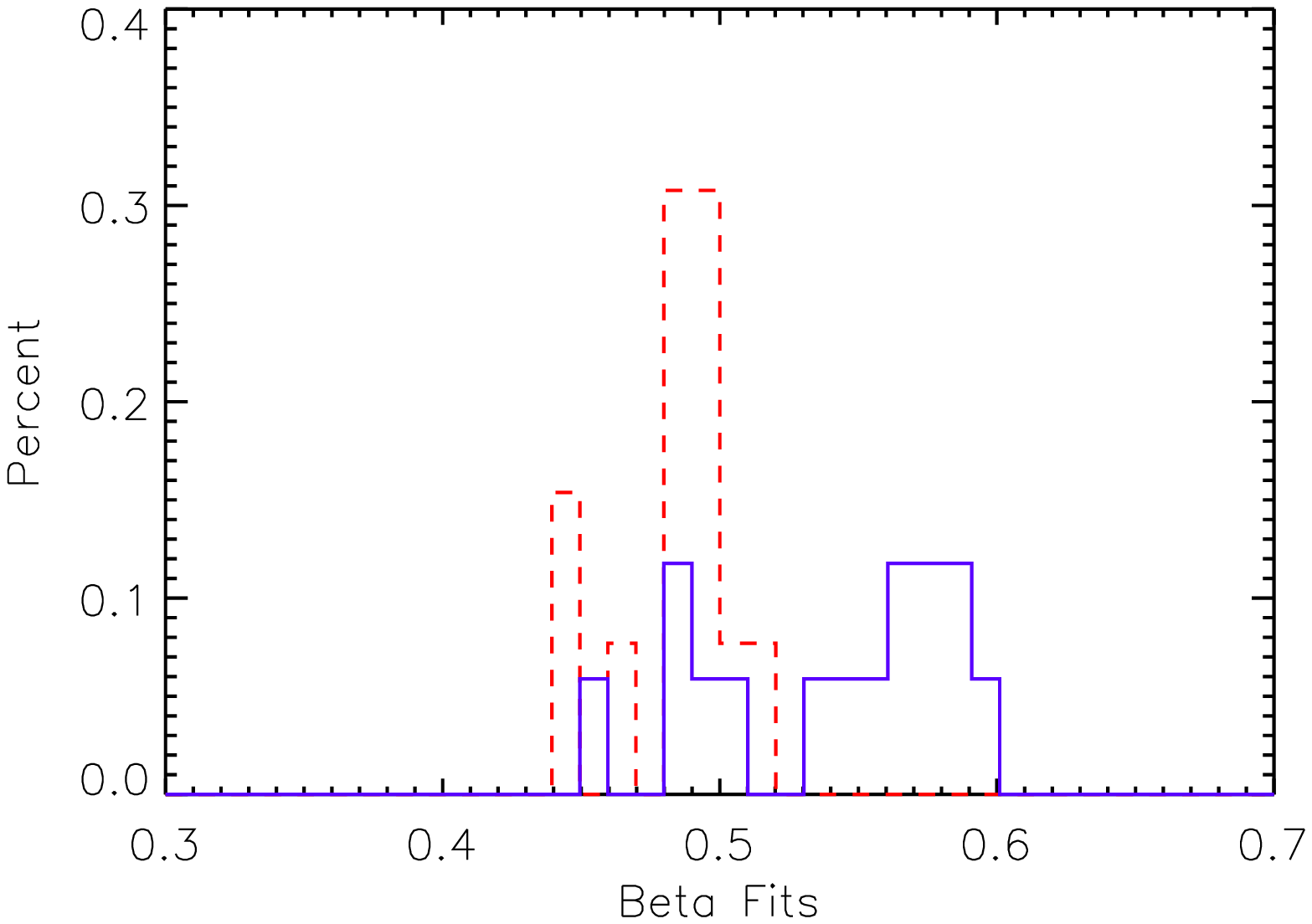}
\caption{Histograms of $\beta$-model fit parameter values.  Red is for NCC clusters and blue for CC clusters.  (Top) Outer $r_{core}$ fits.  (Bottom) $\beta$ fits.  The fits for A426 and A478 are far outliers, and are not included here.  It is clear from the histograms that CC and NCC $r_{core}$ fits are nearly identical, while there is a significant difference in $\beta$ slopes.}
\label{obs_betamod_histograms}
\end{figure}

Some differences in surface brightness profiles between CC and NCC clusters are evident in Figures \ref{obs_surbri} and \ref{obs_surbri_avg}.  CC clusters have considerably higher emission in their cores due to the cooler, denser gas.  NCC clusters have much flatter emission in the inner cluster region that begins to drop off at $\sim 0.1\,r_{200}$.  The overall shapes/slopes of the profiles beyond the cores appear to be different for these two cluster types.  To help quantify these differences, Figure \ref{obs_betamod_histograms} shows histograms of the $\beta$-model parameter fits for each cluster sample.  The averages of the individual parameter fits are given in Table \ref{beta_table}, and the individual best-fit values and fit statistics are presented in Table \ref{all_beta_table}.  Two-sample, one-tailed Kolmogorov-Smirnov (K-S) tests  \citep{wall_jenkins_2003} were also performed to measure the probabilities that CC and NCC model fits come from the same parent distributions.  We draw two important conclusions from the data in Figure \ref{obs_betamod_histograms} and Table \ref{beta_table}.  The first is that the outer slopes of the surface brightness profiles are statistically different between CC and NCC clusters, with a probability they come from the same distribution of only 0.002.  Could this be caused by the different evolution history of CC and NCC clusters?  This will be discussed further in \S \ref{summary}.  The second conclusion we draw is that the $r_{core}$ values are statistically indistinguishable ($Prob=0.209$) between CC and NCC clusters.  This differs from previous results \citep{vik_doublebeta_2006, paper_one} where single $\beta$-models are used for CC clusters.  One expects the average $r_{core}$ value for CC and NCC clusters to be similar if the data were completely corrected for the cooler cores.  If cool cores are not completely removed, however, then single $\beta$-model fits to CC clusters may bias the $r_{core}$ results low, resulting in different $r_{core}$ results for the two cluster types.

\begin{figure}[hbp]
\plottwo{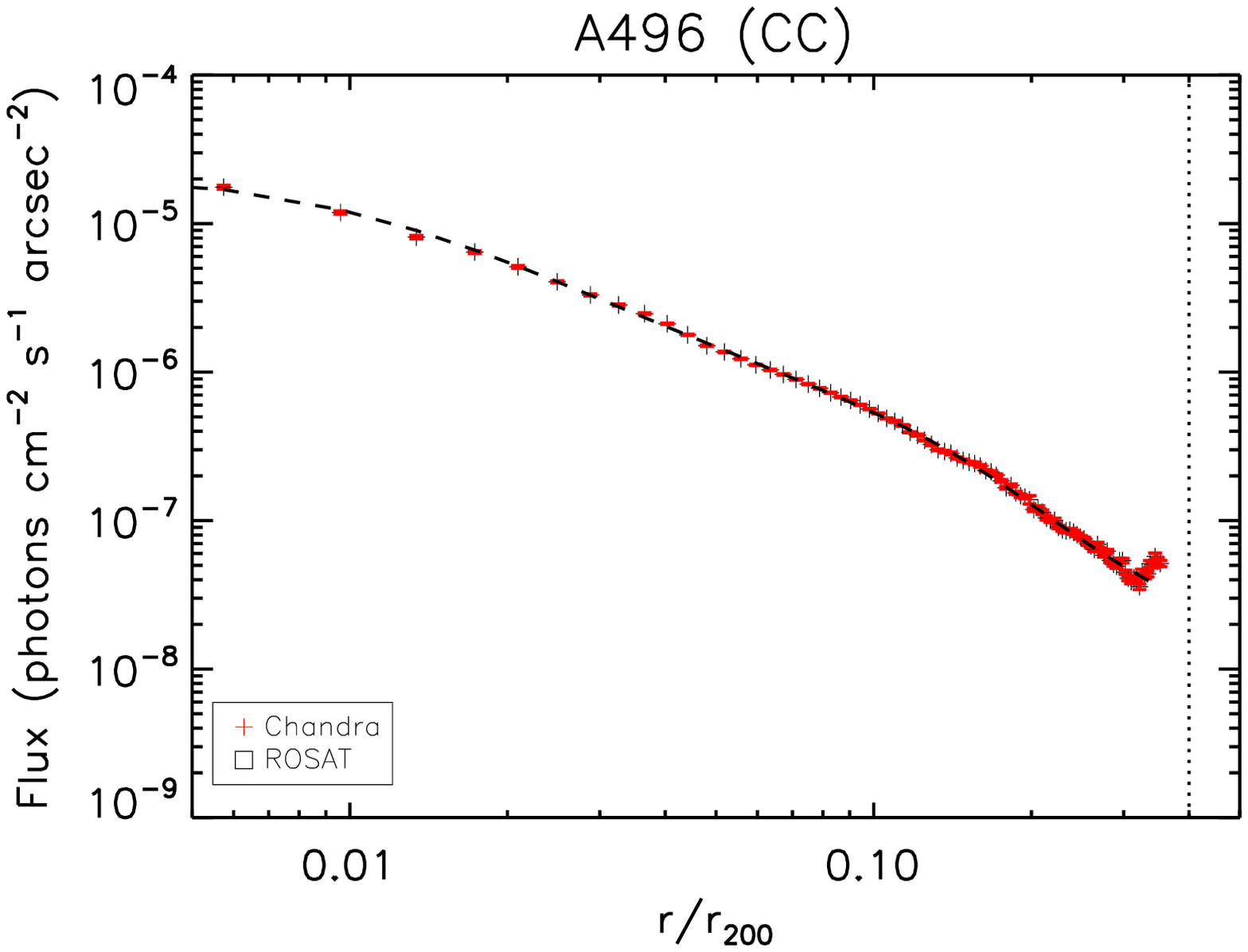}{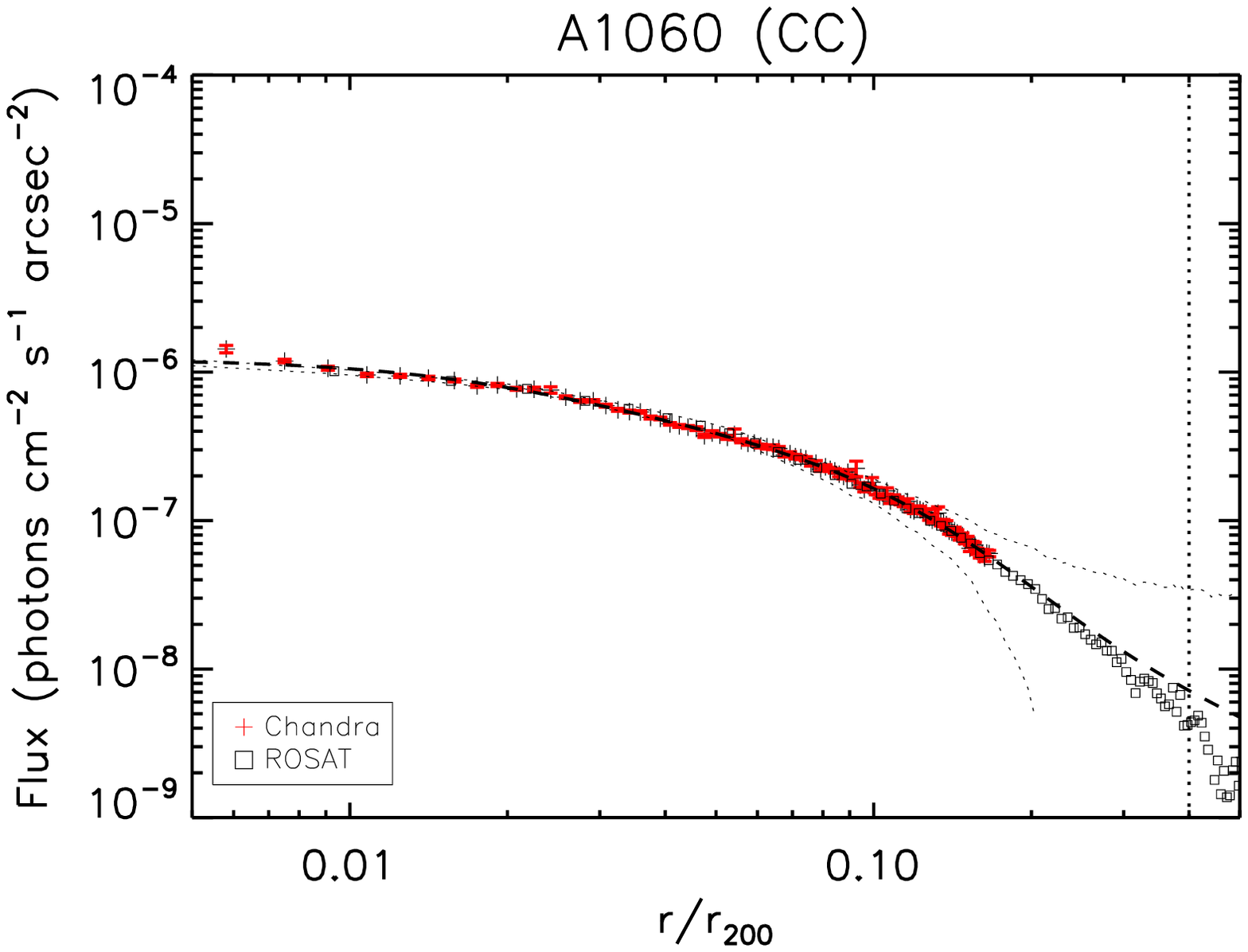}
\plottwo{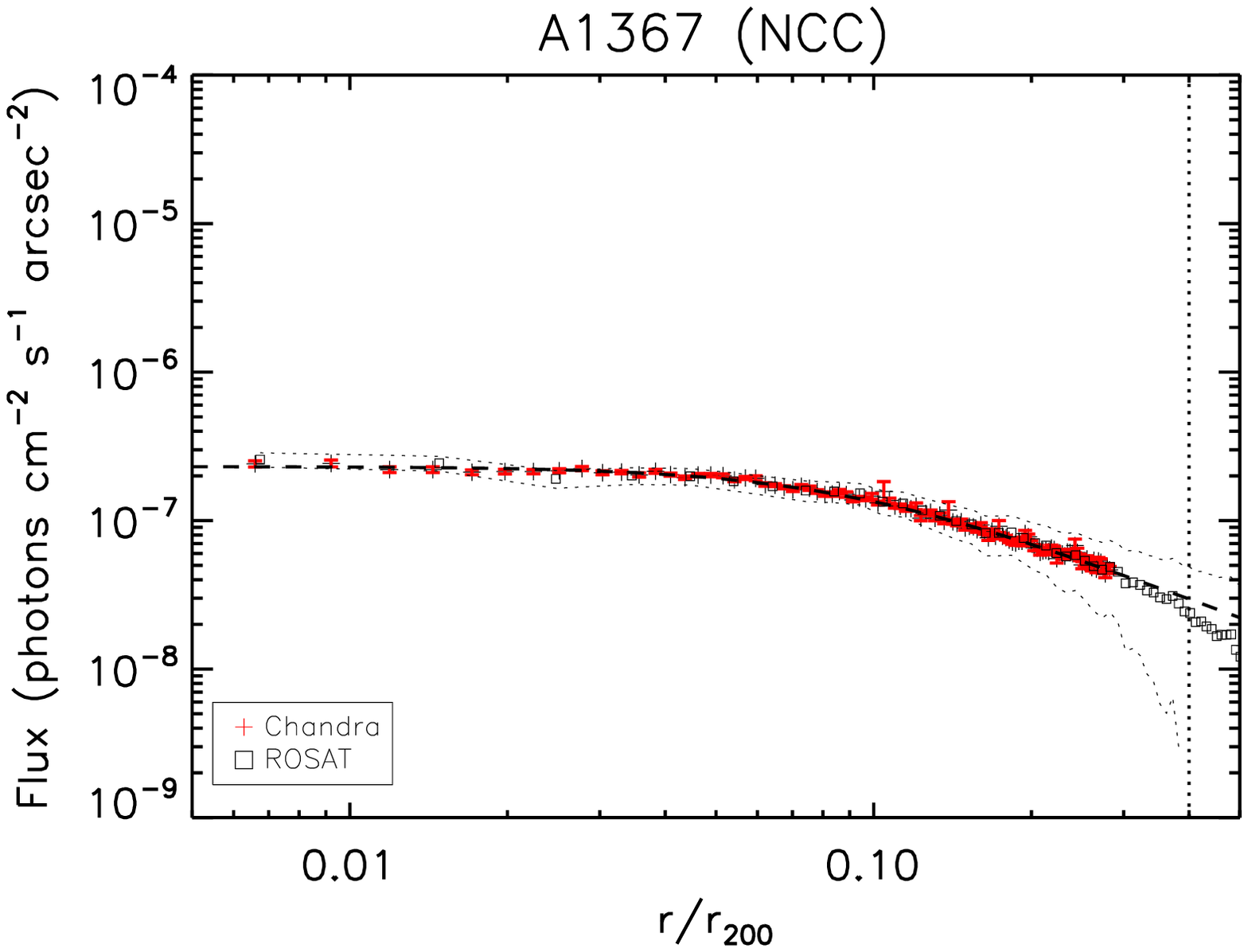}{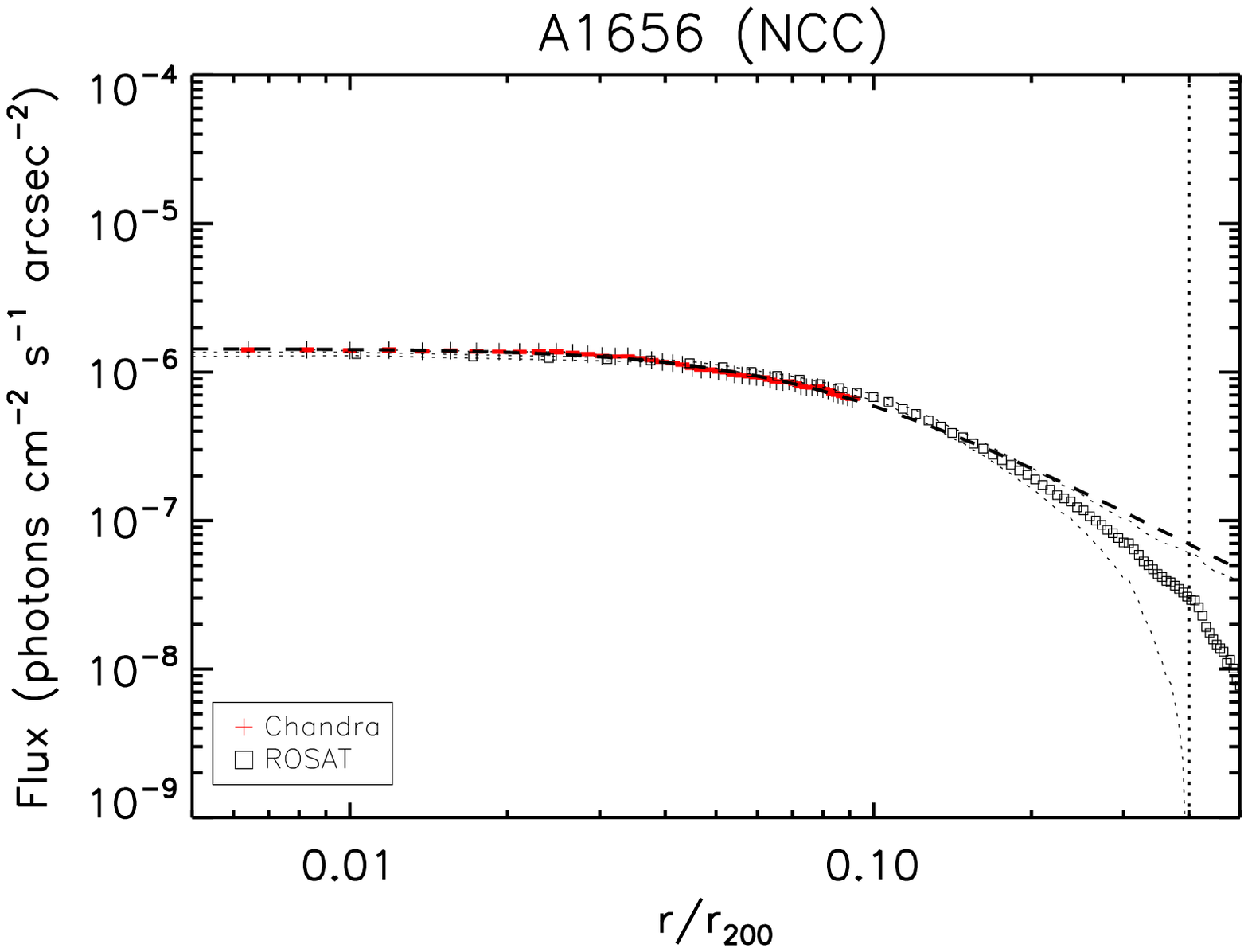}
\plottwo{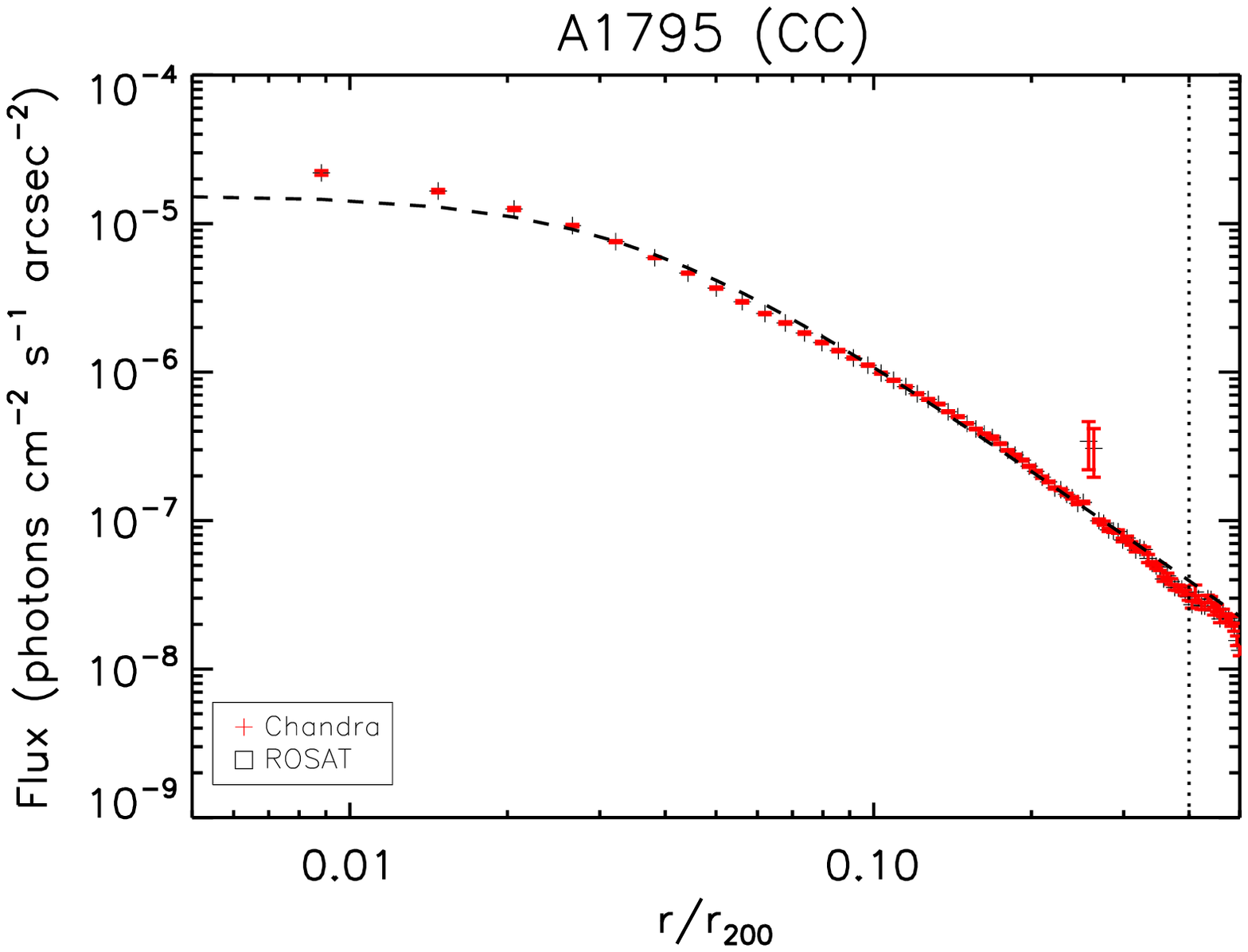}{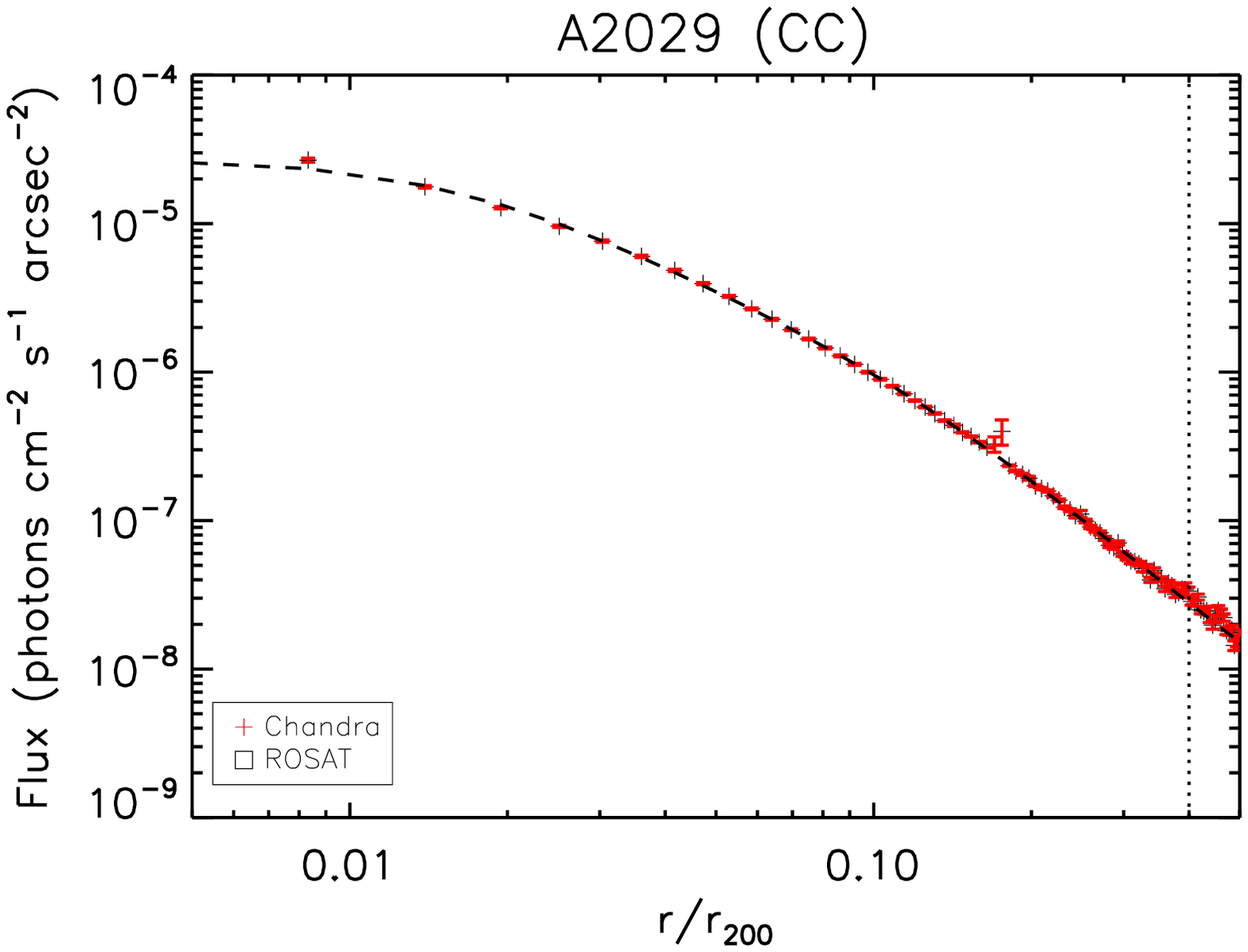}
\plottwo{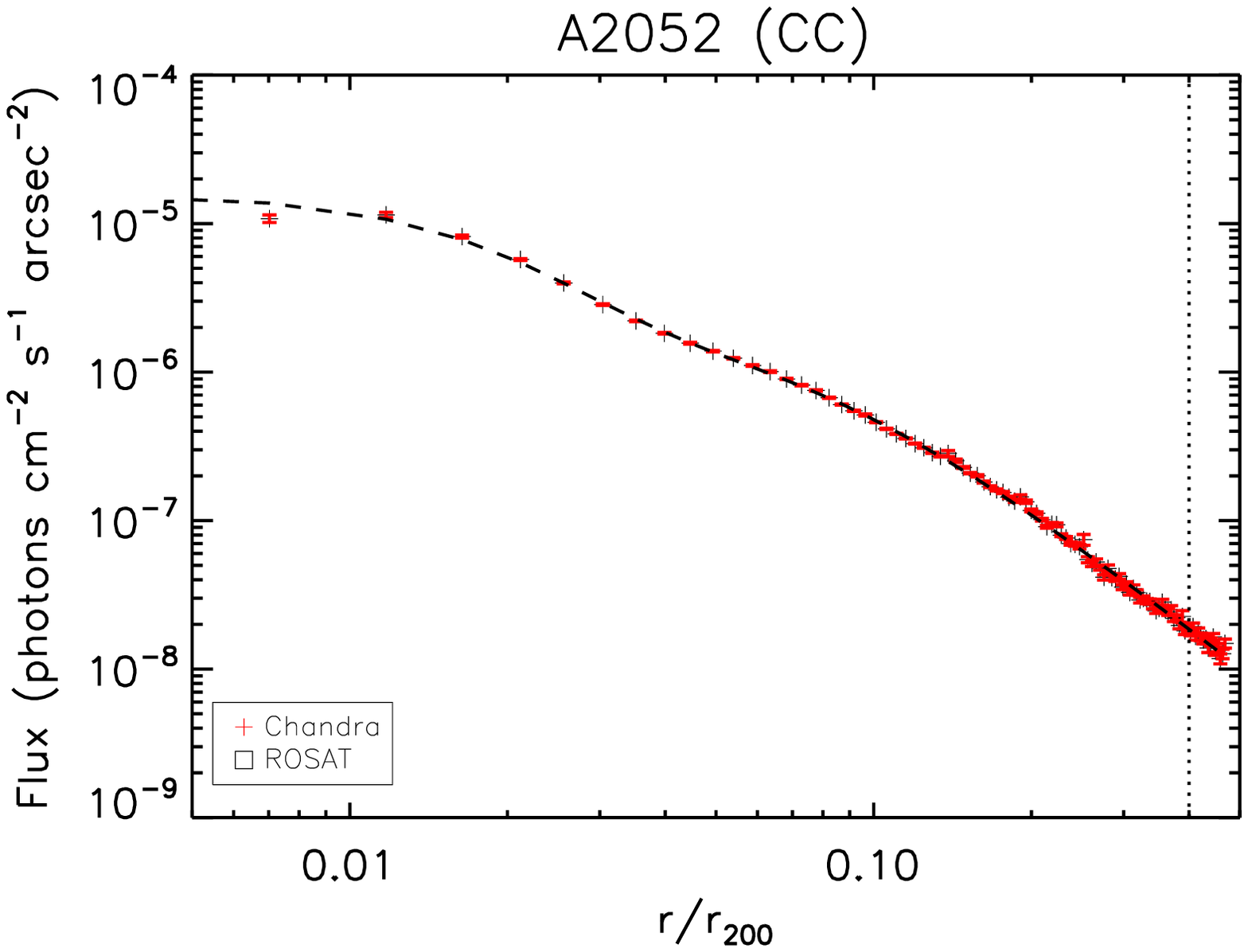}{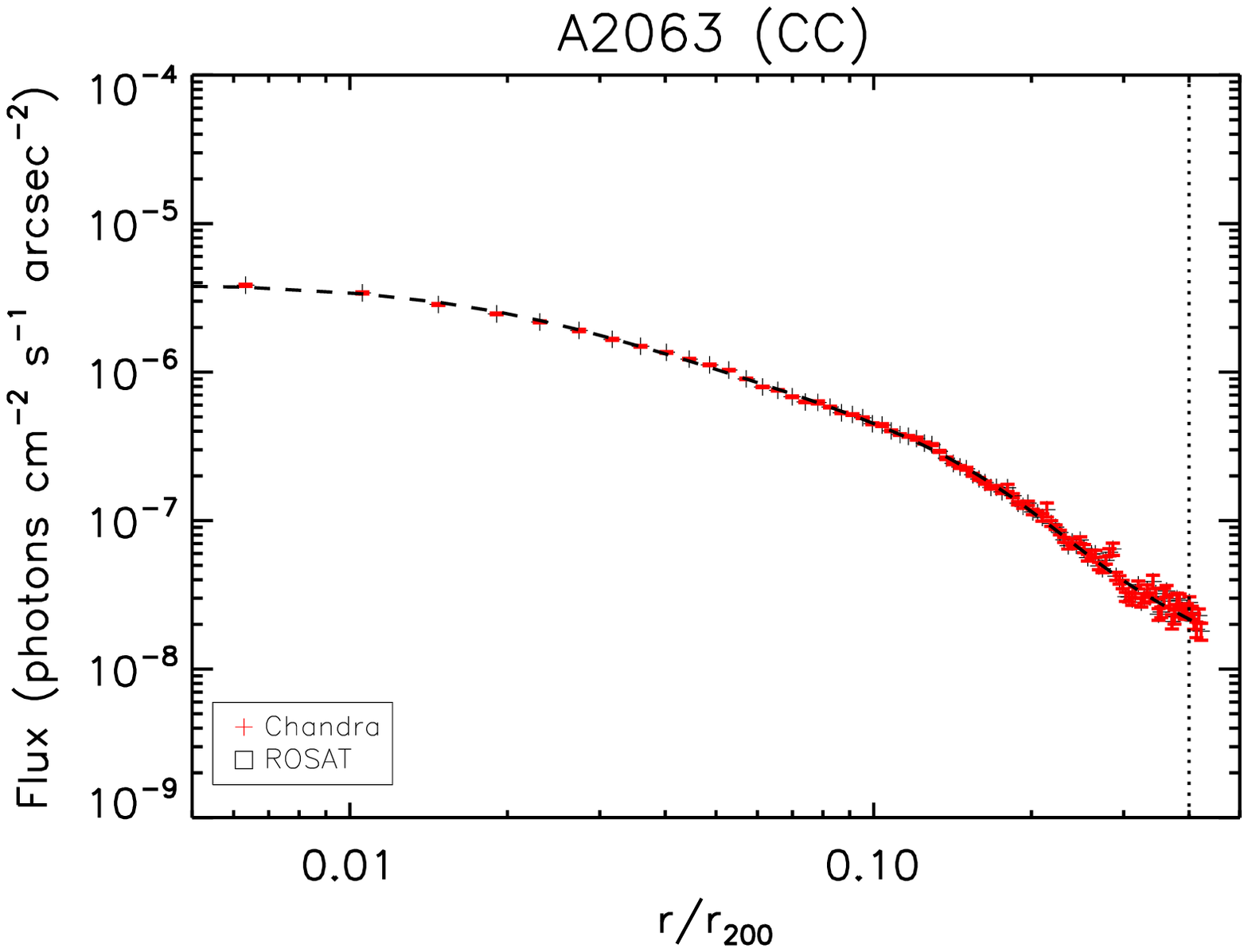}
\end{figure}

\begin{figure}[hbp]
\plottwo{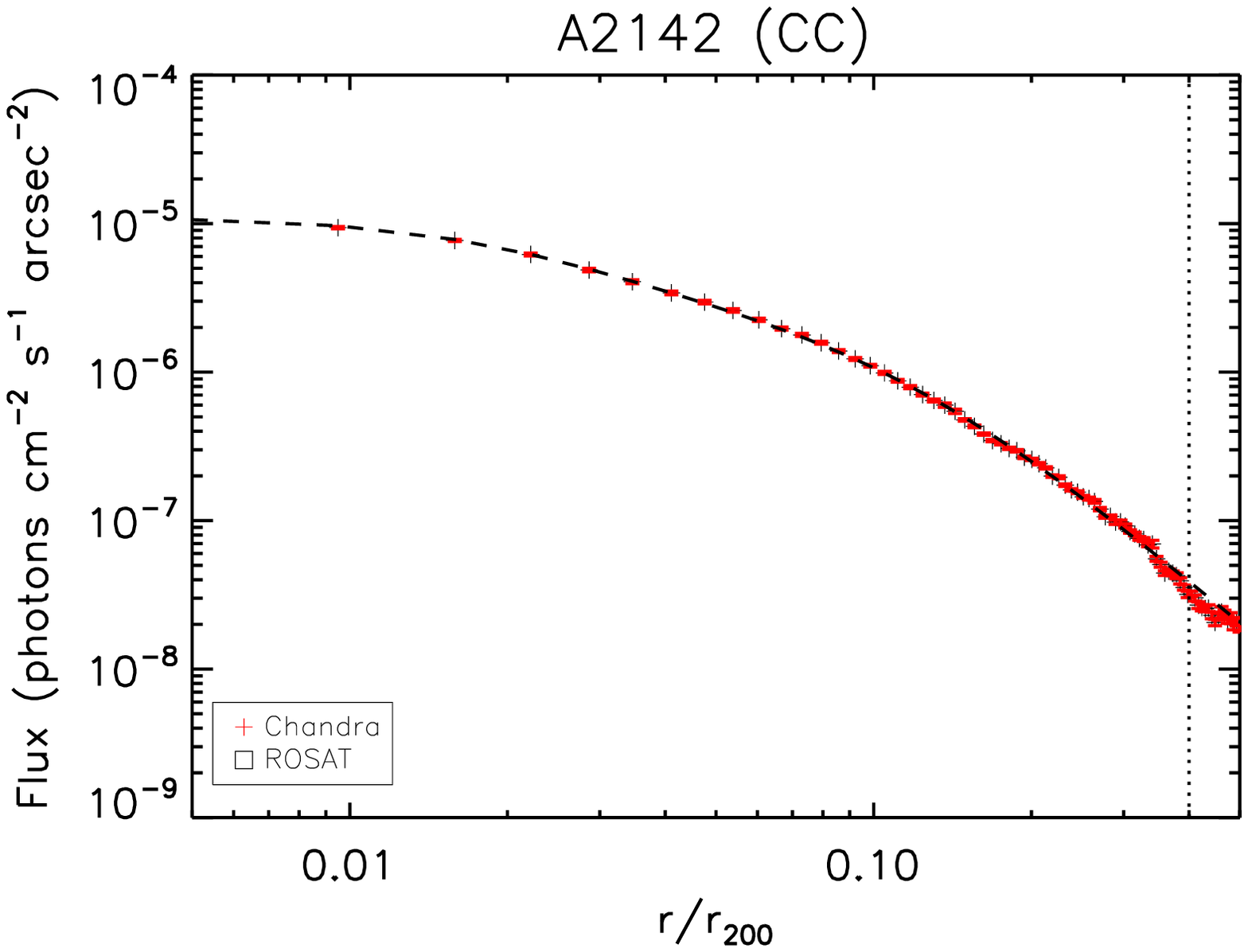}{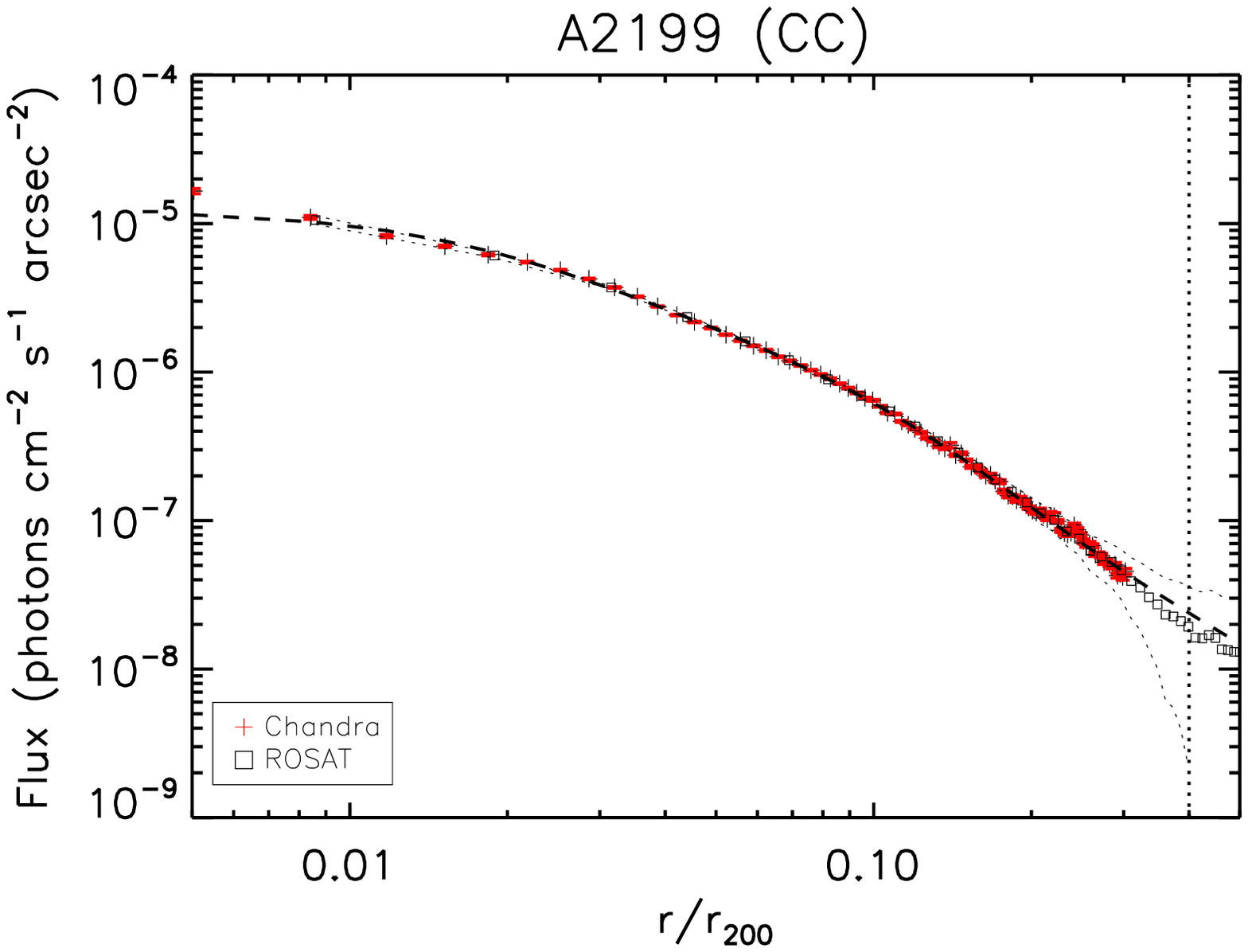}
\plottwo{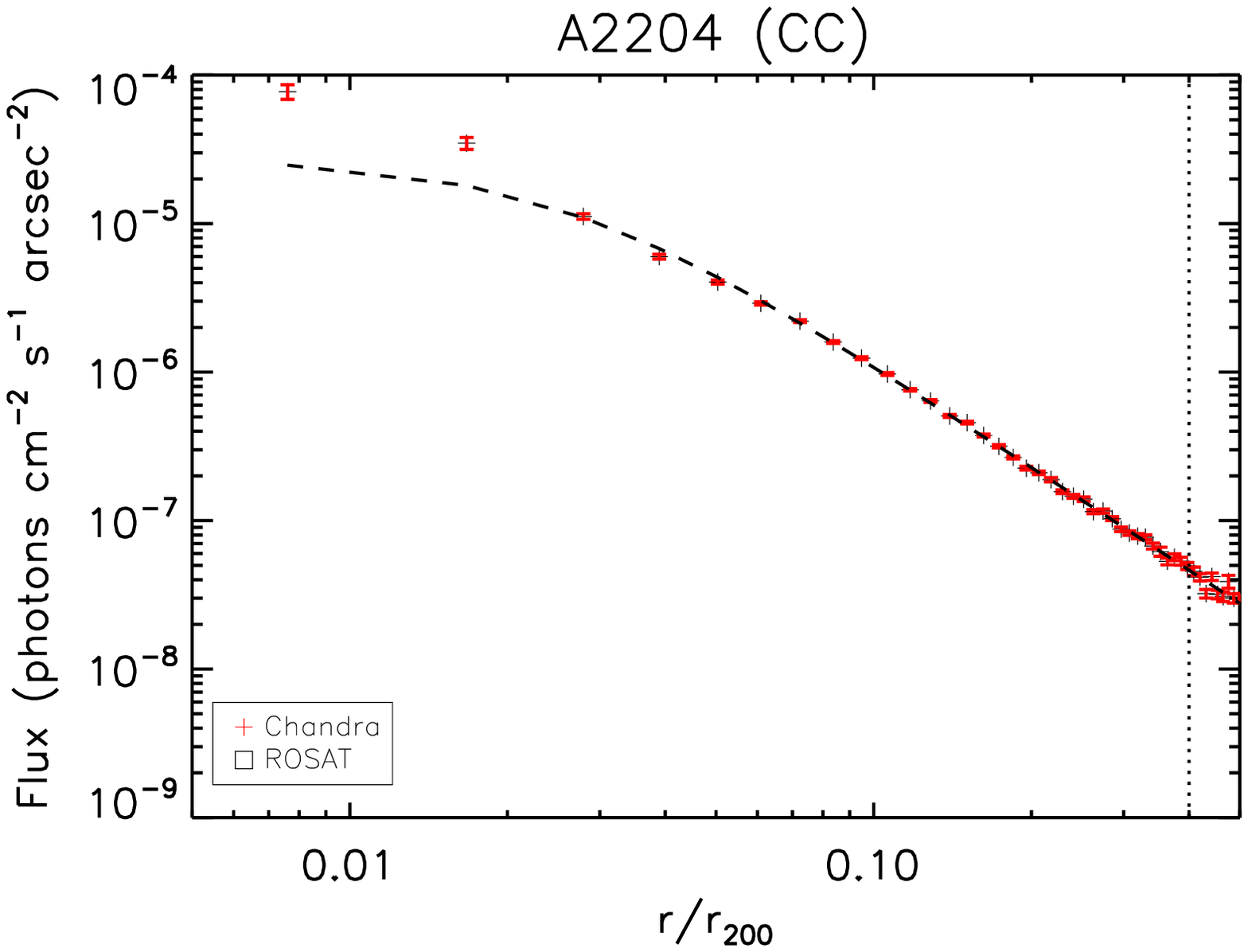}{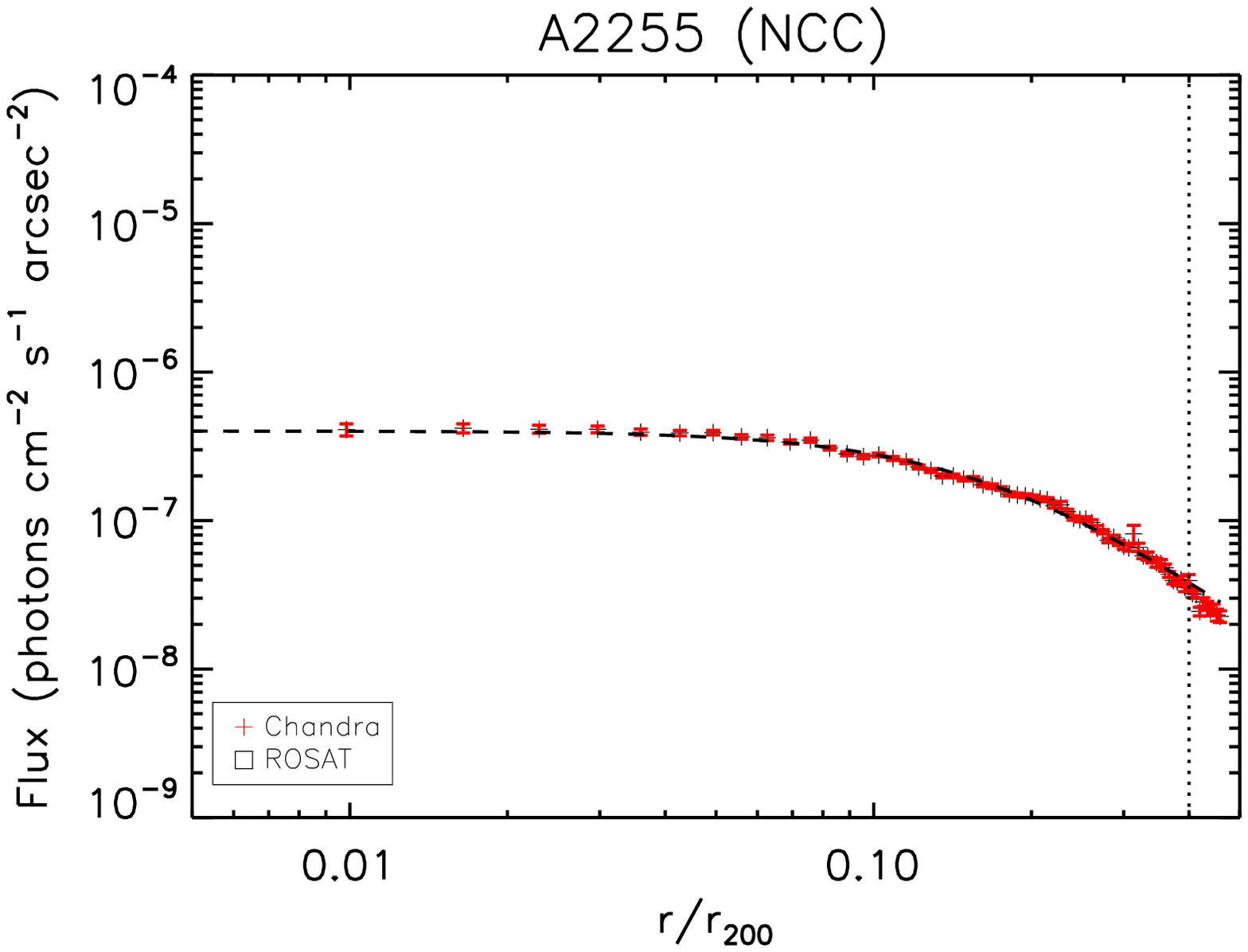}
\plottwo{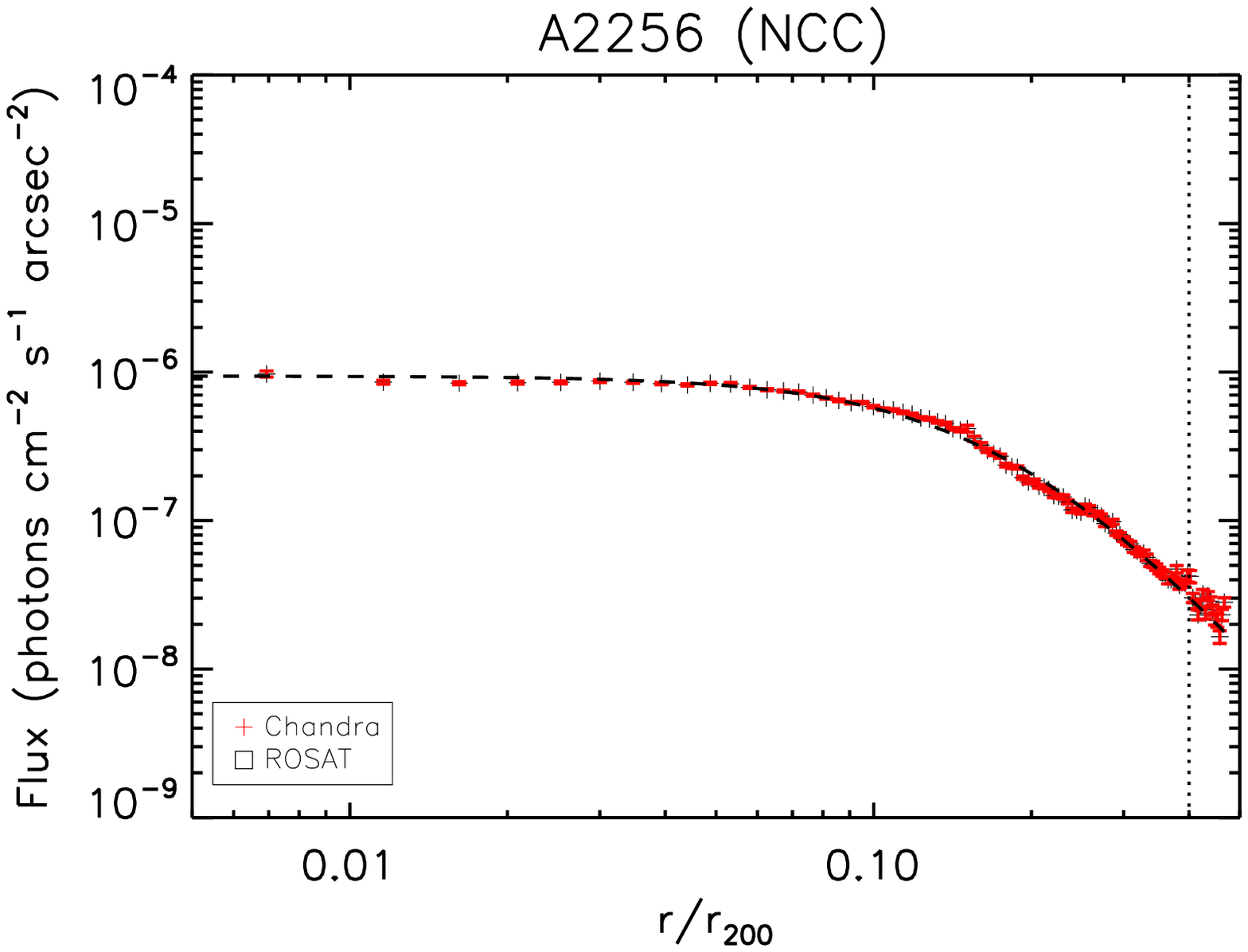}{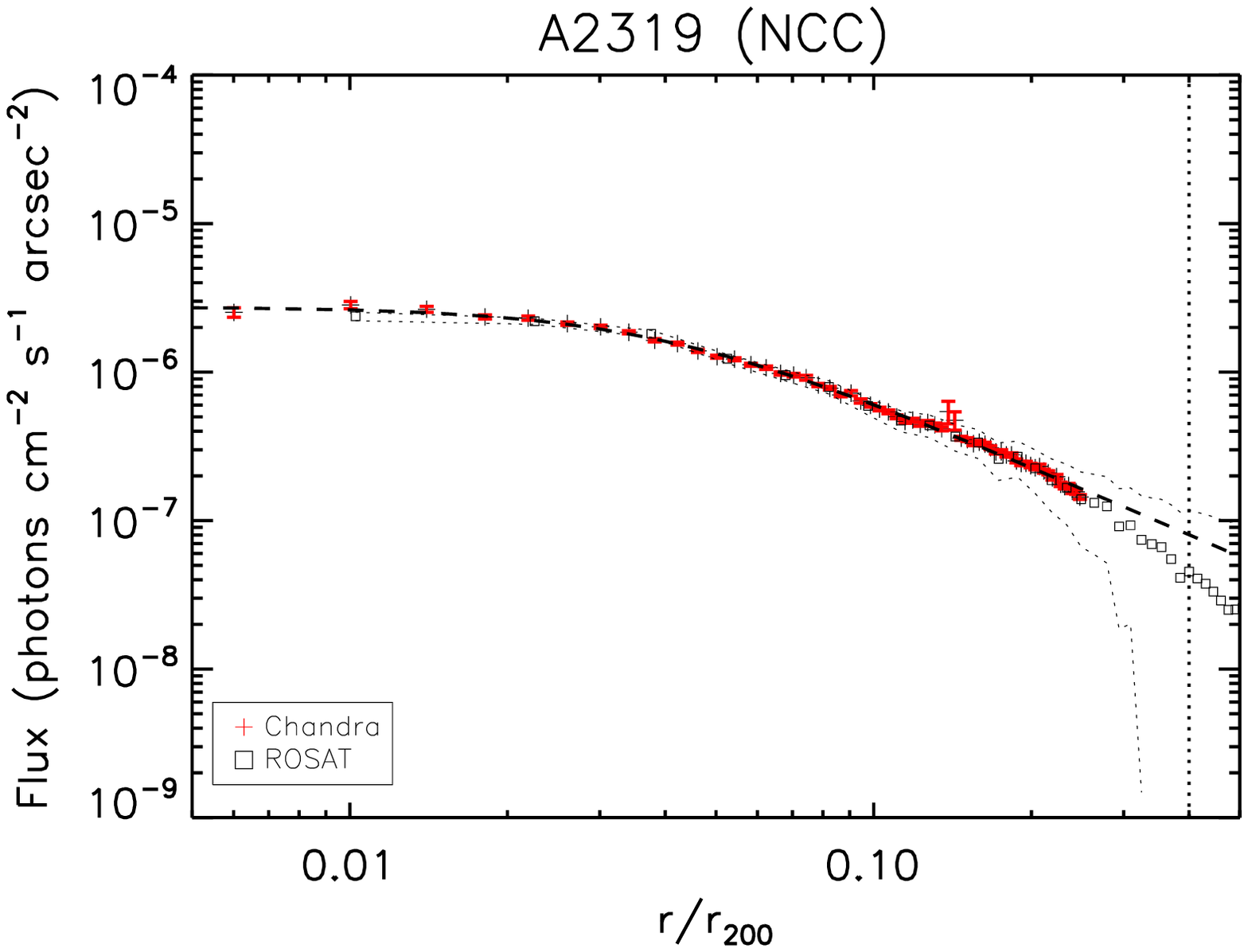}
\plottwo{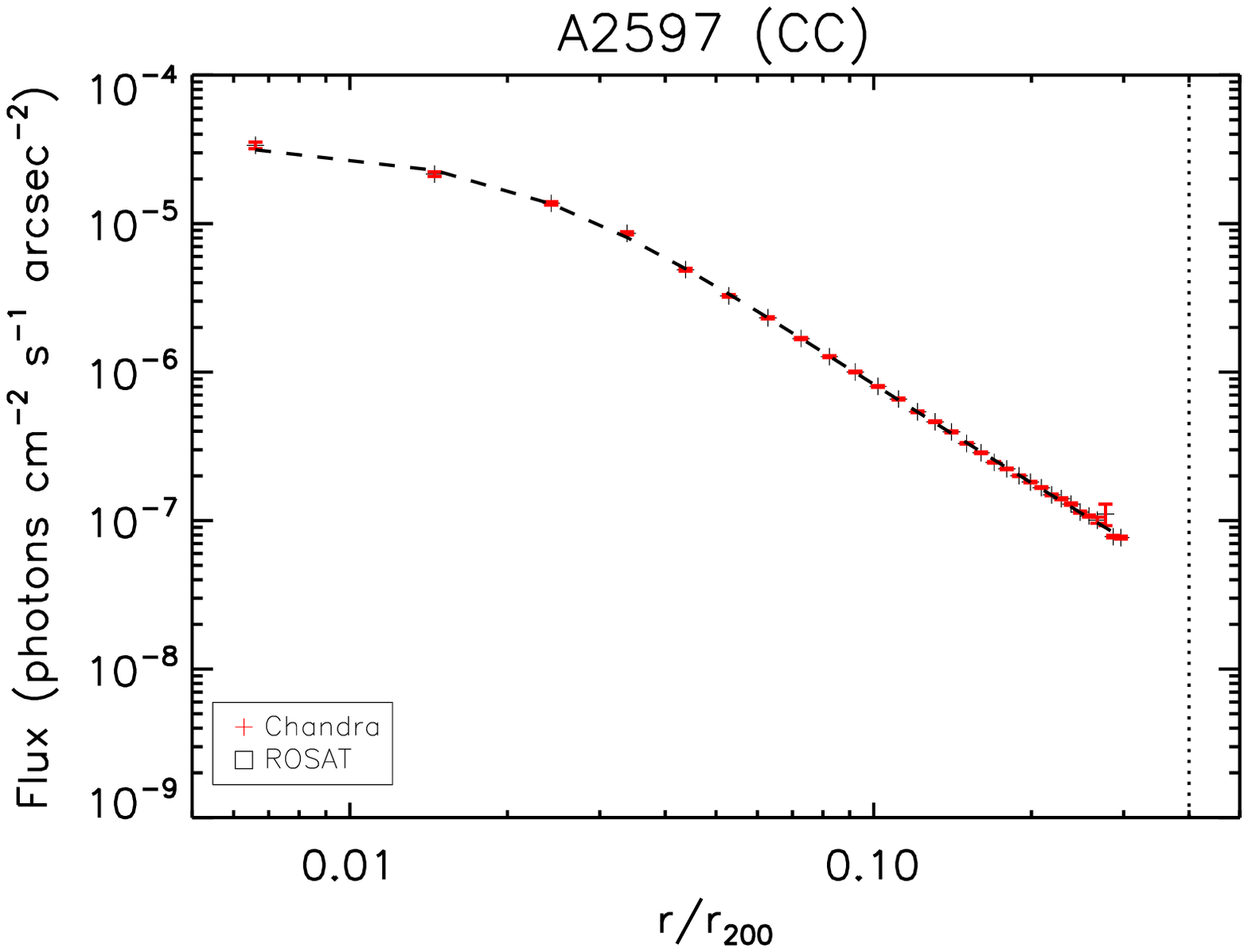}{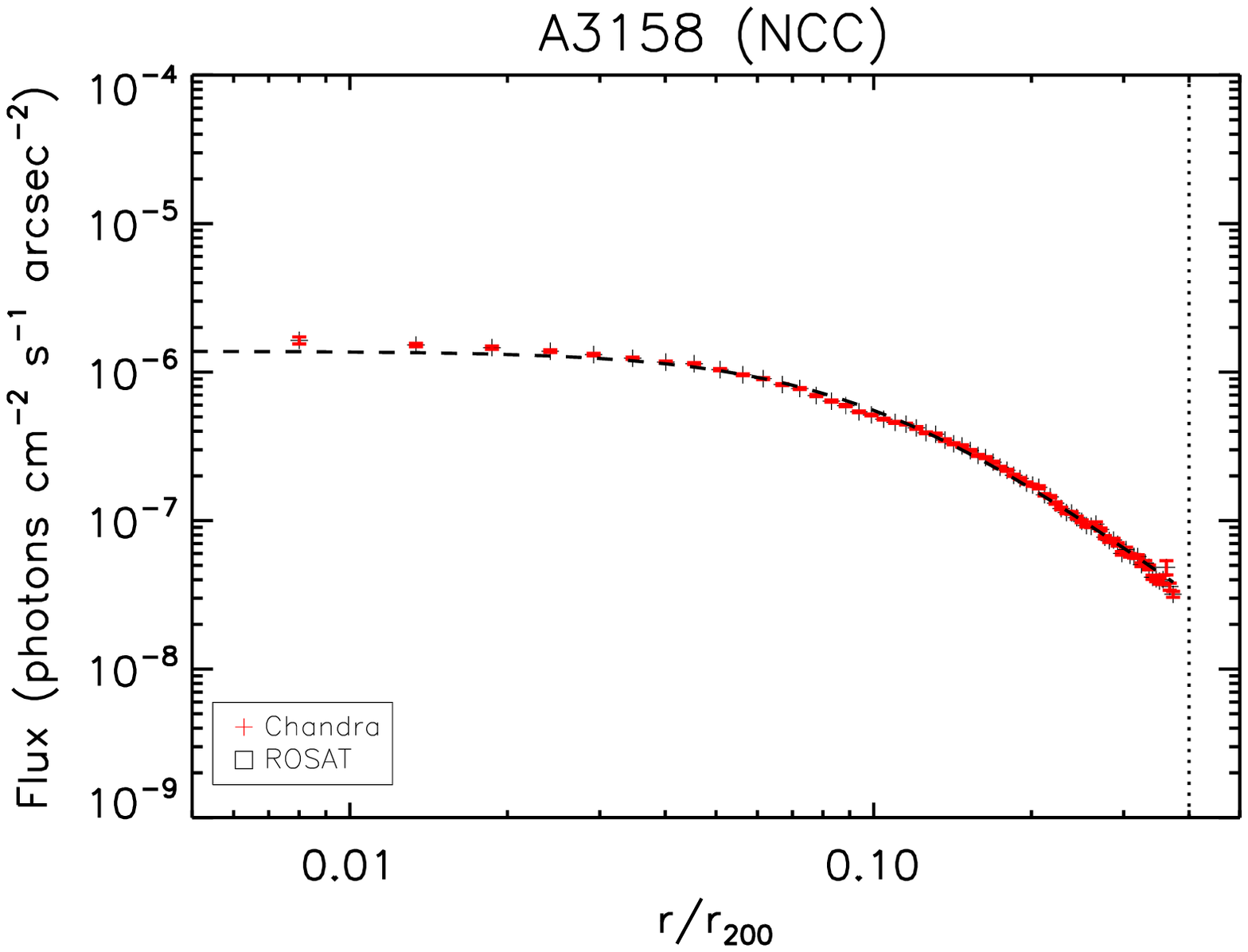}
\end{figure}

\begin{figure}[hbp]
\plottwo{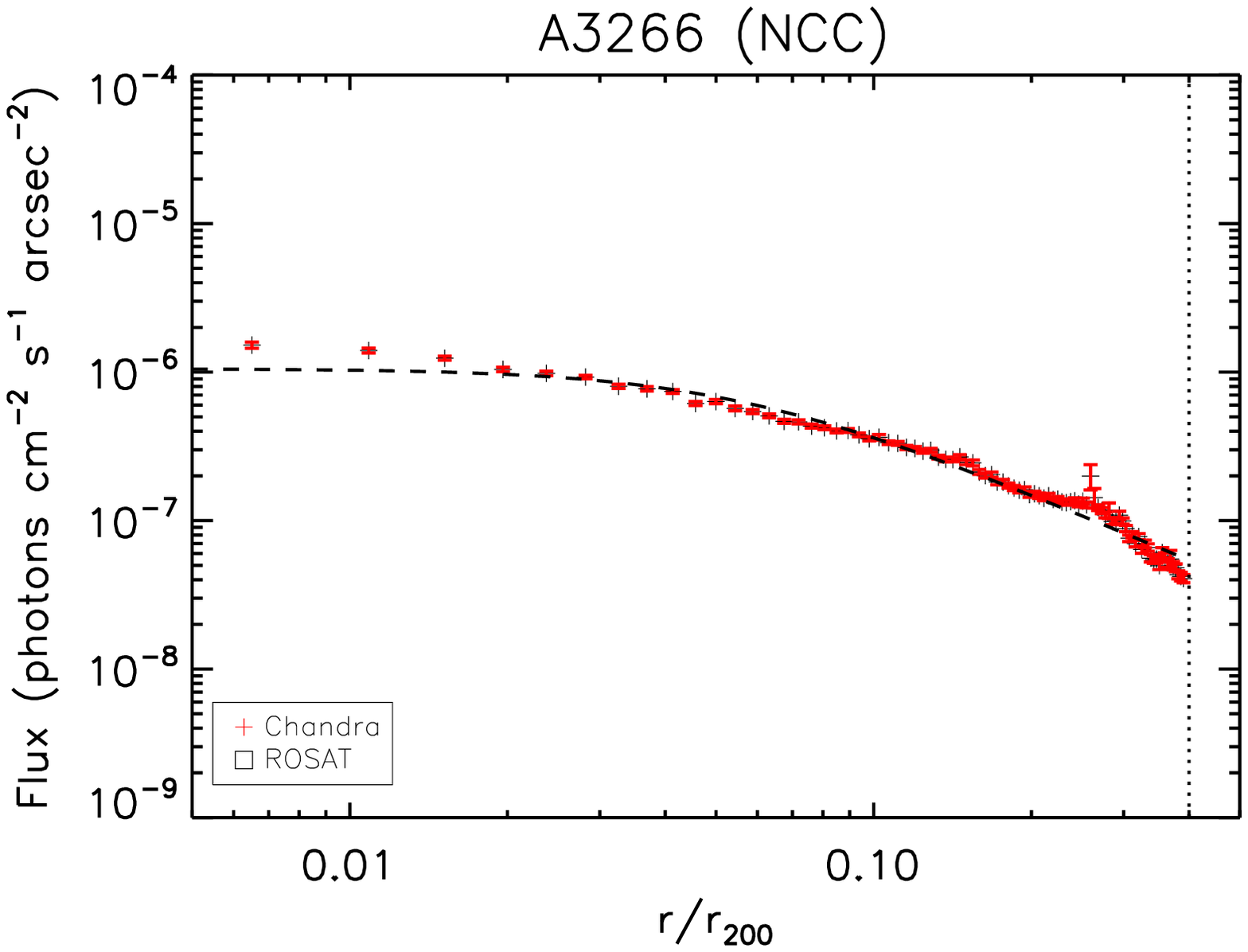}{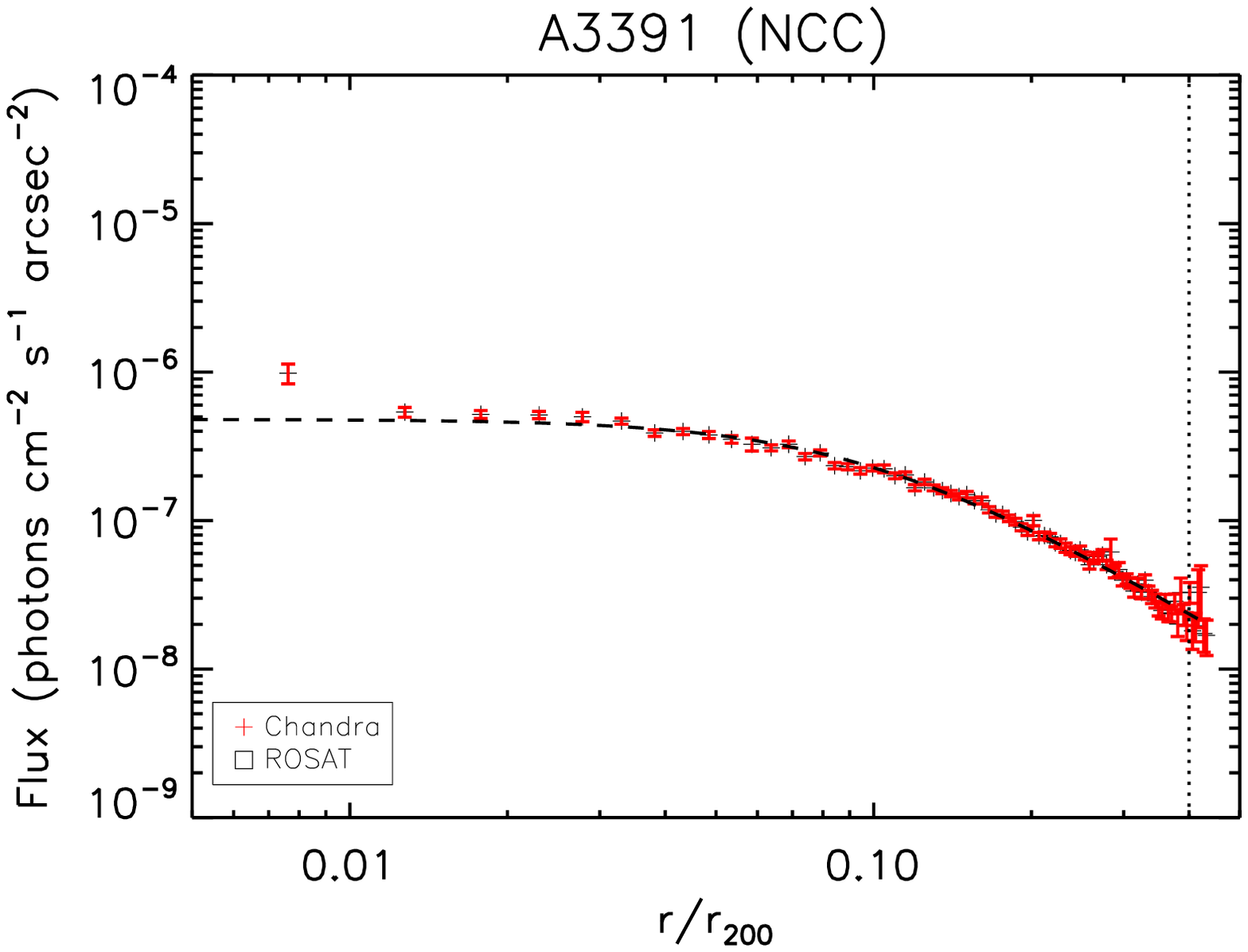}
\plottwo{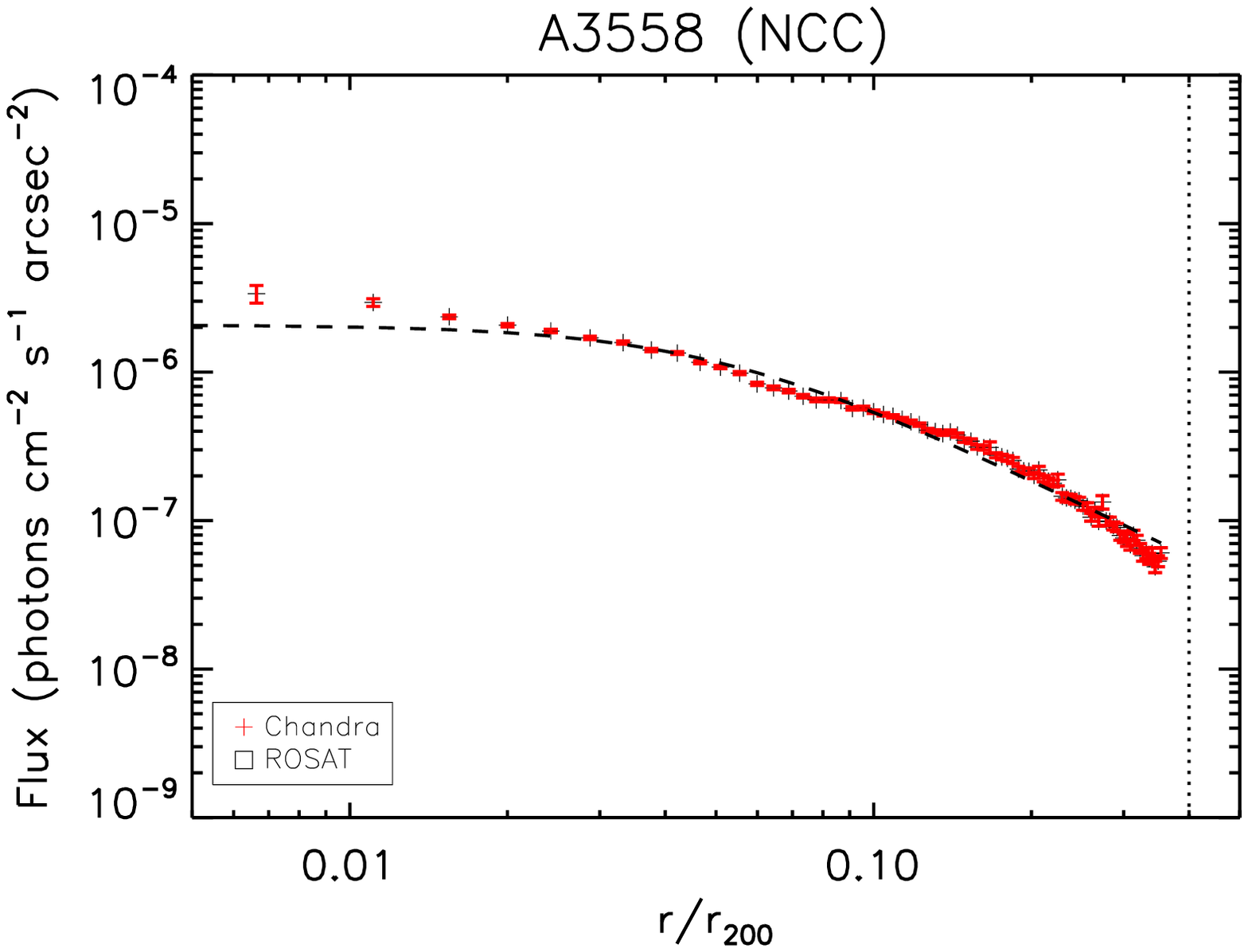}{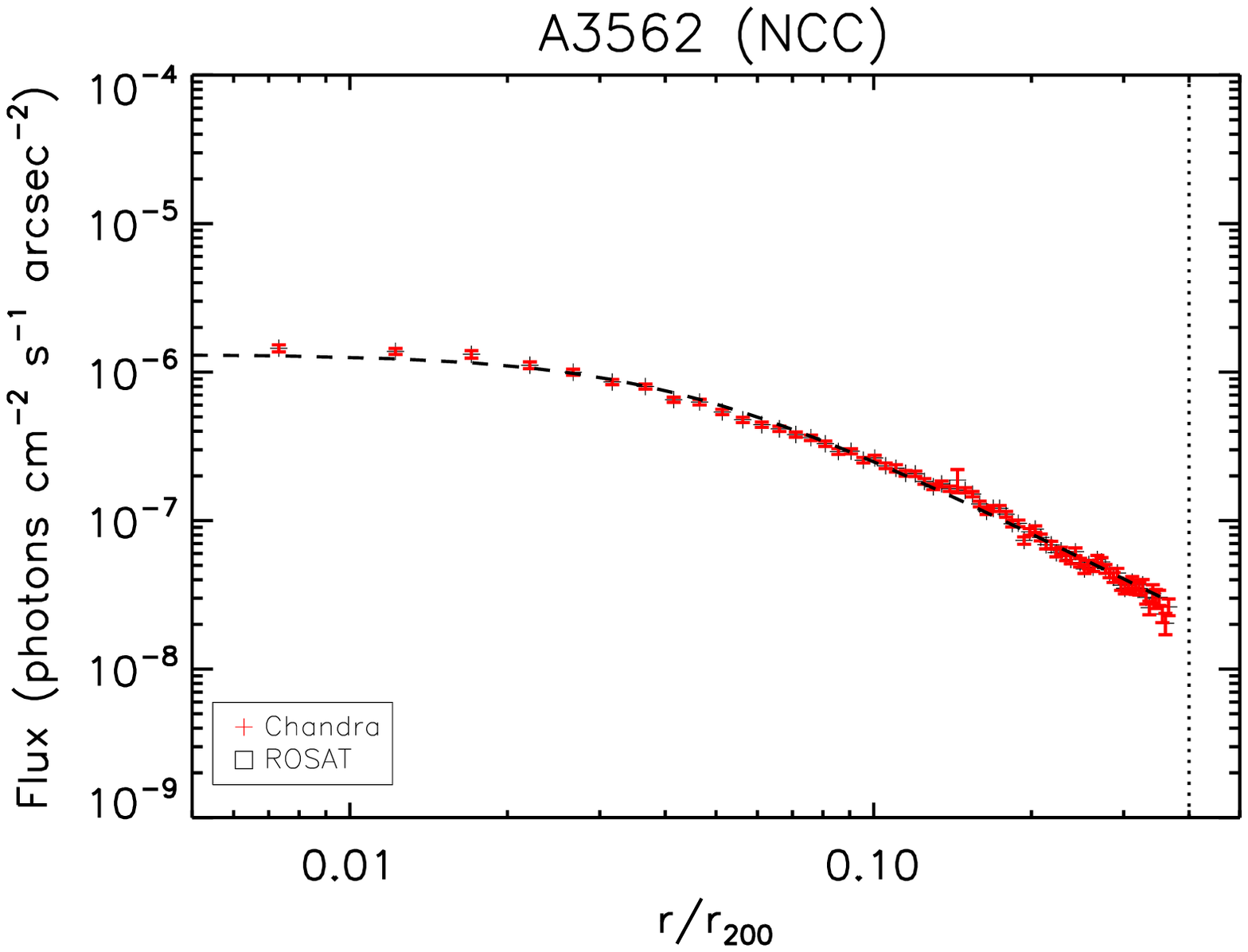}
\plottwo{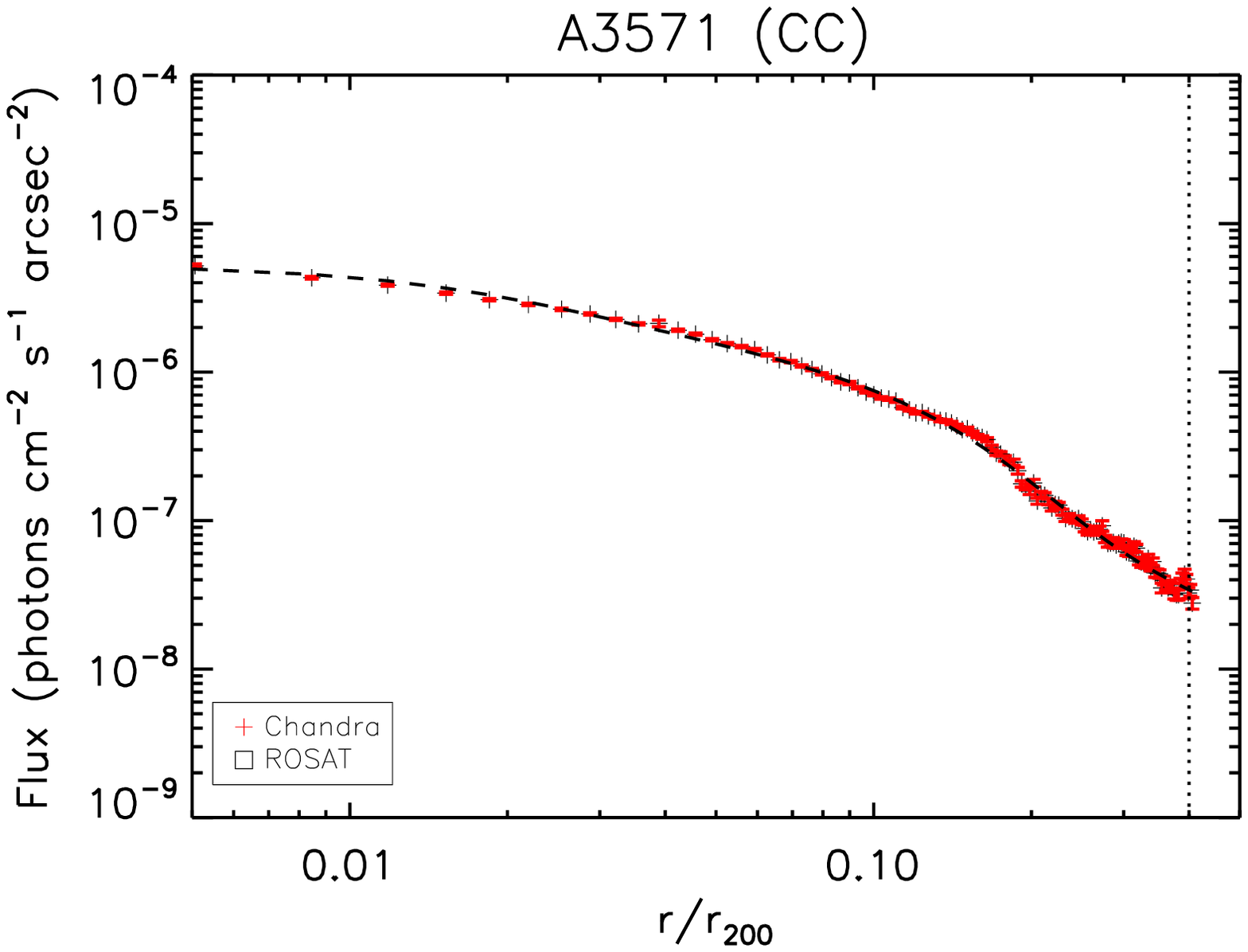}{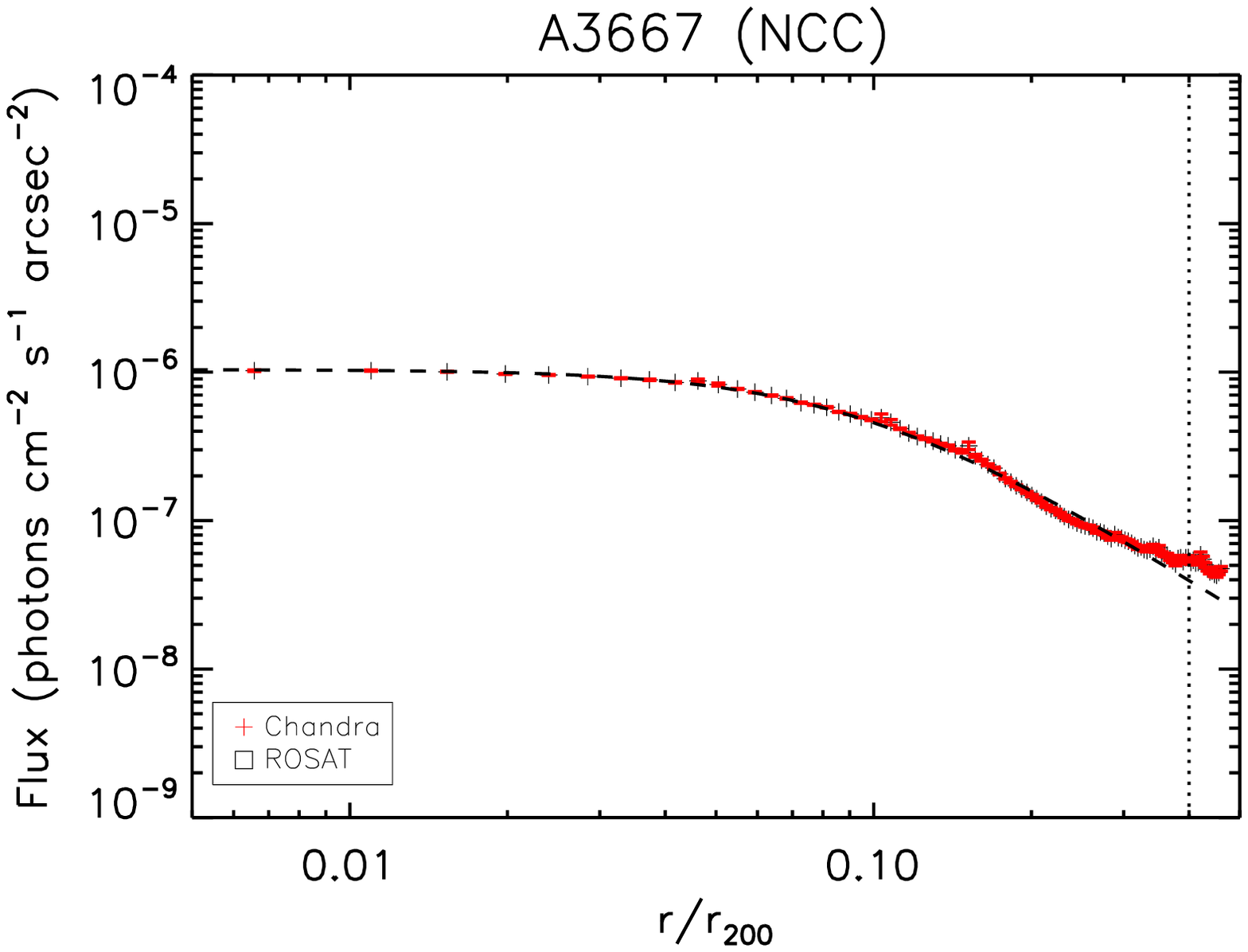}
\plottwo{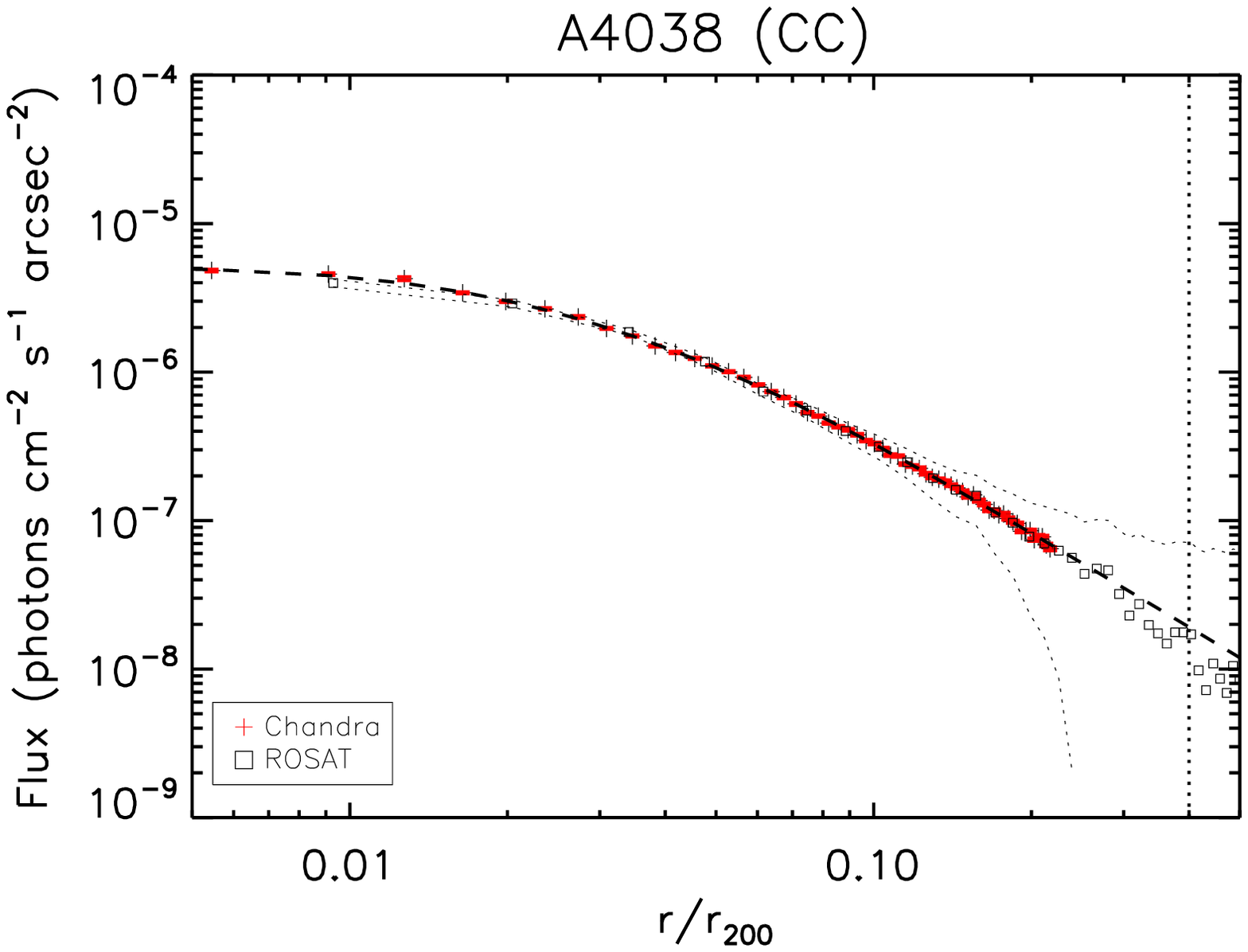}{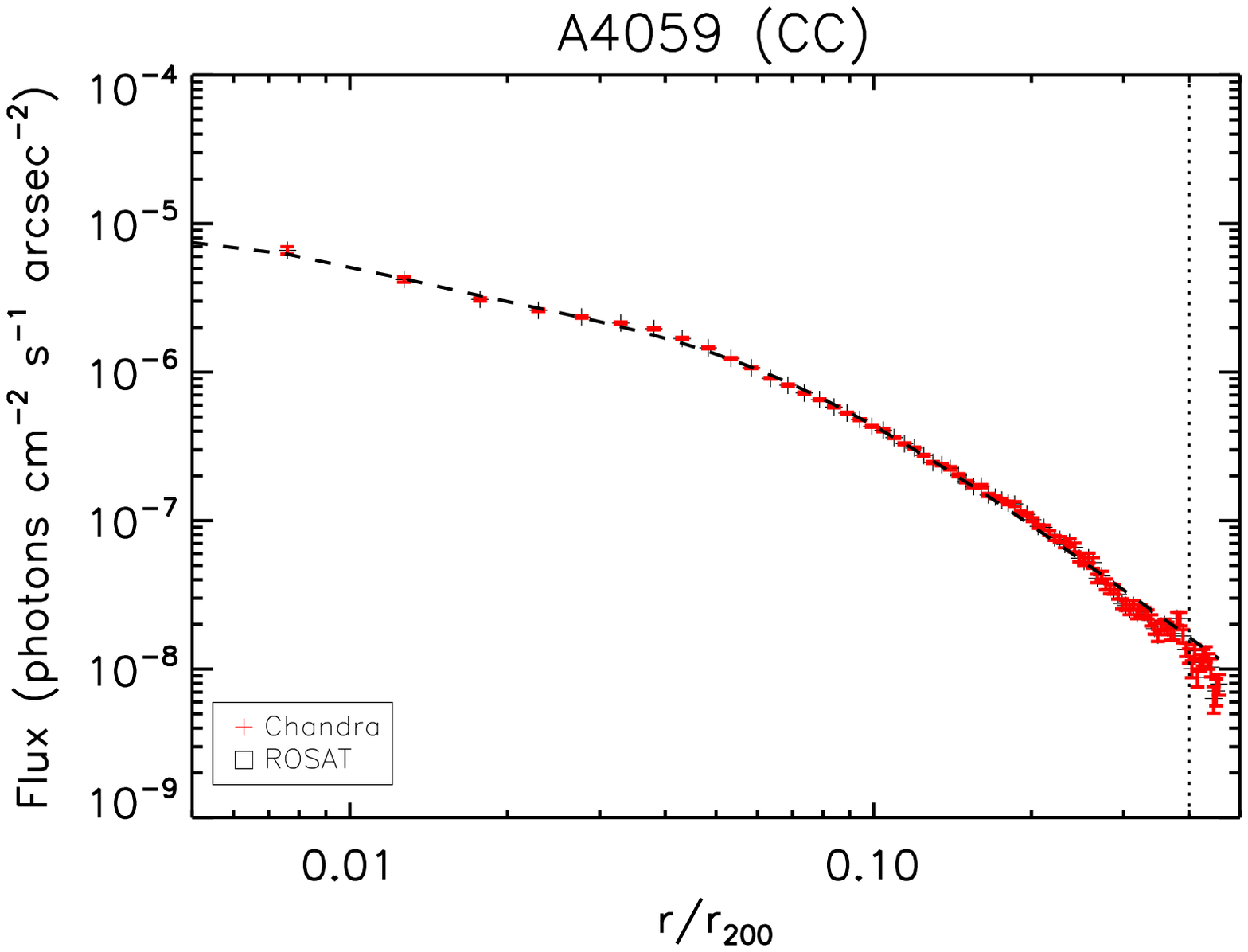}
\end{figure}

\subsection{Hardness Ratio Maps}\label{hr_maps}

Motivated by the predictions of simulations from B08, we looked for evidence of differences in cluster evolution between CC and NCC clusters by exploring the region beyond cool cores to $\sim 0.3\,r_{200}$.  The simulations indicate that CC clusters contain $\sim 40\%$ more cool gas in the region beyond $\sim 0.1\,r_{200}$ than NCC clusters.  One way to estimate the cluster temperatures in this region is with the use of hardness ratio maps.  A two-band hardness ratio map allows one to probe farther from the cluster core since it requires fewer counts than spectral temperatures for a given signal-to-noise.  The hardness ratio of a cluster's X-ray emission is also a proxy for temperature because a significant component of cluster X-ray emission is thermal bremsstrahlung radiation.  Recall that for thermal bremsstrahlung
\begin{equation}
S _x \propto \int{n_e^2\Lambda(E,T)dE} \propto \int_{E_1}^{E_2}{n_e^2T^{-1/2}\mbox{e}^{-E/kT}G(E,T)dE},
\end{equation}
where $S_x$ is the X-ray luminosity in the energy band $E_1-E_2$, $G(E,T)$ is the quantum correction Gaunt factor, and $n_e$ is the electron number density \citep[e.g][]{peacock_1999}.  From the temperature dependence of $S_x$, one sees that the hardness ratio $S_{x2}/S_{x1}$ is just related to the ratio of temperatures $T_2$ and $T_1$, which after integration is roughly proportional to $(T_2/T_1)^{1/2}$.  Thus, a higher hardness ratio corresponds to harder X-ray emission, which in turn implies a higher temperature.

\begin{figure}[h]
\epsscale{1.0}
\plotone{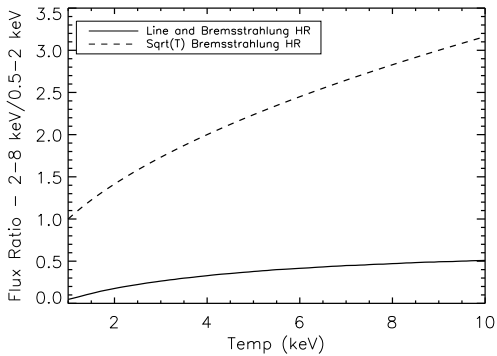}
\caption{Hardness ratio (2.0-8.0 keV / 0.5-2.0 keV) for modeled clusters of temperatures $1 - 10$ keV, including thermal bremsstrahlung and line emission.  Also plotted is the expected hardness ratio including only thermal bremsstrahlung.  The hardness ratio clearly increases with temperature, but not as quickly as expected for pure bremsstrahlung emission.}
\label{fig_total_emission}
\end{figure}

The X-ray spectrum of a cluster, however, has other contributing factors besides thermal bremsstrahlung continuum emission.  Line emission is an important spectral component, especially if the cluster is relatively cool.  To demonstrate that the hardness ratio is in fact proportional to cluster temperature, we used the X-ray spectral fitting package XSPEC to model spectra of clusters of increasing temperatures from 1 to 10 keV.  Metallically was assumed to be $0.3\,Z_{_{\bigodot}}$.  The modeled spectra were broken into two standard \textit{Chandra} bands, $0.5-2.0$ keV and $2.0-8.0$ keV, the photon flux in these bands was integrated, and the ratio of hard band to soft band was calculated.  The resulting temperature dependence of the hardness ratios of the modeled spectra is plotted in Figure \ref{fig_total_emission}.  The temperature itself has little dependence on metallicity; we found only a few percent change in hardness ratio calculated in this way when varying the abundance between 0.3 and $0.7\,Z_{_{\bigodot}}$.  The square root dependence of the hardness ratio on the temperature for a bremsstrahlung model is also plotted.  It is clear that hardness ratio does increase with temperature, but not as fast as pure bremsstrahlung emission predicts.  

The hardness ratio images for the observed cluster subsample were created by dividing two \textit{Chandra} images, the 2.0-8.0 keV bandpass image by the 0.5-2.0 keV bandpass image.  These images are counts-based and reduced as described previously.  They lack corrections for the effective area, quantum efficiency, pointing deviations and exposure durations, which are corrected by other means described below.

First, the actual exposure time is inconsequential as each bandpass image was drawn from the same observation and share the same exposure duration.  As pointing was corrected in each event when recorded, the exposure map accounts for the changes of effective area as the telescope was dithered across the sky.  This means that in theory each pixel could have a different effective area.  We took the hardness ratio maps, however, out to only $\sim 0.3\,r_{200}$ ($\sim500$ kpc), a range over which the effective area of the detector is reasonably flat and where there are sufficient counts when the cluster is centered in the field of view.  

We next accounted for the absolute difference between the energy bands.  The difference in quantum efficiency and effective area between the two energy bands was removed by normalizing each to the idealized spectrum of the cluster.  This idealized spectrum was found by folding the instrument response into the expected spectrum of the cluster at the temperature given by Equation \ref{eq_chen_M_T_relation}, an $N_H$ absorption model, and an APEC plasma model with 0.3 $Z_{_{\bigodot}}$.  Using WebSpec\footnote{http://heasarc.gsfc.nasa.gov/webspec/webspec.html}, an internet interface version of XSPEC, we combined the instrument response with the APEC model spectrum, providing the expected count rate in each energy band for gas at the $M-T$ average temperature from Equation \ref{eq_chen_M_T_relation}.  The ratio of the two fluxes gives the hardness ratio of gas at this average temperature; each cluster's hardness ratio map was then normalized to this ratio.  Thus, if we assume the average temperature is in fact the virial temperature for the cluster, a hardness ratio of unity would correspond to a pixel at the virial temperature; less than one would be cooler, and greater than one would be warmer.  

\begin{figure}[h]
\epsscale{1.0}
\plotone{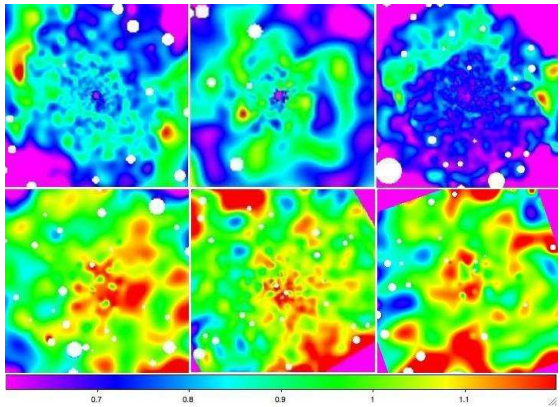}
\caption{Six observational hardness ratio maps.  Each is the two \textit{Chandra} images, 2-8 keV/0.5-2 keV, normalized by the hardness ratio corresponding to the cluster virial temperature.  White circles are locations of removed point sources.  Each image is cut to a radial metric field-of-view of $0.3\,r_{200}$.  (Top)  CC clusters A478, A2204, A2029.  (Bottom)  NCC clusters A2255, A3158, and A3391.  Note the clear difference in normalized hardness ratio between CC and NCC clusters.}
\label{Chandra_HR_examples}
\end{figure}

Before division, we adaptively smoothed the images with a gaussian kernel using the CIAO tool \textbf{csmooth}.  The gaussian size scales at each pixel using a Poisson expectation of noise to create an image with a signal-to-noise ratio of 3 in each pixel within the 0.5-2.0 keV band.  The same smoothing kernel was applied to the high energy band, the two images were divided, and the resulting hardness ratio image was normalized to the average cluster temperature using the WebSpec expected flux ratio.  The final image gives an approximate visual representation of the temperature across the cluster with respect to the virial temperature.  This is directly comparable to the hardness ratio maps created from the simulations, which are also normalized to their virial temperatures.  Figure \ref{Chandra_HR_examples} shows examples of hardness ratio maps for six observed clusters.

As a check on this process and in order to probe the histories of CC and NCC clusters, we examined clusters paired as close as possible in mass.  This also translates into similar virial temperatures according to the mass-temperature relation (Figure \ref{fig_temp_mass}).  In this way, observed differences in the clusters cannot be attributed to different cluster masses.  Matching mass to within $10\%$, we found three pairs of observed CC and NCC clusters from the subset of our sample for which hardness ratio maps were produced (see below for more details).  One pair included A2204 and so was disregarded.  The remaining pairs, along with one more unmatched pair for illustration, and their bulk temperature fits from \S \ref{bulk_temperatures} for the region in question are found in Table \ref{table_bulk_temperatures}.

Not every cluster has sufficient \textit{Chandra} coverage to calculate hardness ratios out to our standard $0.3\,r_{200}$.  As a result, we used only 8 CC and 10 NCC clusters in the hardness ratio data analysis.  We also excluded Abell 3571 and 2597 from the analysis even though they have sufficient coverage.  While both we and \citet{ohara_2006} label A3571 as a CC cluster, \citet{chen_2007} lists its cooling time and mass inflow rate as $0.84\times10^{10}$yr and 35 M$_{\odot}$/yr respectively.  The cooling time is high and the mass inflow rate is low compared to other CC clusters.  It also falls as an extreme outlier compared to the other CC clusters' HR values in our analysis.  It is our belief that A3571 is best described as an ``in-between'' cluster, not an NCC, but not fully a CC either.  Thus, we disregarded it for the HR analysis so as not to significantly skew the CC HR mean.  A2597 is also a peculiar CC outlier, with a rapidly increasing hardness ratio profile, which matches neither the typical CC nor the typical NCC result.  Again, to avoid confusion from potentially including an ``in-between" cluster, A2597 was removed from the hardness ratio analysis.

\begin{figure}[h]
\epsscale{0.75}
\plotone{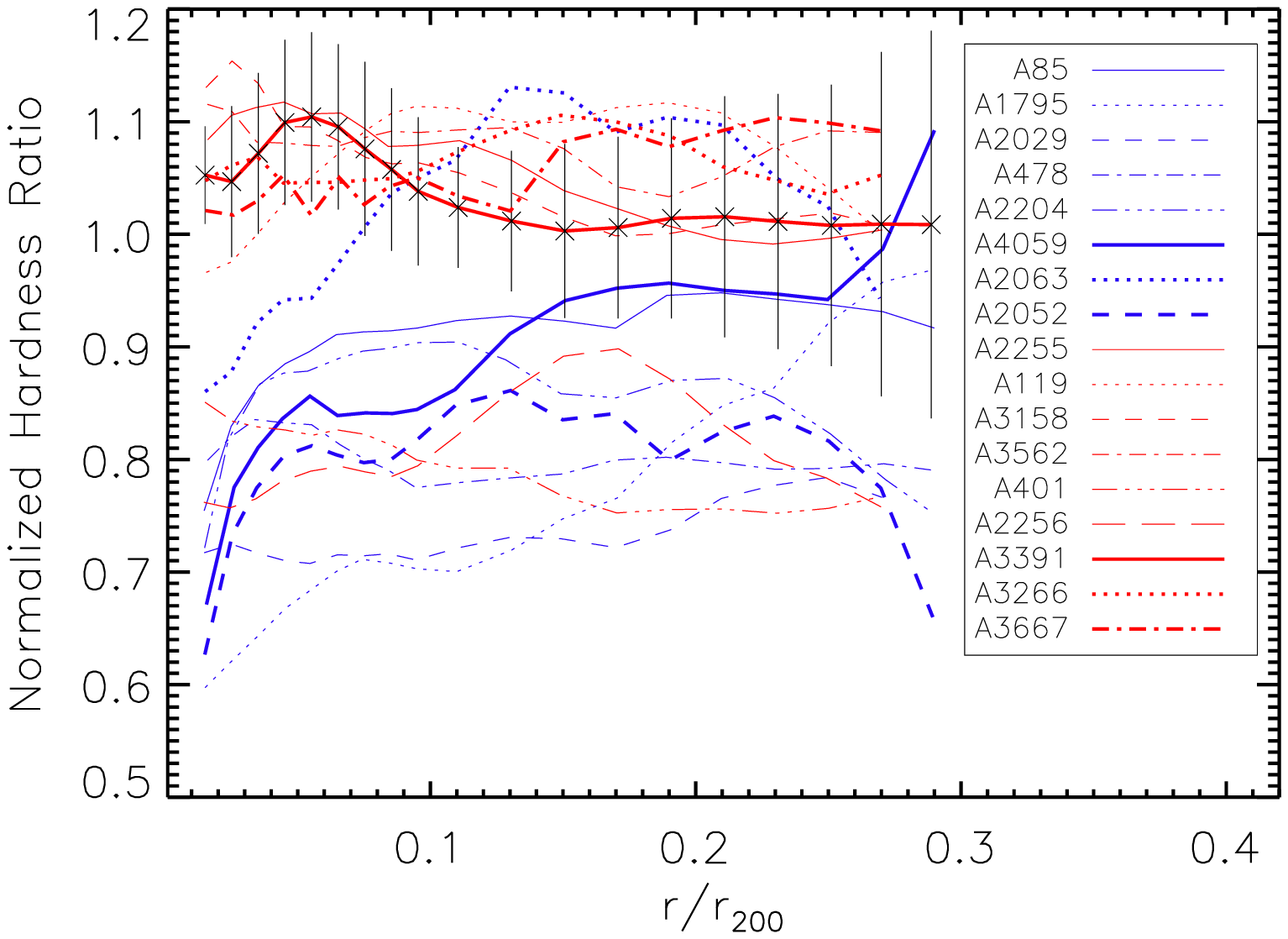}
\plotone{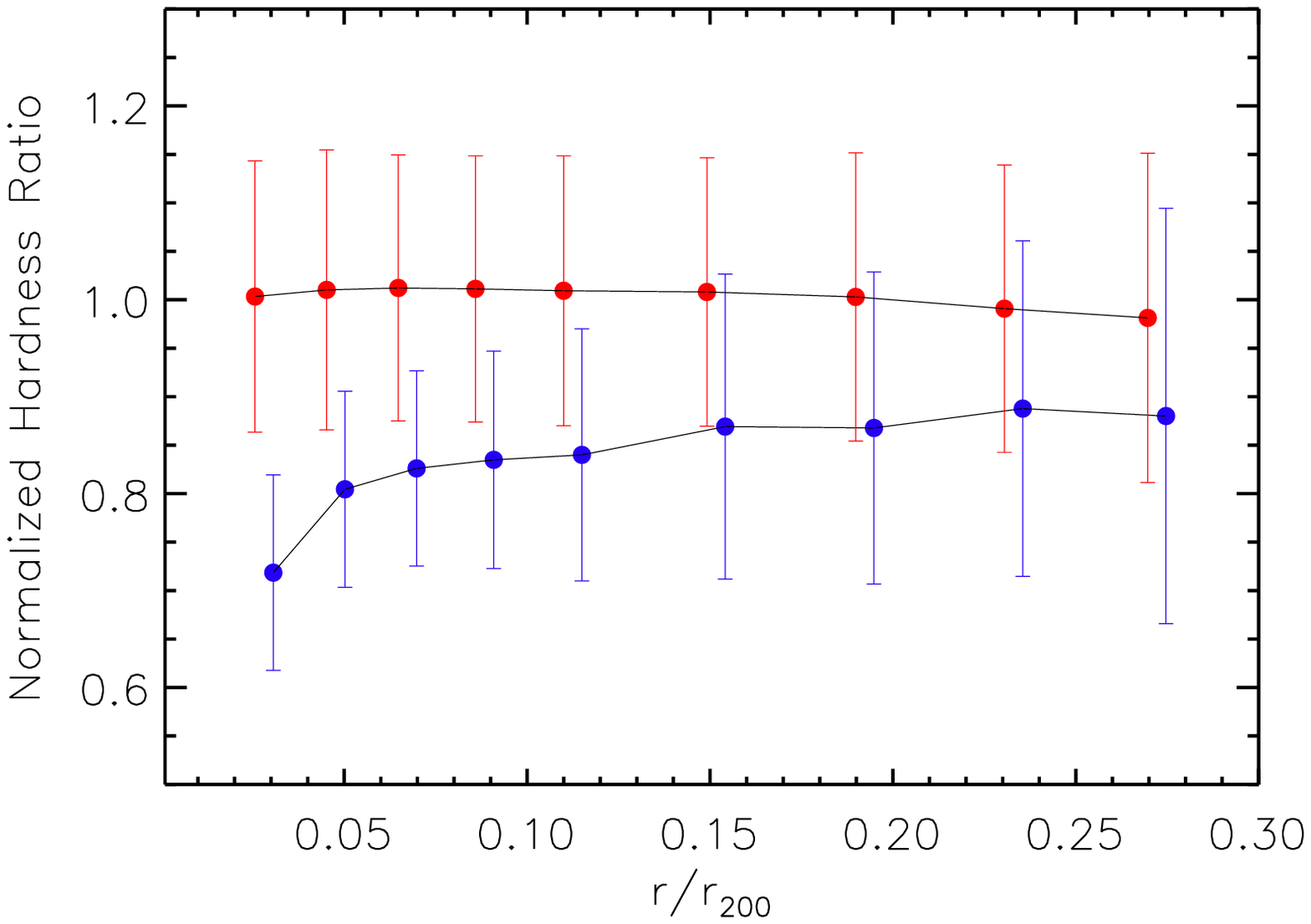}
\caption{(Top) Individual observed normalized hardness ratio radial profiles.  Blue lines are CC clusters, whereas red lines are NCC clusters.  Standard deviation errors are plotted for one cluster that are representative of all the clusters.  (Bottom) Averaged CC and NCC hardness ratio profiles, with standard deviation error bars.  Here and throughout the paper, CC and NCC clusters have been slightly offset radially for clarity.}
\label{HR_profiles}
\end{figure}

The top of Figure \ref{HR_profiles} shows the individual normalized hardness ratio profiles for the observed CC (blue) and NCC (red) clusters.  Standard deviation error bars for A3391, which are representative of typical errors in all the cluster profiles, are included for illustration.  With the exception of the innermost regions of the CC clusters, the hardness ratio profiles for all clusters tend to be relatively flat out to $\sim 0.3\,r_{200}$ despite the modest decline in temperature with radius (see \S \ref{bulk_temperatures}).  This flatness can be attributable to the hardness ratio's modest dependence on temperature.  In general, NCC clusters have higher normalized hardness ratio profiles than the CC clusters.  To quantify this offset, we averaged the CC and NCC profiles, and these averaged profiles can be found in the bottom of Figure \ref{HR_profiles}, with standard deviation error bars.  Though the scatter is somewhat large, there is an average offset in hardness ratio well beyond the cool cores between CC and NCC clusters, with NCC clusters having larger ratios by $\sim 0.1$. 

\begin{figure}[h]
\epsscale{1.0}
\plotone{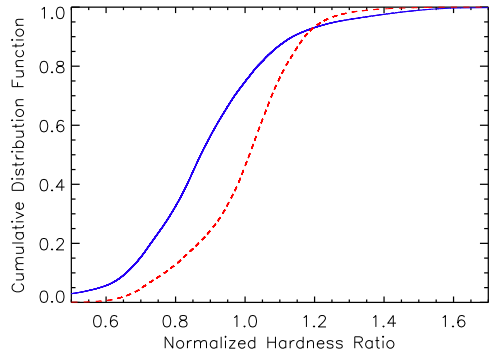}
\caption{Cumulative distribution functions for all independent and unbinned observed CC (blue) and NCC (red) hardness ratio map pixels in the range $0.15 - 0.3\,r_{200}$.  While the distributions are quantifiably very different, as evident in \S \ref{hr_maps}, the distributions are quite visibly different as well.  The median hardness ratio value for the CC clusters is $\sim 0.85$ while for NCC clusters it is $\sim 1.0$ in this range.}
\label{fig_KS_observed}
\end{figure}

To determine if the observed hardness ratio offsets are statistically significant, we performed a K-S test \citep{wall_jenkins_2003} on the unbinned hardness ratio map pixels from all cluster hardness ratio maps.  We separated the sample sets into CC and NCC groups, and constricted the hardness ratio maps to the region $(0.15 - 0.3)r_{200}$ to be certain we did not bias the distributions by including any cool core pixels.  Cumulative distribution functions of the individual hardness ratio map pixel values are displayed in Figure \ref{fig_KS_observed}, where it is clear there is a significant offset between CC and NCC hardness ratios.  Including unbinned map pixels, the two sample sizes in the K-S test are quite large.  As a result, the significance of the K-S D-statistic follows the significance of a $\chi^2$ distribution with two degrees of freedom \citep{wall_jenkins_2003}.  The $\chi^2$ statistic corresponding to the obtained D-statistic is $\chi^2 = 29163$, which means the probability that the CC and NCC hardness ratio pixel sets come from the \textit{same} distribution is $\lll1\%$.  We conclude that the hardness ratio offset seen in the averaged profiles results from intrinsic differences in the properties of clusters of very similar mass.  Given that hardness ratios track overall cluster temperature, this suggests that CC clusters have more cool gas than NCC clusters well beyond their cool cores.

\subsection{Bulk Temperatures}\label{bulk_temperatures}

The results of the hardness ratio analysis suggest that the offset seen in hardness ratios should also be present in mass-matched cluster temperature profiles.  To test this hypothesis, we defined four regions for a sub-sample of mass-matched clusters evenly spaced between $\sim(0.1 - 0.3) r_{200}$ and found bulk temperatures for each of these regions by fitting their X-ray spectra using XSPEC v.12.3.0.  The cores of CC clusters are much cooler than the outer cluster regions, and similarly, the central gas in an NCC cluster is often warmer than gas farther out.  To ensure we removed all the central gas so as not to bias our cluster bulk temperatures low or high for CC or NCC clusters, respectively, we looked at the surface brightness profiles (\S\ref{surface_brightness_profs}) of the CC clusters in question, found where they begin to turn over at the edges of the cool cores, and doubled this radius, which was then used as an inner radius cutoff.  These cutoffs were independently defined for each CC cluster,  but have an average of $0.11\,r_{200}$, which we used as the inner radius cutoff for all NCC clusters.

We removed point sources with the CIAO tool \textbf{wavdetect}, searching over five scales, 1, 2, 4, 8, and 16.  With the point sources removed, we then used the tool \textbf{specextract} to extract spectra with no grouping.  The spectra include energies between $0.5$ and $8$ keV from the point source-cleaned level 2 event files in the defined regions for fitting in XSPEC.  Since we need the bulk temperatures as a proof of principle and not for use with detailed cluster temperature analyses, if a cluster has more than one observation, we only utilized the longest exposure event files; using the spectrum from such a large region of each cluster provided us with more than enough counts for high S/N in each energy channel in the fitting process.  Additionally, the spectra for front- and back-illuminated chips were extracted separately since they have different sensitivities.

Appropriately subtracting the background spectrum is vital if an accurate and believable cluster spectral fit is to be achieved.  The same procedure was followed as for determining the background counts for the \textit{Chandra} images discussed above, but we instead used the ACIS Blank-Sky Background files available from the CIAO database.  Our temperature results below agree with published temperatures within our errors.  All our spectral fits used the energy range $(0.5 - 7)$ keV.  The extracted event and normalized background spectra were then taken into XSPEC, where we used an APEC model with photoelectric absorption to fit the background subtracted cluster spectra.  Weighted average values for galactic hydrogen column density \citep{dickey_lockman_1990, kalberla_2005} were obtained from the online HEASARC $N_H$ tool\footnote{http://heasarc.gsfc.nasa.gov/cgi-bin/Tools/w3nh/w3nh.pl} but were allowed to change during fitting.  Redshifts were taken from Table \ref{clusters_table} and initial temperatures were from the  $M - T$ relation of Equation \ref{eq_chen_M_T_relation}.

The results for three paired clusters can be found in Table \ref{table_bulk_temperatures}.  The masses of the CC and NCC clusters in the first two matched pairs are within $10\%$ of each other, while the third (unmatched) pair is given for further illustration.  The regions over which the spectra were extracted are provided, along with fit results for hydrogen column, metallicity, and the fraction of the observed count rates attributable to the source instead of the background.  The reduced $\chi^2$ and degrees of freedom for each fit are also provided.   Most fitted hydrogen column values are within a factor of two or three from those obtained by the HEASARC online tool, discrepancies attributable to the use of the ACIS blank-sky backgrounds for the background subtractions.  These backgrounds include data from several pointings and so likely include data from a region of the sky with slightly different galactic hydrogen than the regions of our particular cluster observations.  A normalized temperature is also given, which is simply the spectroscopically fit temperature $T_{fit}$ divided by the bulk cluster temperature $T_{Chen}$ \citep{chen_2007} in Table \ref{clusters_table}.  It is important to recall that $T_{Chen}$ has been corrected for the influence of cooler gas in CC cluster cores.  In addition, recall that CIAO and CALDB 4.1.1 were used to update the analyses for the two mass-matched cluster pairs while the analyses for A4059 and A3391 (slightly cooler clusters at 3.94 and 5.89 keV, respectively) were completed with CIAO and CALDB 3.3.0.  Lastly, a fourth cluster pair was also identified, including the CC cluster A2204.  Measured temperatures for A2204 were exceedingly high, and a literature search revealed that these \textit{Chandra} data over-estimate the temperature of A2204 compared to the \textit{XMM} and \textit{Suzaku} satellites \citep{Reiprich_2008}, possibly caused by uncorrected PSF smearing in the \textit{XMM} data and inaccurate \textit{Chandra} calibration at high temperatures \citep[see][and references therein]{Reiprich_2008}.  In light of this discovery, we removed A2204 and its pair from the bulk temperature analysis.

In general, an offset similar to that found in the observed normalized hardness ratios is present in the normalized temperatures.  The observed normalized hardness ratio and temperature profiles are plotted for each paired cluster in Figure \ref{fig_HR_profiles}.  In most cases, the temperature profiles track the observed hardness profiles reasonably well, indicating that the hardness ratio offset does indeed originate in an offset in cluster temperatures and lending further confidence to our assumption that hardness ratio maps can act as first-order proxies for cluster temperature.  The temperature and hardness ratio profiles do not agree in every instance, but we believe the discrepancies are well explained by temperature substructure in some of the clusters (see below), which could easily be the result of minor cluster mergers.  Relatively large and numerous pockets of temperature substructure create significant deviations in the average temperature profile.  The hardness ratios do not manifest the substructure as strongly owing to the hardness ratio's derivation from broad-band images, which tend to suppress temperature substructure due to the $n^2$ dependence of surface brightness.  Additionally, \citet{cavagnolo_2008} found that the ratio of hard to soft band spectroscopically fitted temperatures tends to be higher in systems known to be merging, where a deviation from unity indicates a band-dependence to the temperature fit.  This is evidence suggesting unresolved clumps can alter the fitted temperatures in a band-dependent fashion.  In their study, the hard band ratios for A401 and A2255 were both considerably greater than 1.1 (1.37 and 1.32, respectively), which seems to corroborate our suggestion that temperature substructure in our clusters explains the differences between their normalized hardness ratio and fitted temperature profiles.

For each cluster, Figure \ref{fig_HR_profiles} also contains expected normalized hardness ratios corresponding to the spectroscopically-fit temperatures, which are plotted as dotted lines.  These expected normalized hardness ratios were calculated using WebSpec and the temperature results from fitting the extracted spectra with XSPEC.  They include no metals since we are interested in probing the effects of the temperature alone on the hardness ratios.  The hardness ratios were normalized by the corresponding hardness ratio for a cluster with zero metals exhibiting the same virial temperature as the cluster in question.  As can be seen in Figure \ref{fig_HR_profiles}, in most cases the expected normalized hardness ratios agree well with both the observed normalized hardness ratios and the normalized spectral temperatures, except where significant temperature substructure is apparent (e.g., A401 see below).

One should not be concerned with the increased unevenness in the temperature profiles as compared to the relatively smooth hardness ratio profiles.  Recall from Figure \ref{fig_total_emission} that the hardness ratio has a clear but modest dependence on temperature.  This has the effect of ``flattening" the hardness ratio profile compared to the temperature profile from which it is derived, removing a great deal of the unevenness from pockets of cooler and/or warmer gas. Below, we discuss the remaining discrepancies between temperature and hardness ratio profiles for each cluster separately with the aide of published temperature maps.

\subsubsection{A478}
The normalized temperature of A478 shows a large spike roughly three arcminutes from the cluster core.  A temperature map from \citet{Bourdin_2008} contains a ring of hotter material between two and four arcminutes from the cluster core, $\sim 2 - 3$ keV warmer than the material just inside it.  This is enough substructure to explain the large spike in our temperature profile.  But, no comparable substructure is seen in the X-ray surface brightness map.

\subsubsection{A401}
A temperature map from \citet{Bourdin_2008} of A401 shows large blotches of gas both cooler and warmer than the average cluster temperature.  There is one large warm region roughly two arcminutes in radius between three and six arcminutes from the core $\sim 3$ keV warmer than the average cluster temperature.  Several cool pockets the same radial distance from the core as the warm blotch are present as well, but are only $\sim 1.5$ keV cooler than the average temperature.  It is possible radially averaging the temperature in this region results in an overall increase in temperature compared to the average, which is what is observed in the temperature profile for A401 in Figure \ref{fig_HR_profiles}, relative to the hardness ratio profile.

\subsubsection{A85}
A ring of small warm gas clumps nearly twice the average temperature of A85 is evident in a temperature map in \citet{Durret_2005}.  These small clumps appear between three and seven arcminutes from the cluster core, and line up nicely with the slightly elevated temperatures in our temperature profile.  Unfortunately, A85 is an example of the limitations of looking beyond the core region; we completely ignore the fairly relaxed central region in hopes that cool core gas will not bias our results, yet we are confronted with more temperature substructure.  Again, there is no comparable substructure in the surface brightness and hardness ratio maps.

\subsubsection{A2255}
Our temperature profile for A2255 matches very well with the observed hardness ratio profile, although there is a small dip in temperature around three arcminutes from the core.  A temperature map from \citet{Sakelliou_2006} reveals a patch of cooler gas roughly three to four arcminutes from the core that accounts for the observed temperature dip.  The temperature profile shows evidence it is beginning to increase four or five arcminutes from the core, and the \citet{Sakelliou_2006} temperature map does contain several significantly warmer patches at roughly this radial distance.
 
\subsubsection{A4059}
A4059 is another cluster whose normalized temperature profile agrees well with the observed normalized hardness ratio profile.  A slightly elevated temperature at roughly four arcminutes from the core is observed, however.  The temperature map of \citet{Reynolds_08} shows some patches of warmer gas a few arcminutes from the core, but it is hard to reliably correlate these warm gas pockets with the structure in our temperature profile since the \citet{Reynolds_08} temperature map does not have the same angular extent.  Their map focuses on the inner cluster region, which we have excluded.

\subsubsection{A3391}
The only discrepancy between the hardness ratio and temperature profiles for A3391 is a rapid drop-off in temperature beyond six arcminutes.  Although the resolution is poor, an ASCA temperature map from \citet{Markevitch_1998} reveals larger cooler regions beyond roughly six arcminutes, again matching our temperature results very well.
\\
\\
\\

Overall, the results of the hardness ratio and temperature analysis appear to indicate that there are both similarities and differences between CC and NCC clusters well outside the cluster cores.  A statistically significant offset in normalized hardness ratios between CC and NCC clusters suggests that on average CC clusters tend to have more cool gas well beyond their cores than NCC clusters.  In addition, there is general agreement between hardness ratio and spectral temperature profiles, further suggesting there is in fact a difference in ICM temperature between CC and NCC clusters that drives the offset in hardness ratios.  Although the hardness ratio profiles are quite flat, cluster temperature maps reveal the presence of temperature substructure, which is manifest in the much more irregular temperature profiles.  Such temperature substructure is expected from our simulations in not only NCC but CC clusters as well since CC clusters suffer late minor mergers.  In the next section, we compare these results to an analogous analysis of simulated clusters.

\begin{figure}[h]
\epsscale{1.1}
\plottwo{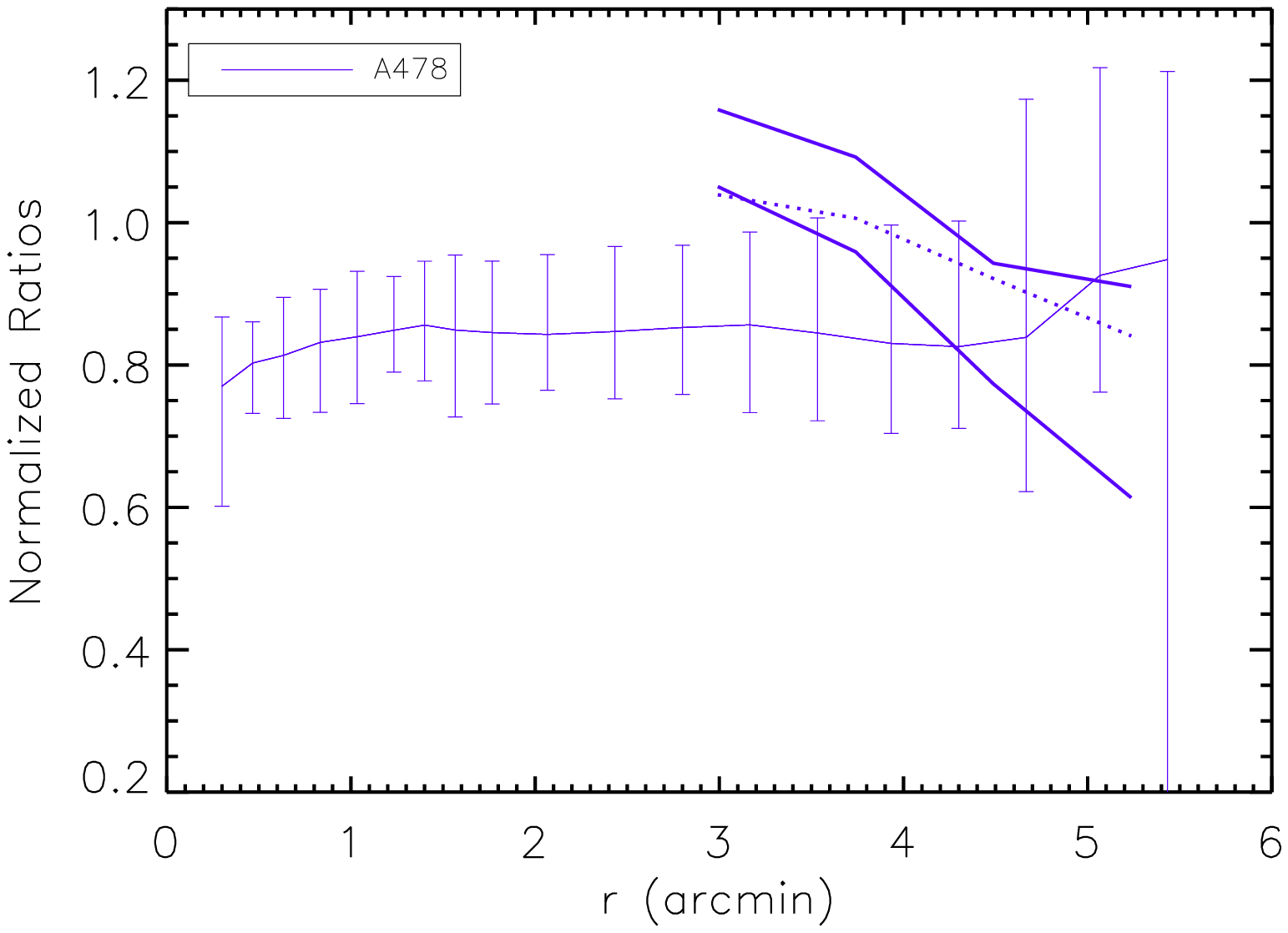}{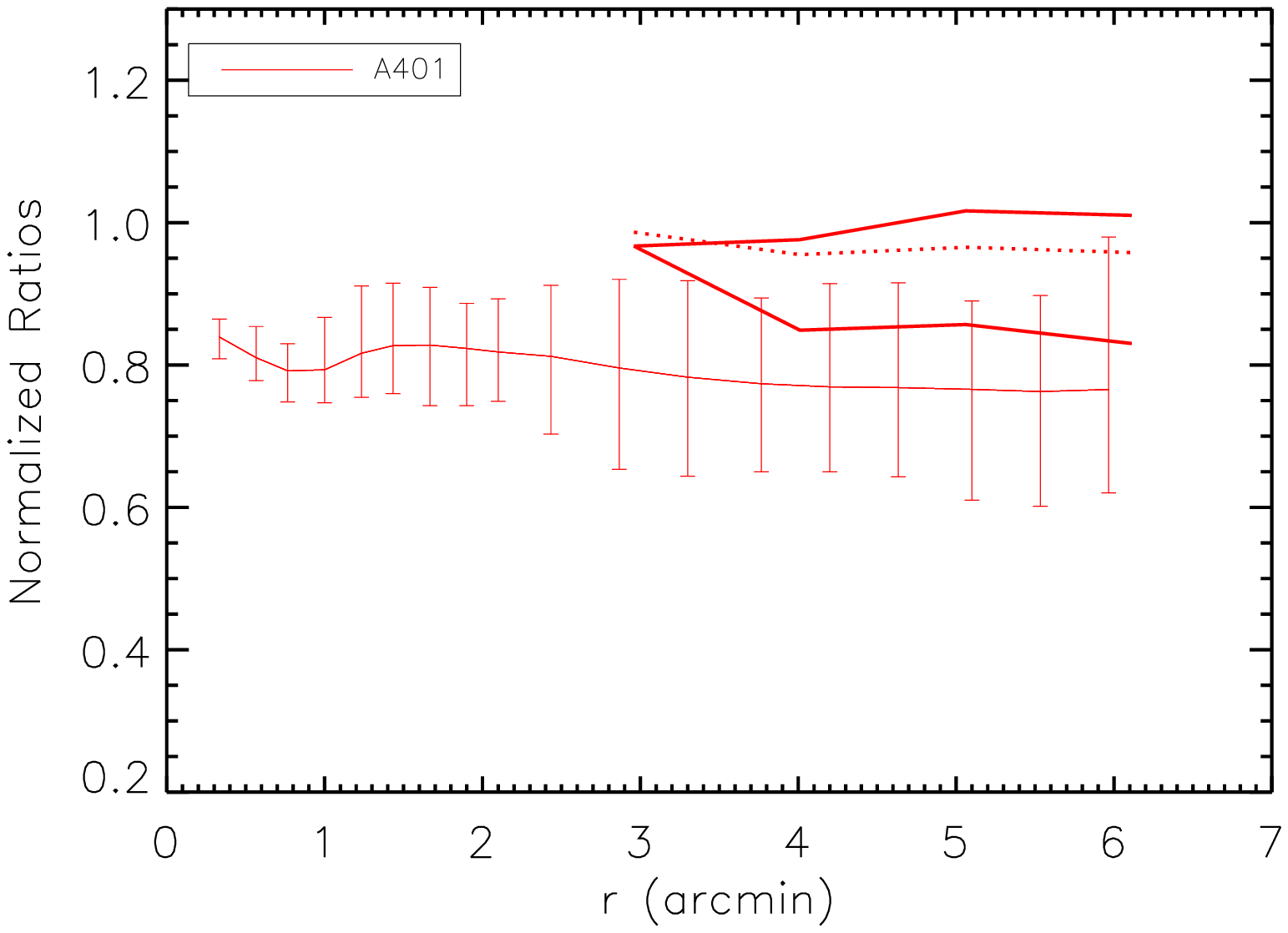}
\plottwo{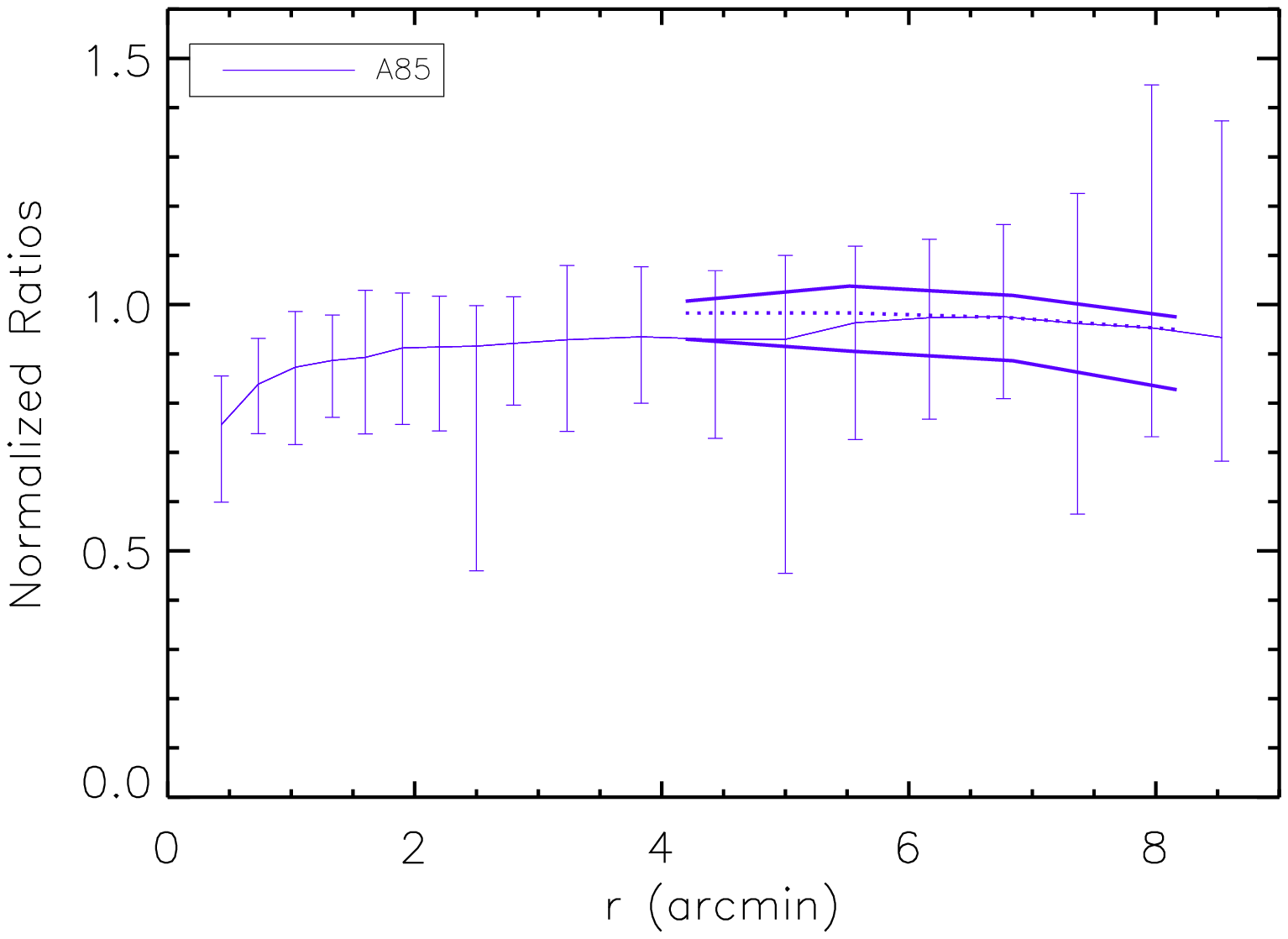}{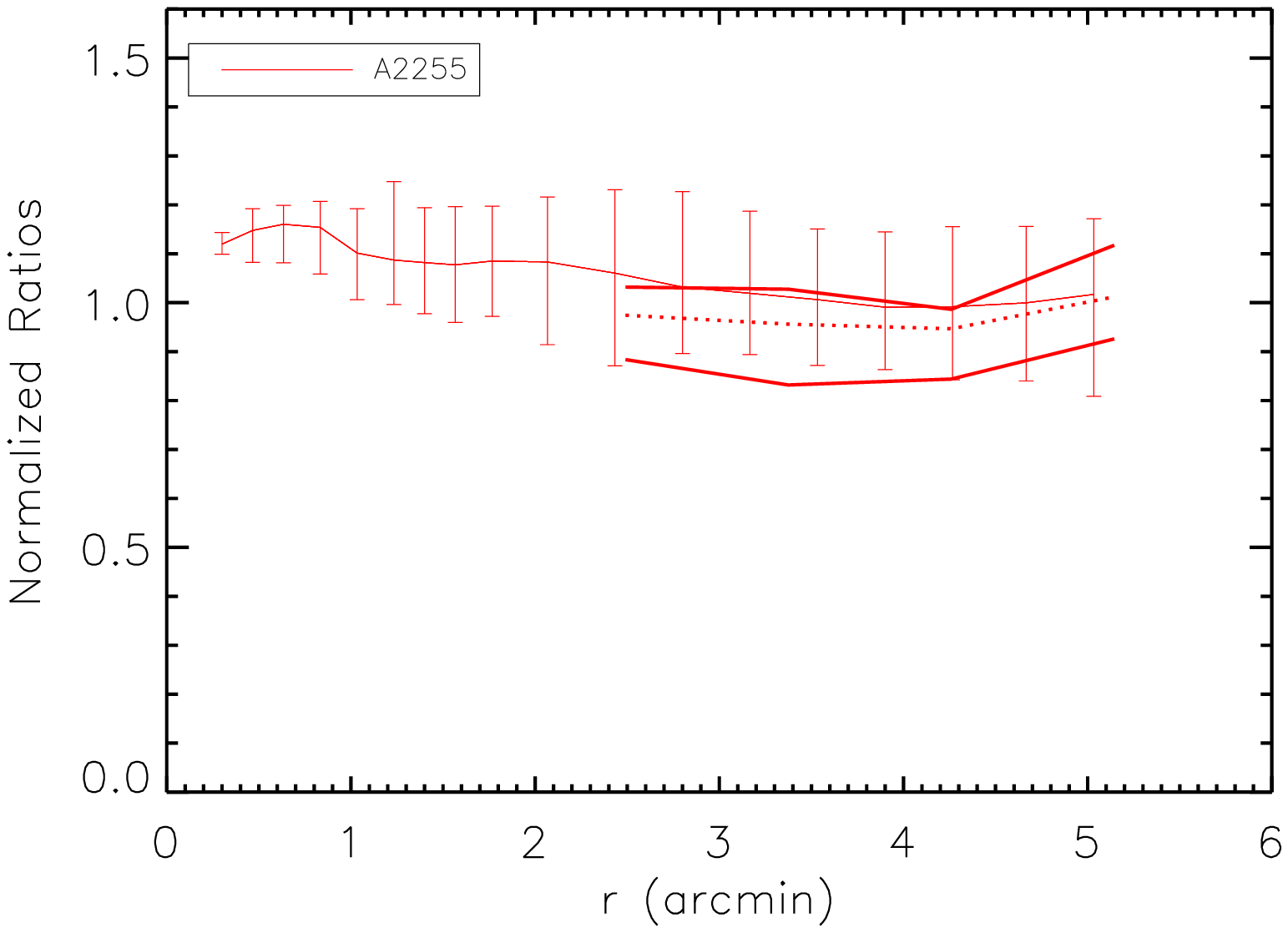}
\plottwo{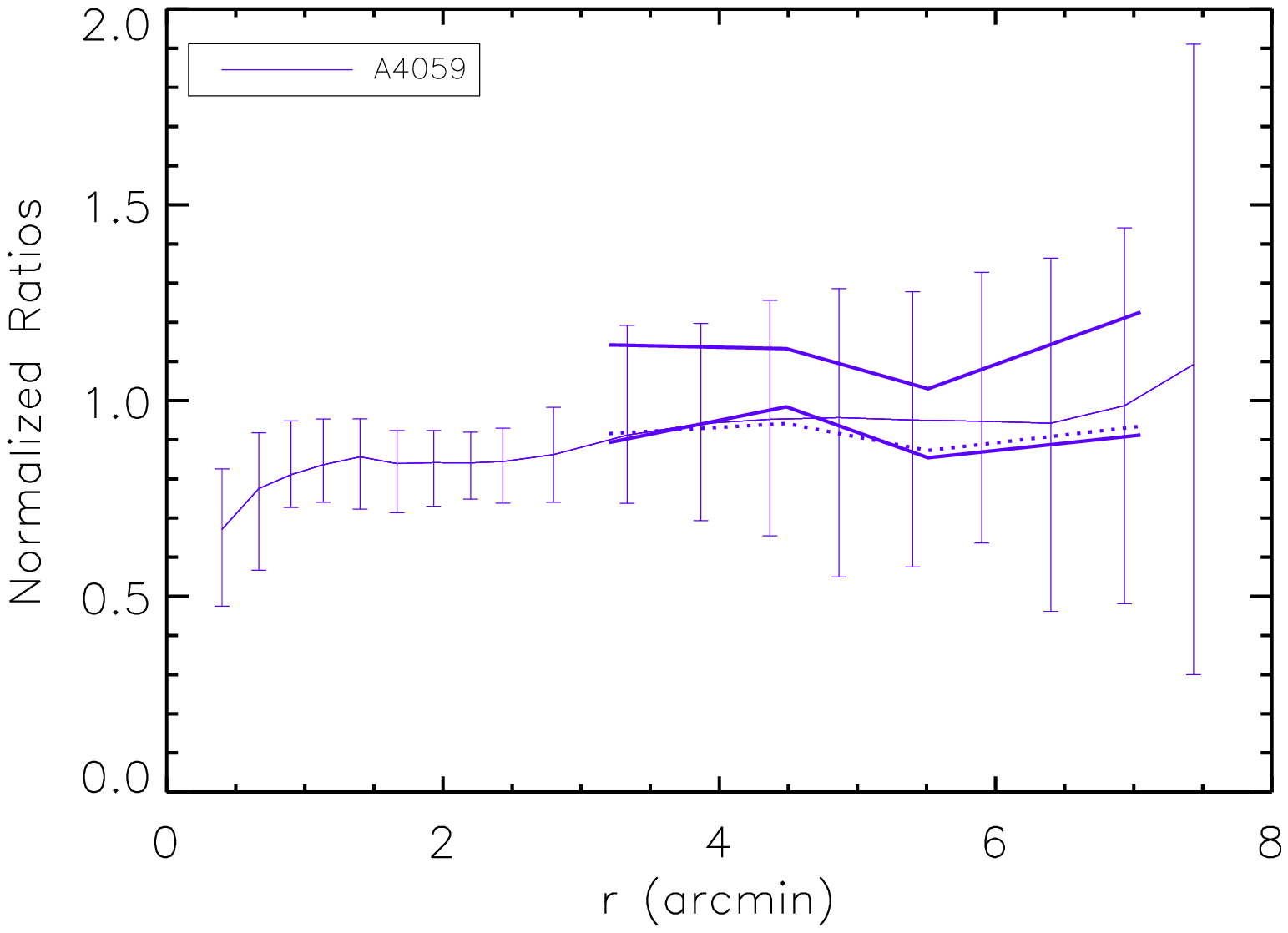}{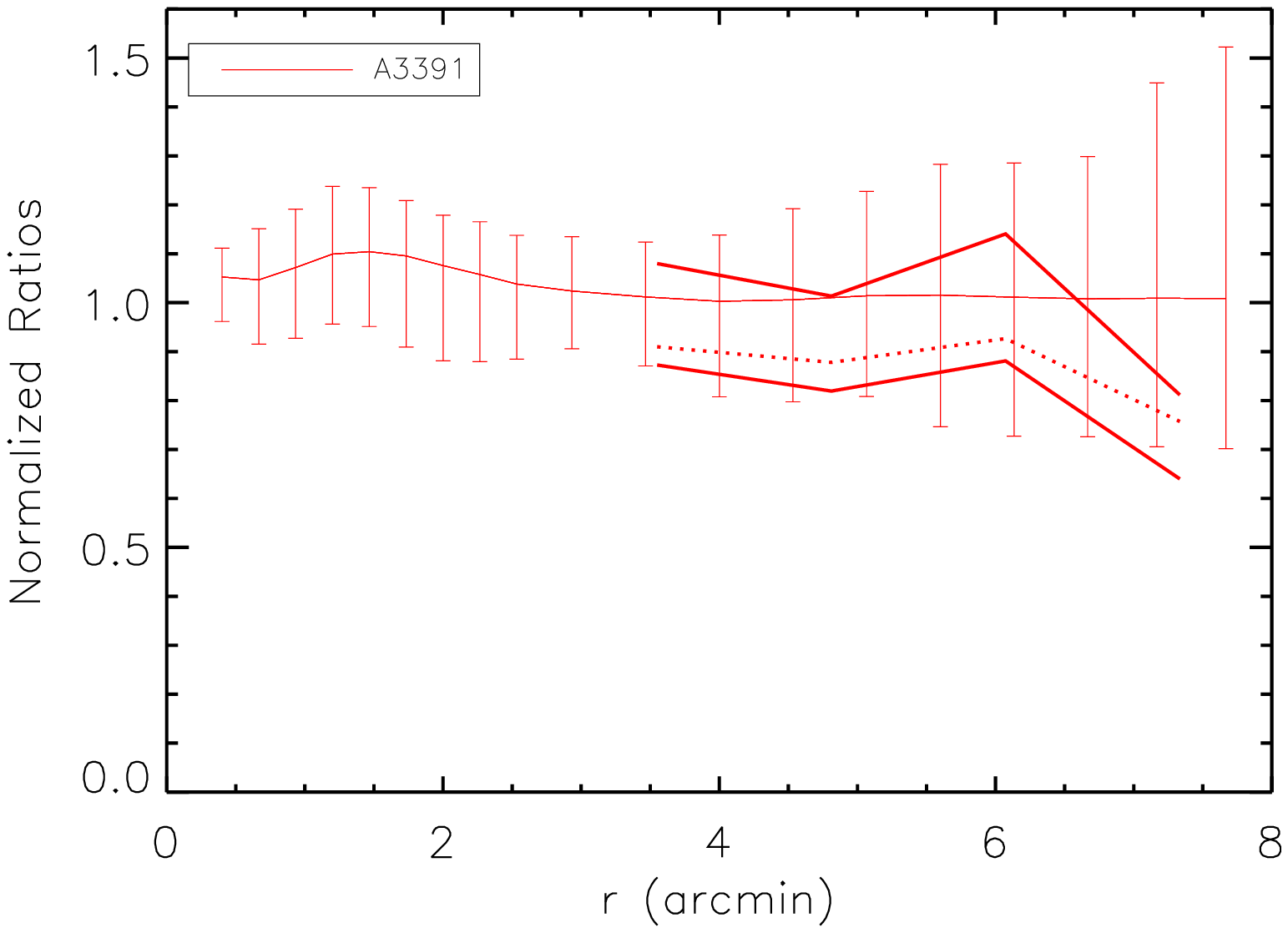}
\caption{Normalized hardness ratio and normalized spectroscopically fit temperature profiles for the six mass-matched clusters.  CC clusters are in blue while NCC clusters are in red, with mass-matched clusters in the same row.  The thick solid lines with error bars are the observed hardness ratio values with 90\% regions.  The two solid lines delineate the 90\% regions for the temperatures, which have been normalized by the cluster temperatures in Table \ref{clusters_table}.  Dotted lines correspond to expected model hardness ratios for clusters with 0.3 solar metallicity and the redshifts and temperatures from Table \ref{table_bulk_temperatures}.  The first four clusters were analyzed with CIAO and CALDB 4.1.1, while the last two clusters were analyzed with CIAO 3.3.0.}
\label{fig_HR_profiles}
\end{figure}


\section{Simulation Comparisons}\label{sim_comparisons}

With the analysis of our observed cluster sample in hand, we next compare these results to those of the simulated clusters from B08.  What can we say about the excess cool gas found in CC clusters outside their cores?  What are the similarities and differences between the observed and simulated properties of CC and NCC clusters?  How might we improve our simulations so that they better represent the physics of observed clusters?  We compare the results of the simulated clusters to those of the real clusters to answer these questions and learn how to achieve more realistic clusters in future simulations.  We begin by providing a quick description of the simulations from B08, then discuss the results of the surface brightness profile analysis, and conclude with an overview of simulated temperature and hardness ratio profiles.

\subsection{Simulation Description}

The simulations were performed using the Enzo code \citep{oshea_2004}.  This code combines an N-body algorithm that evolves collisionless dark matter particles using an Eulerian hydrodynamics scheme (Piecewise Parabolic Method or PPM) that employes adaptive mesh refinement (AMR) \citep[see][]{burns_2004, motl_2004, paper_one}.  These simulations include radiative cooling, star formation and energy feedback.  Radiative cooling is based on a Raymond-Smith plasma emission model \citep{brickhouse_1995} that assumes a constant metallicity of 0.3 solar.  Star formation helps soften the cool cores by allowing excessively cool gas to drop out into ``star particles.''  These star particles also allow energy to be reinserted into the system by means of supernovae by a prescription outlined in \citet{cen_ostriker_1992} and \citet{burns_2004}.  This is explained in detail in B08.  

In B08, we examined the simulation volumes and successfully compared bulk X-ray properties to those observed in surveys of galaxy clusters.  For the first time, our simulations produced both CC and NCC clusters in the same numerical volume.  B08 used comparisons of spectroscopic-like temperatures, $f_{gas}$ (the fraction of gas mass over total mass), $M_{500,Gas}$ (the gas mass within $r_{500}$) and individual $\beta$-model fits to similar observational collections in \citet{chen_2007}.  We showed that our simulations match observed bulk properties of clusters well enough for use in making testable predictions that we examine in this paper.  

CC and NCC clusters experience different merger histories according to our simulations.  NCC clusters suffer early major mergers that destroy any cool cores that have begun to form, while CC clusters remain undisturbed during early epochs.  It is important to note that star formation and feedback play a crucial role in ``softening" the cool cores (i.e., heating, expanding, and lowering the gas density), thus allowing collisions with comparable mass halos to ram pressure strip and destroy nascent cool cores.  At later times, CC clusters may be involved in mergers, but their cool cores are large and ``strong" enough to not be disrupted by the event.  In addition, fewer larger mergers in the histories of CC clusters means that the gas through the clusters is shock-heated less and so remains cooler than gas in NCC clusters.

\subsection{Simulated Surface Brightness Profiles}\label{sec_sim_surface_brightness}

\begin{figure}[h]
\epsscale{1.0}
\plotone{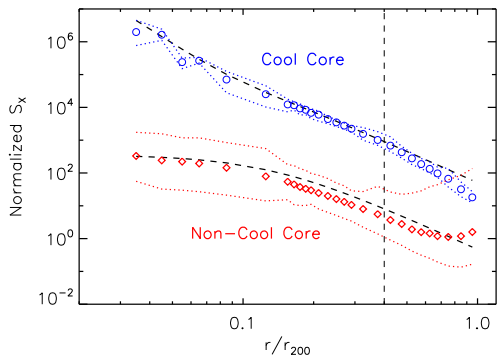}
\caption{The averaged surface brightness profiles of 10 CC (blue) and 78 NCC (red) simulated clusters.  The data have been averaged across all the clusters of each type and uncertainties are displayed as 90\% regions demarcated as dotted lines.  A $\beta$-model has been fit to the region $0-0.4\,r_{200}$ separately for both cluster types and then extrapolated over the full range of the plot.  The vertical dashed line indicates the edge of the fitting region.  The NCC profile has been arbitrarily rescaled by a factor of 10 to separate the curves for clarity.  The $\beta$-model somewhat overestimates the surface brightness at the virial radius, particularly in CC clusters.}
\label{avg_sim_sx_profile}
\end{figure}

The azimuthally averaged surface brightness profiles of 10 CC and 78 NCC simulated clusters, each normalized to its expected luminosity ($M^{3/2}$), are displayed in Figure \ref{avg_sim_sx_profile} with 90\% regions.  This is a statistically complete sample that includes every cluster at $z=0$ with $M > 10^{14}$ M$_{\odot}$ that can be clearly identified as a CC or NCC cluster and that lack extreme recent merger effects; see B08 for more details.  Double and single $\beta$-models were fitted to the averaged CC and NCC cluster profiles, respectively, and over-plotted.  The $\beta$-model fits to the simulations indicate that fitting observed surface brightness profiles to the farthest edge of observations (for \textit{Chandra}, often to $\sim0.3-0.5\,r_{200}$) may overestimate the surface brightness at large radii, but more so in CC than NCC clusters.  This overestimation of surface brightness in turn leads to an overestimation of the extrapolated cluster gas mass at $r_{200}$ as discussed in \citet{hallman_2006}.  Clearly, effects such as this complicate the use of galaxy clusters for precision cosmology and suggest that care must be taken when experiments are based on their extrapolated properties. 

\begin{figure}[h]
\epsscale{1.0}
\plotone{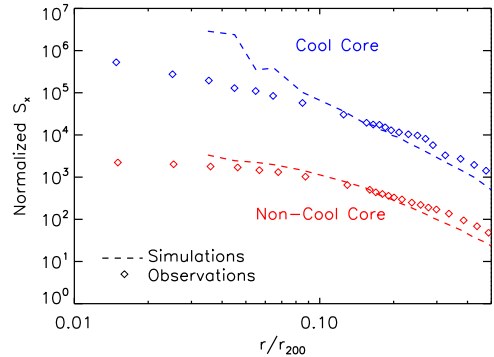}
\caption{A comparison of simulated and observed averaged surface brightness profiles.  The simulated clusters include 10 CC (blue) and 78 NCC (red), all the available clusters at $z=0$.  The NCC clusters have been arbitrarily divided by 10 to separate the lines for clarity.  Uncertainties have been removed for clarity.  The excess central luminosity of the simulated CC clusters is due to overcooling as discussed in B08.  The outer profiles for the simulated clusters, especially the cool cores, are somewhat steeper than the observed clusters (see Table \ref{beta_table}).}
\label{compare_obs_sim_surbri}
\end{figure}

Figure \ref{compare_obs_sim_surbri} shows the averaged observational data compared to the averaged surface brightness profiles of our simulated clusters; 90\% regions have been removed for clarity.  While the simulated CC clusters show an excess luminosity at the core relative to observations, the overall agreement between simulations and observations is reasonable, though it is clear more work is required to better model observed clusters.  The CC core excess luminosity is a known limitation due to the overcooling of high density regions within the simulations and is discussed in B08.  Also, the larger errors associated with the simulations at small radii in Figure \ref{avg_sim_sx_profile} are a result of limited resolution for the simulated clusters (16 kpc peak resolution).  Pixel sizes at the simulated cluster centers are relatively large and provide only a few points with which to average across, resulting in 90\% regions that appear to underestimate the precision of the simulated data below $\sim0.1\,r_{200}$. 

\begin{figure}[h]
\epsscale{0.75}
\plotone{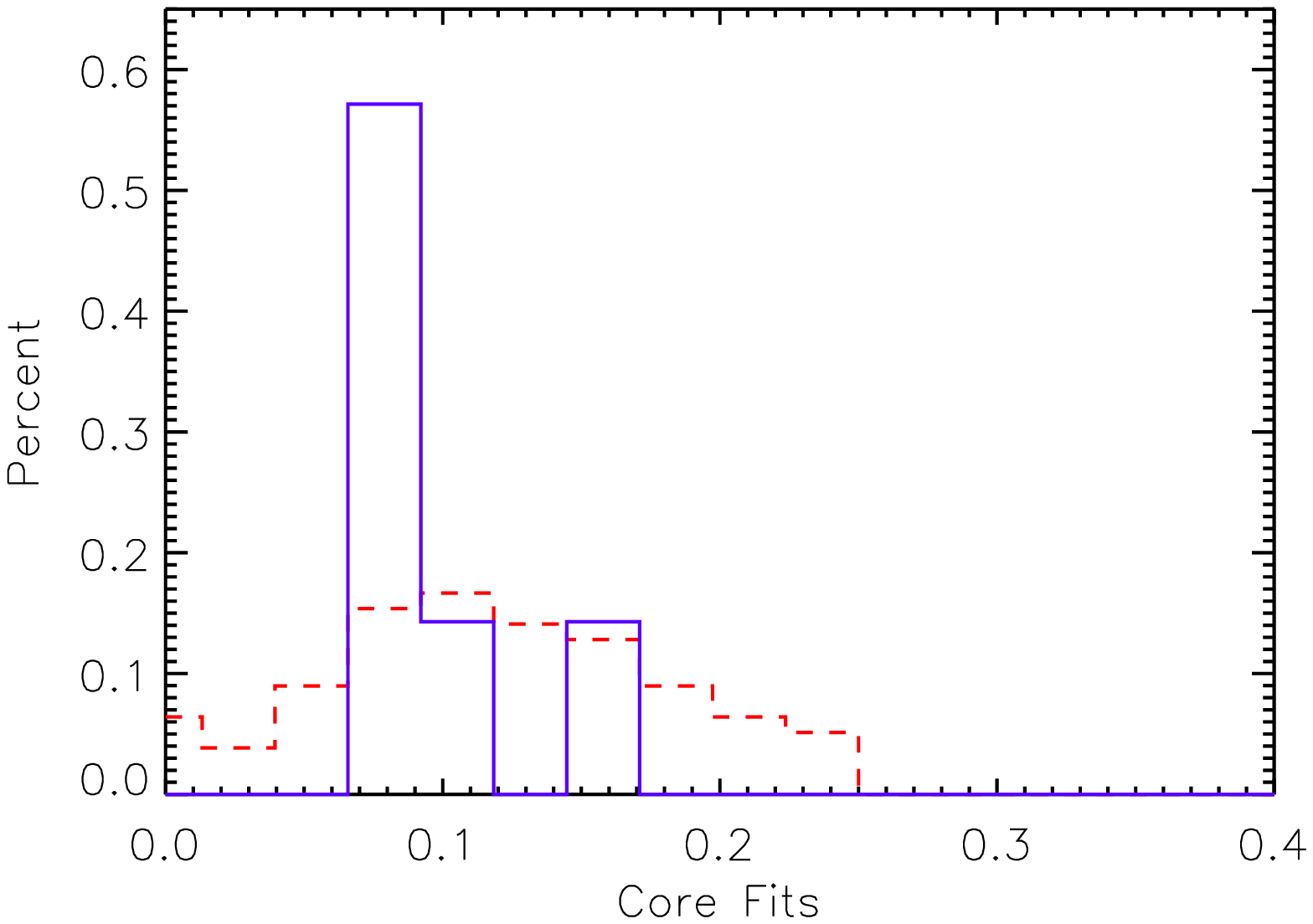}
\plotone{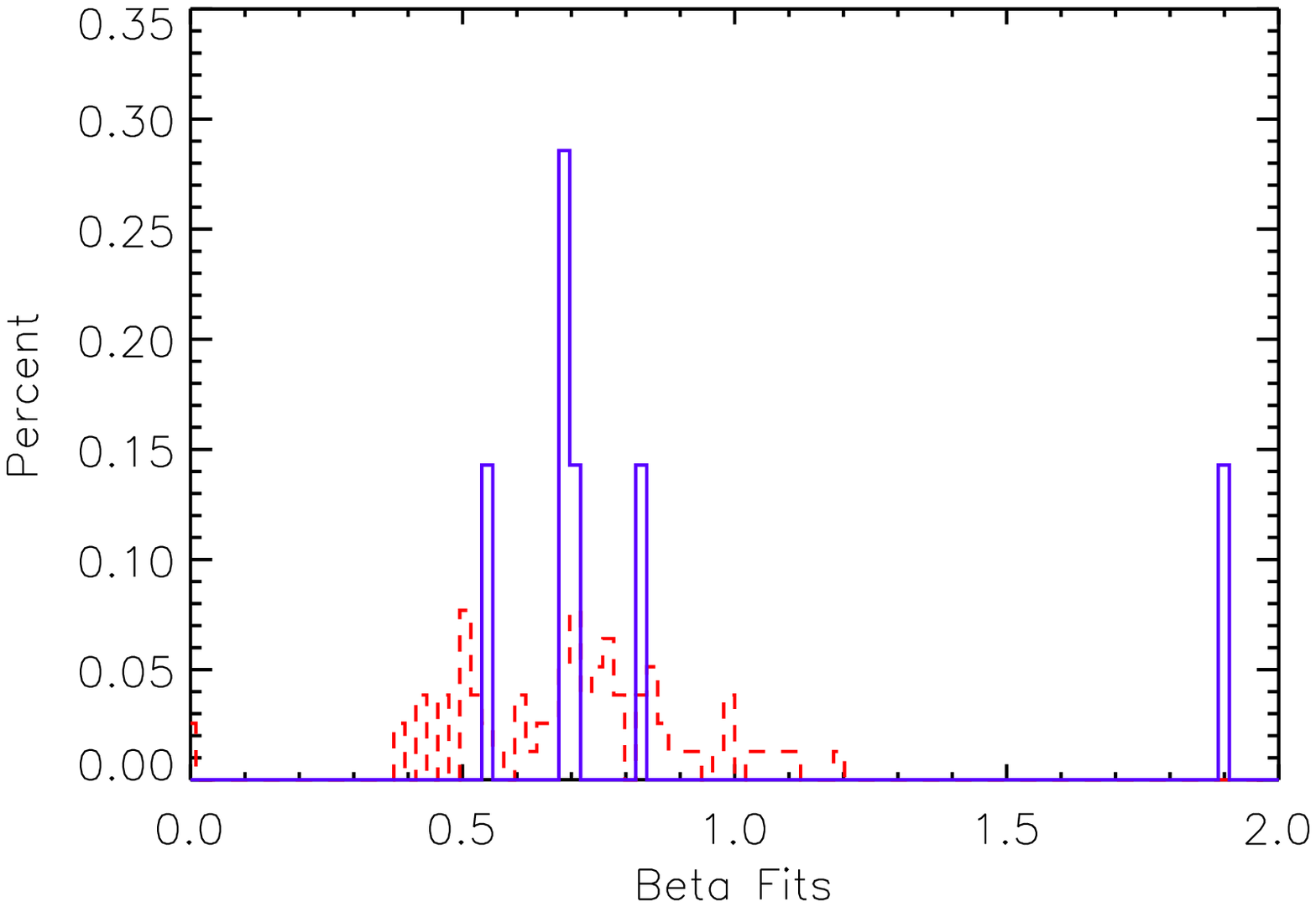}
\caption{Histograms of simulated $\beta$-model fit parameter values.  (Top) Outer $r_{core}$ fits.  (Bottom) $\beta$ fits.  It is clear from the histograms that both $r_{core}$ and $\beta$ fits for CC (blue) and NCC (red) clusters are statistically indistinguishable.}
\label{sim_betamod_histograms}
\end{figure}

As was done for the observations, double and single $\beta$-models for CC and NCC clusters, respectively, were fitted to the cluster profiles individually between 0 and $0.4\,r_{200}$.  These individual fits were then grouped into cluster category, averaged and displayed in Table \ref{beta_table}.  Histograms of the individual cluster fits are provided in Figure \ref{sim_betamod_histograms}.  Much can be said by comparing the observed and simulated results in Table \ref{beta_table} and Figure \ref{compare_obs_sim_surbri}, and there are some clear similarities.  K-S test results confirm that there is no statistical differences between outer CC and NCC $r_{core,2}$ fits, neither for observed nor simulated clusters, a result that contrasts with those found using solely single $\beta$-models \citep{vik_doublebeta_2006, paper_one} where the cool core is excluded from the fit.  Additionally, the size of the averaged observed and simulated CC cluster cores ($r_{core,1}$) are consistent.  The profiles also clearly indicate that simulated NCC clusters do a good job of capturing the flat inner profile inside cluster cores.

There are important differences in $\beta$-model parameters and surface brightness profiles, however.  The effect of overcooling the simulated cluster cores is to drastically raise the inner $\beta$ slope for the CC clusters compared to the observed CC clusters.  In the outer profiles, and as evident in the $\beta$ slopes in Table \ref{beta_table}, the observed clusters exhibit flatter profiles than the simulations.  This may be evidence for additional heating and feedback not captured by our ``star particle" technique in the outer regions of observed clusters, perhaps puffing up the outer profile and making them brighter - another clue that our feedback prescriptions must be updated.  This also suggests that the entropy in the outer regions of observed clusters may be higher than in our simulated clusters.  All of these differences point to corrections that must be made in future iterations of the simulations if we wish to accurately model observed clusters.

\subsection{Simulated Temperatures and Hardness Ratios}\label{sec_sim_HR}

\begin{figure}[h]
\epsscale{1.0}
\plotone{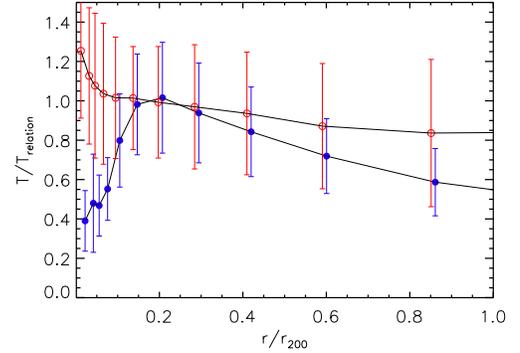}
\caption{Averaged simulated temperature profiles for CC (blue) and NCC (red) clusters, normalized by the $M-T$ relation temperatures.  To mimic what is done for observations, the CC cluster temperatures have been ``corrected" for their cool cores.  Error bars delineate the standard deviation at each radial point.}
\label{sim_profile_with_Baldi}
\end{figure}

Unlike for the observed clusters, we have full knowledge of simulated temperatures and so we calculated averaged temperature profiles for the simulated clusters.  These give further confidence in the use of observed hardness ratios as a first-order proxy for cluster temperatures, and provide another test of the assumption that the offset seen between the observed and simulated CC and NCC hardness ratio profiles is a consequence of differing cluster temperatures beyond the cores of mass-matched cluster pairs.  Figure \ref{sim_profile_with_Baldi} shows the averaged temperature profiles, again with standard deviation error bars.  It is clear in the figure that the simulated temperature profiles exhibit the same offset seen in the hardness ratio profiles one would expect if in fact the temperatures drive the hardness ratio values; CC clusters are cooler and NCC clusters are warmer well beyond the cluster cores.  Another K-S test over the region $0.3 - 1.0\,r_{200}$ for the averaged temperature profiles yields a probability of $8.4\%$ that the temperature profiles come from the same distribution.  This result is compelling evidence that the offset does indeed exist in temperature.

We also performed a hardness ratio analysis on the simulated clusters for comparison with observations.  Following the same procedure for simulated data as we used with the \textit{Chandra} data (detailed in Section \ref{hr_maps}), we split the synthetic X-ray images of the simulated clusters into two energy bands, 2-8 keV and 0.5-2 keV.  We then divided the two images (2-8 keV/0.5-2 keV) and normalized them to the hardness ratio that corresponds to the ``virial" temperature to produce normalized hardness ratio images.  As in the observed hardness ratio maps, pixel values less than one represent gas with $T < T_{vir}$ and pixel values greater than one represent $T > T_{vir}$.   Also, as in the observed clusters an $M - T$ relation is used to find the cluster temperatures for normalization.  Plotting the log of simulated cluster masses $ M_{500}$ versus log of the cluster temperatures, a best-fit line yields the corresponding mass-temperature relation:
\begin{equation}
\log _{10}\left( \frac{M_{500}}{5\times10^{14} \mbox{M}_{\odot}} \right) = 0.629 + 1.55 \log_{10}\left( \frac{T}{4 \mbox{keV}}\right).
\end{equation}
The observed relation from \citet{chen_2007} and the simulated cluster relation agree well, where both are consistent with the 3/2 slope expected if the clusters are scale-free.  In an effort to mimic what is done with observations as much as possible, we ``corrected" the simulated CC cluster temperatures for the presence of their cool cores.  We first fit for the $M-T$ relation for simulated NCC clusters only and found the deviation from this relation for each CC cluster separately.  We then fit a line to the resulting temperature offset-cluster mass plot to define an average temperature correction for a given mass.  Finally, we applied this temperature correction to all simulated CC clusters. 

\begin{figure}[h]
\epsscale{0.75}
\plotone{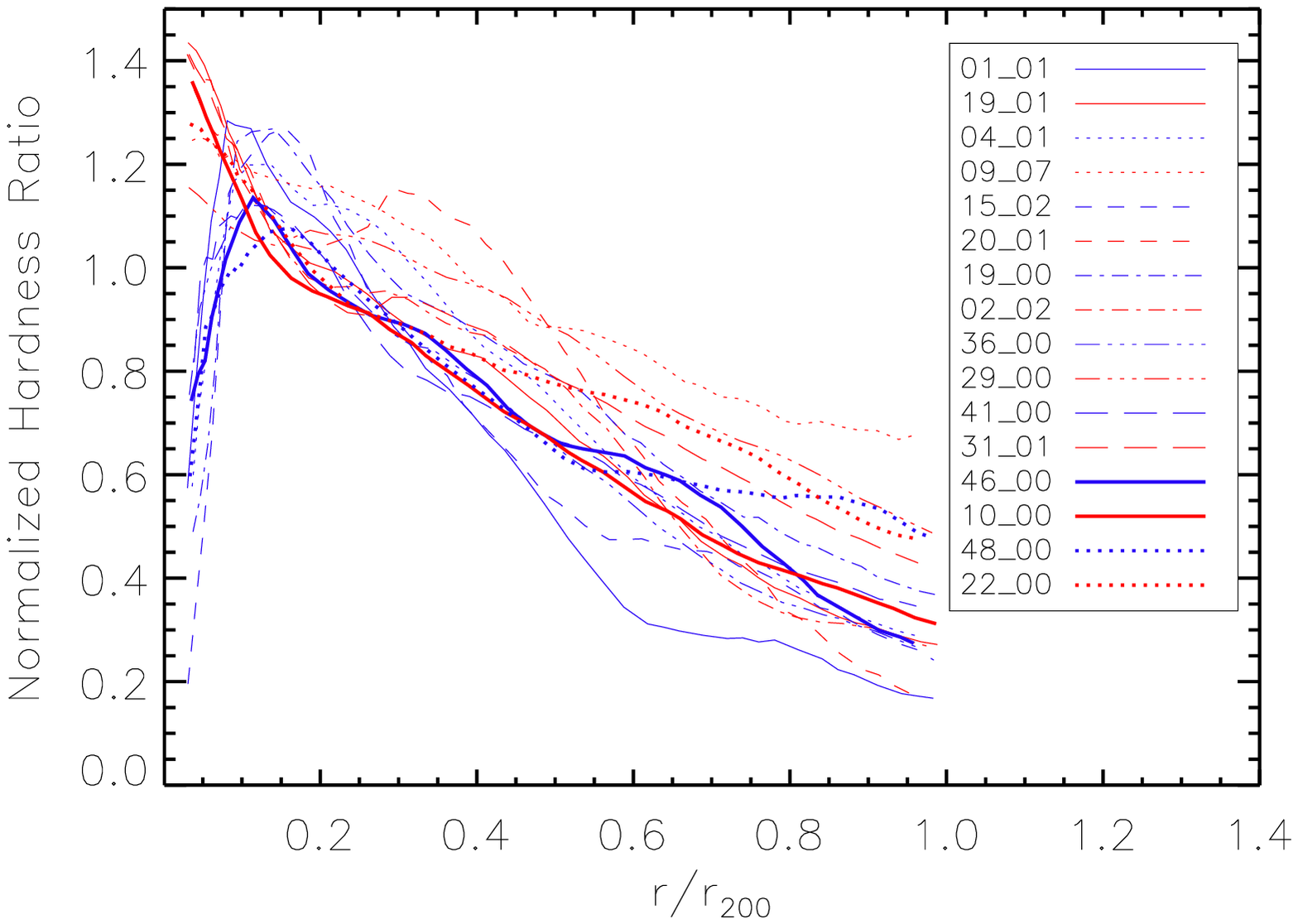}
\plotone{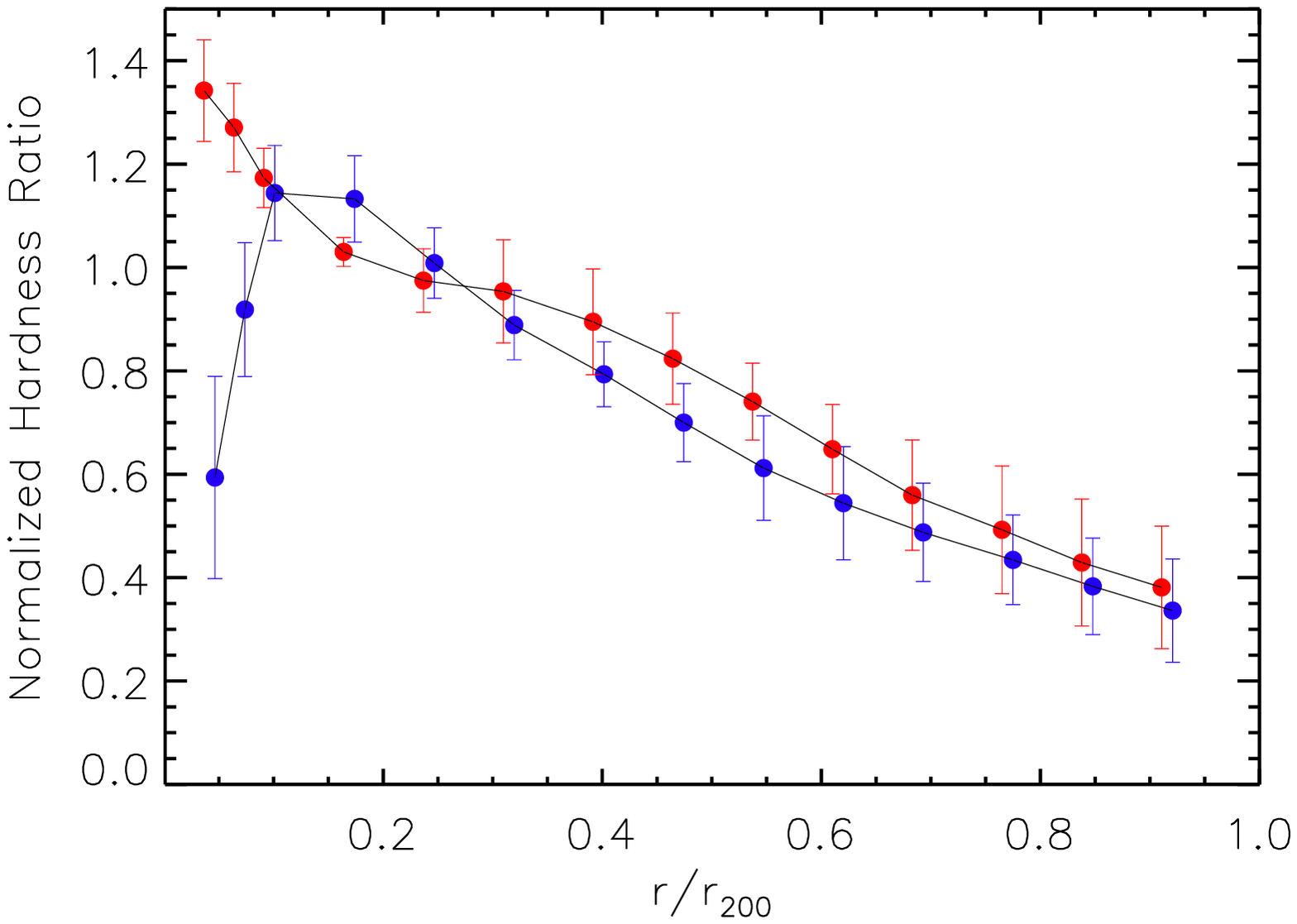}
\caption{(Top) Individual simulated hardness ratio radial profiles for CC (blue) and NCC (red) clusters.  The mass-matched pairs are next to each other in the legend and have the same line style and line thickness.  (Bottom) The average hardness ratio profiles.  Error bars are standard deviations at each point.}
\label{fig_sim_HR_profiles}
\end{figure}

As we did for the observed clusters, simulated clusters were mass-matched to within $10\%$.  The hardness ratio profiles for the individual simulated clusters are found in the top of Figure \ref{fig_sim_HR_profiles}.  Mass-matched pairs are plotted with the same line style and line thickness.  The bottom of Figure \ref{fig_sim_HR_profiles} contains the corresponding averaged CC and NCC hardness ratio profiles.  In the outer regions of the clusters, we qualitatively see the same hardness ratio offset as was observed in the real cluster profiles: the NCC clusters are harder than the CC clusters, despite very similar masses.  Performing the same K-S test as was done for the observed clusters, but for the region $(0.3 - 1.0)\,r_{200}$ to avoid the region of simulated clusters most likely affected by the overcooled cores, the outcome was identical.  Namely, the probability that the simulated CC and NCC hardness ratio pixel sets come from the same distribution is $\lll1\%$.  However, there are also differences in the hardness ratio profiles between observations and simulations.  In comparing Figure \ref{HR_profiles} to Figure \ref{fig_sim_HR_profiles}, the observed clusters have a larger separation in hardness ratio between CC and NCC clusters in the radii 0.1 to $0.4\,r_{200}$.

\section{Summary and Conclusions}\label{summary}

Using an optimized data pipeline to reduce \textit{Chandra} and \textit{ROSAT} data, we analyzed 30 nearby, rich Abell galaxy clusters split nearly equally between CC and NCC clusters.  While the expected increase in radial information from the \textit{ROSAT} data rarely appeared, the additional data did help reaffirm the data reduction pipeline by matching well across instruments.  The observational data were analyzed by examining radial surface brightness profiles and hardness ratio maps.

Utilizing the ENZO AMR simulations, we also analyzed 88 simulated clusters at $z=0$.  To the best of our knowledge, these are the first simulations to produce both CC and NCC clusters within the same cosmological volume.  These simulations were used to create data products that are directly comparable to observed clusters and initial efforts indicate that the simulations are reasonable first-order representations of actual clusters, but several modifications must be made for the future.  The simulations were then used to make testable predictions about the differences between CC and NCC clusters.  Specifically, simulations suggest that NCC clusters suffer early major mergers (roughly defined as $>50\%$ increase in mass in less than one Gyr) that disrupted any nascent cool core that may have begun to form.  CC clusters, on the other hand, settle down and strengthen for longer periods of time and suffer generally smaller mergers; this allows their cores to survive subsequent accretion events.

Our major results include
\begin{itemize}
\item Observed CC and NCC clusters, fit with single and double $\beta$-models, respectively, have comparable cluster core radii, in contrast to that claimed from single $\beta$-model fits where cool cores were excluded.  Apparently, this technique for CC clusters is not completely successful and biases the fits to smaller core radii.
\item There is a statistically significant difference in the slopes (i.e. $\beta$ values) of the outer surface brightness profiles between observed CC and NCC clusters.  This may be consistent with a different evolutionary history between these cluster types.
\item CC clusters show a significant offset in normalized hardness ratio values beyond their cool cores in comparison to NCC clusters, which is consistent with cooler gas extending out to at least $0.4\,r_{200}$ in CC clusters.
\item Simulated clusters have steeper outer surface brightness profile slopes ($\beta$ values) than observed clusters (Table \ref{beta_table}).  This may suggest some missing physics in the simulations beyond the cluster core (i.e., preheating?).
\end{itemize}

Although our simulations generally agree qualitatively with observations, comparisons with observations also suggest a number of inconsistencies that must be resolved if we desire to use simulated clusters as statistical proxies for real clusters in precision cosmological studies.  Inadequacies include overcooling in the CC cores, overly similar and overly large $\beta$ slopes between simulated CC and NCC clusters at intermediate cluster radii ($\sim0.4\,r_{200}$) compared to observed clusters, and a smaller hardness ratio offset between CC and NCC clusters than found in observations.  There is also evidence that observed clusters may have higher entropy than our simulated clusters via their flatter outer surface brightness profiles.  

Recently, \citet{Arieli_2008} suggest the use of Galcons, or galaxy constructs, in cosmological simulations instead of ``star particles" for regulating heat feedback.  \citet{Arieli_2008} find that star particles tend to group in dense and cold regions of the ICM gas, namely, the cores of CC clusters where the gas conditions are more favorable to star formation.  Any feedback heating from these star particles has little effect on the cluster beyond the inner regions of the core where the stars form.  Galcons, however, are capable of feedback heating over much more of the cluster in the form of ``galactic winds."  These winds can fill the role of any number of proposed feedback/heating mechanisms that may be at work in real clusters.  The use of Galcons in our simulations has the potential to ameliorate our CC cluster overcooling problem while at the same time increasing the effectiveness of feedback heating in in-falling halos, which could flatten the overly steep surface brightness profiles, and perhaps other unexpected yet positive consequences.

We plan to extend this direction of research with additional simulations that include Galcons and AGNs \citep[e.g.][]{Hoeft_2008,Puchwein_2008} to take advantage of potential AGN feedback mechanisms.  In addition, the clusters selected for this study were chosen because they all have at least 3 ksec of observations with \textit{XMM}, so we plan to include \textit{XMM} data in our observational analysis as well.  With additional observations and new simulations, we hope to further understand the differences between CC and NCC clusters and their use in precision cosmological studies.

\acknowledgments

This work was supported in part by grants from NASA ADP (NNX07AH53G) and the National Science Foundation (AST-0407368 and AST-0807215).  EJH thanks the NSF for support through AAPF AST-0702923.  We thank Brian O'Shea, T. Reiprich, and M. Markevitch for stimulating discussions.\clearpage

\bibliography{cc_bibdesk}

\begin{thebibliography}{124}
\expandafter\ifx\csname natexlab\endcsname\relax\def\natexlab#1{#1}\fi

\bibitem[{{Allen} {et~al.}(2008){Allen}, {Rapetti}, {Schmidt}, {Ebeling},
  {Morris}, \& {Fabian}}]{allen_2008}
{Allen}, S.~W., {Rapetti}, D.~A., {Schmidt}, R.~W., {Ebeling}, H., {Morris},
  R.~G., \& {Fabian}, A.~C. 2008, \mnras, 383, 879

\bibitem[{{Allen} {et~al.}(2004){Allen}, {Schmidt}, {Ebeling}, {Fabian}, \&
  {van Speybroeck}}]{allen_2004}
{Allen}, S.~W., {Schmidt}, R.~W., {Ebeling}, H., {Fabian}, A.~C., \& {van
  Speybroeck}, L. 2004, \mnras, 353, 457

\bibitem[{{Arieli} {et~al.}(2008){Arieli}, {Rephaeli}, \&
  {Norman}}]{Arieli_2008}
{Arieli}, Y., {Rephaeli}, Y., \& {Norman}, M.~L. 2008, \apjl, 683, L111

\bibitem[{{Baganoff} {et~al.}(2003){Baganoff}, {Maeda}, {Morris}, {Bautz},
  {Brandt}, {Cui}, {Doty}, {Feigelson}, {Garmire}, {Pravdo}, {Ricker}, \&
  {Townsley}}]{asca_grades}
{Baganoff}, F.~K., {Maeda}, Y., {Morris}, M., {Bautz}, M.~W., {Brandt}, W.~N.,
  {Cui}, W., {Doty}, J.~P., {Feigelson}, E.~D., {Garmire}, G.~P., {Pravdo},
  S.~H., {Ricker}, G.~R., \& {Townsley}, L.~K. 2003, \apj, 591, 891

\bibitem[{{Bautz} {et~al.}(2007){Bautz}, {Miller}, {Arnaud}, {Porter},
  {Hayashida}, {Henry}, {Hughes}, {Kawaharada}, {Makishima}, {Sato}, {Sanders},
  \& {Tamura}}]{A1795_1}
{Bautz}, M., {Miller}, E., {Arnaud}, K., {Porter}, S., {Hayashida}, K.,
  {Henry}, P., {Hughes}, J.~P., {Kawaharada}, M., {Makishima}, K., {Sato}, K.,
  {Sanders}, J., \& {Tamura}, T. 2007, Progress of Theoretical Physics
  Supplement, 169, 20

\bibitem[{{Binney} {et~al.}(2007){Binney}, {Bibi}, \& {Omma}}]{Binney_2007}
{Binney}, J., {Bibi}, F.~A., \& {Omma}, H. 2007, \mnras, 377, 142

\bibitem[{{Blanton} {et~al.}(2007){Blanton}, {Douglass}, {Sarazin}, {Clarke},
  \& {McNamara}}]{A2052_1}
{Blanton}, E.~L., {Douglass}, E.~M., {Sarazin}, C.~L., {Clarke}, T.~E., \&
  {McNamara}, B.~R. 2007, in Heating versus Cooling in Galaxies and Clusters of
  Galaxies, ed. H.~{B{\"o}hringer}, G.~W. {Pratt}, A.~{Finoguenov}, \&
  P.~{Schuecker}, 109

\bibitem[{{Blanton} {et~al.}(2004){Blanton}, {Sarazin}, {McNamara}, \&
  {Clarke}}]{A262_3}
{Blanton}, E.~L., {Sarazin}, C.~L., {McNamara}, B.~R., \& {Clarke}, T.~E. 2004,
  \apj, 612, 817

\bibitem[{{Bourdin} \& {Mazzotta}(2008)}]{Bourdin_2008}
{Bourdin}, H., \& {Mazzotta}, P. 2008, \aap, 479, 307

\bibitem[{{Bregman} {et~al.}(2006){Bregman}, {Fabian}, {Miller}, \&
  {Irwin}}]{A426_2}
{Bregman}, J.~N., {Fabian}, A.~C., {Miller}, E.~D., \& {Irwin}, J.~A. 2006,
  \apj, 642, 746

\bibitem[{{Brickhouse} {et~al.}(1995){Brickhouse}, {Raymond}, \&
  {Smith}}]{brickhouse_1995}
{Brickhouse}, N.~S., {Raymond}, J.~C., \& {Smith}, B.~W. 1995, \apjs, 97, 551

\bibitem[{{Briel} {et~al.}(2004){Briel}, {Finoguenov}, \& {Henry}}]{A3667_1}
{Briel}, U.~G., {Finoguenov}, A., \& {Henry}, J.~P. 2004, \aap, 426, 1

\bibitem[{{Burns} {et~al.}(2008){Burns}, {Hallman}, {Gantner}, {Motl}, \&
  {Norman}}]{paper_one}
{Burns}, J.~O., {Hallman}, E.~J., {Gantner}, B., {Motl}, P.~M., \& {Norman},
  M.~L. 2008, \apj, 675, 1125

\bibitem[{{Burns} {et~al.}(2004){Burns}, {Motl}, {Norman}, \&
  {Bryan}}]{burns_2004}
{Burns}, J.~O., {Motl}, P.~M., {Norman}, M.~L., \& {Bryan}, G.~L. 2004, in The
  Riddle of Cooling Flows in Galaxies and Clusters of Galaxies, ed.
  T.~{Reiprich}, J.~{Kempner}, \& N.~{Soker}, 291

\bibitem[{{Cattaneo} \& {Teyssier}(2007)}]{Cattaneo_Teyssier_2007}
{Cattaneo}, A., \& {Teyssier}, R. 2007, \mnras, 376, 1547

\bibitem[{{Cavagnolo} {et~al.}(2008){Cavagnolo}, {Donahue}, {Voit}, \&
  {Sun}}]{cavagnolo_2008}
{Cavagnolo}, K.~W., {Donahue}, M., {Voit}, G.~M., \& {Sun}, M. 2008, \apj, 682,
  821

\bibitem[{{Cavaliere} \& {Fusco-Femiano}(1976)}]{cff_1976}
{Cavaliere}, A., \& {Fusco-Femiano}, R. 1976, \aap, 49, 137

\bibitem[{{Cen} \& {Ostriker}(1992)}]{cen_ostriker_1992}
{Cen}, R., \& {Ostriker}, J. 1992, \apj, 393, 22

\bibitem[{{Chen} {et~al.}(2007){Chen}, {Reiprich}, {B{\"o}hringer}, {Ikebe}, \&
  {Zhang}}]{chen_2007}
{Chen}, Y., {Reiprich}, T.~H., {B{\"o}hringer}, H., {Ikebe}, Y., \& {Zhang},
  Y.-Y. 2007, \aap, 466, 805

\bibitem[{{Choi} {et~al.}(2004){Choi}, {Reynolds}, {Heinz}, {Rosenberg},
  {Perlman}, \& {Yang}}]{A4059_2}
{Choi}, Y.-Y., {Reynolds}, C.~S., {Heinz}, S., {Rosenberg}, J.~L., {Perlman},
  E.~S., \& {Yang}, J. 2004, \apj, 606, 185

\bibitem[{{Ciotti} \& {Ostriker}(2007)}]{Ciotti_Ostriker_2007}
{Ciotti}, L., \& {Ostriker}, J.~P. 2007, \apj, 665, 1038

\bibitem[{{Clarke} {et~al.}(2006){Clarke}, {Blanton}, {Sarazin}, {Kassim}, \&
  {Anderson}}]{A262_1}
{Clarke}, T.~E., {Blanton}, E., {Sarazin}, C., {Kassim}, N., \& {Anderson}, L.
  2006, in Bulletin of the American Astronomical Society, Vol.~38, Bulletin of
  the American Astronomical Society, 371

\bibitem[{{Clarke} {et~al.}(2005{\natexlab{a}}){Clarke}, {Blanton}, \&
  {Sarazin}}]{A2029_1}
{Clarke}, T.~E., {Blanton}, E.~L., \& {Sarazin}, C.~L. 2005{\natexlab{a}}, in
  X-Ray and Radio Connections (eds. L.O. Sjouwerman and K.K Dyer) Published
  electronically by NRAO, http://www.aoc.nrao.edu/events/xraydio Held 3-6
  February 2004 in Santa Fe, New Mexico, USA, (E7.08) 7 pages, ed. L.~O.
  {Sjouwerman} \& K.~K. {Dyer}

\bibitem[{{Clarke} \& {Ensslin}(2006)}]{A2256_1}
{Clarke}, T.~E., \& {Ensslin}, T.~A. 2006, \aj, 131, 2900

\bibitem[{{Clarke} {et~al.}(2005{\natexlab{b}}){Clarke}, {Sarazin}, {Blanton},
  {Neumann}, \& {Kassim}}]{A2597_2}
{Clarke}, T.~E., {Sarazin}, C.~L., {Blanton}, E.~L., {Neumann}, D.~M., \&
  {Kassim}, N.~E. 2005{\natexlab{b}}, \apj, 625, 748

\bibitem[{{Cortese} {et~al.}(2004){Cortese}, {Gavazzi}, {Boselli},
  {Iglesias-Paramo}, \& {Carrasco}}]{A1367_1}
{Cortese}, L., {Gavazzi}, G., {Boselli}, A., {Iglesias-Paramo}, J., \&
  {Carrasco}, L. 2004, \aap, 425, 429

\bibitem[{{de Messieres} {et~al.}(2007){de Messieres}, {O'Connell}, {Donahue},
  {McNamara}, {Nulsen}, {Voit}, \& {Wise}}]{A478_1}
{de Messieres}, G., {O'Connell}, R.~W., {Donahue}, M., {McNamara}, B.~R.,
  {Nulsen}, P.~E.~J., {Voit}, M., \& {Wise}, M.~W. 2007, in American
  Astronomical Society Meeting Abstracts, Vol. 211, American Astronomical
  Society Meeting Abstracts, 96.16

\bibitem[{{De Petris} {et~al.}(2002){De Petris}, {D'Alba}, {Lamagna},
  {Melchiorri}, {Orlando}, {Palladino}, {Rephaeli}, {Colafrancesco}, {Kreysa},
  \& {Signore}}]{A1656_2}
{De Petris}, M., {D'Alba}, L., {Lamagna}, L., {Melchiorri}, F., {Orlando}, A.,
  {Palladino}, E., {Rephaeli}, Y., {Colafrancesco}, S., {Kreysa}, E., \&
  {Signore}, M. 2002, \apjl, 574, L119

\bibitem[{{Dickey} \& {Lockman}(1990)}]{dickey_lockman_1990}
{Dickey}, J.~M., \& {Lockman}, F.~J. 1990, \araa, 28, 215

\bibitem[{{Donahue} \& {Voit}(2004)}]{donahue_and_voit_2004}
{Donahue}, M., \& {Voit}, G.~M. 2004, in Clusters of Galaxies: Probes of
  Cosmological Structure and Galaxy Evolution, ed. J.~S. {Mulchaey},
  A.~{Dressler}, \& A.~{Oemler}, 143

\bibitem[{{Dunn} \& {Fabian}(2006)}]{A2063_2}
{Dunn}, R.~J.~H., \& {Fabian}, A.~C. 2006, \mnras, 373, 959

\bibitem[{{Dupke} \& {White}(2003)}]{A496_2}
{Dupke}, R., \& {White}, III, R.~E. 2003, \apjl, 583, L13

\bibitem[{{Dupke} {et~al.}(2007){Dupke}, {White}, \& {Bregman}}]{A496_1}
{Dupke}, R., {White}, III, R.~E., \& {Bregman}, J.~N. 2007, \apj, 671, 181

\bibitem[{{Durret} {et~al.}(2005{\natexlab{a}}){Durret}, {Lima Neto}, \&
  {Forman}}]{Durret_2005}
{Durret}, F., {Lima Neto}, G.~B., \& {Forman}, W. 2005{\natexlab{a}}, \aap,
  432, 809

\bibitem[{{Durret} {et~al.}(2005{\natexlab{b}}){Durret}, {Lima Neto}, \&
  {Forman}}]{A85_2}
---. 2005{\natexlab{b}}, Advances in Space Research, 36, 618

\bibitem[{{Edge} {et~al.}(1990){Edge}, {Stewart}, {Fabian}, \&
  {Arnaud}}]{edge_1990}
{Edge}, A.~C., {Stewart}, G.~C., {Fabian}, A.~C., \& {Arnaud}, K.~A. 1990,
  \mnras, 245, 559

\bibitem[{{Eke} {et~al.}(1998){Eke}, {Navarro}, \& {Frenk}}]{Eke_1998}
{Eke}, V.~R., {Navarro}, J.~F., \& {Frenk}, C.~S. 1998, \apj, 503, 569

\bibitem[{{Fabian}(1994)}]{Fabian_1994}
{Fabian}, A.~C. 1994, \araa, 32, 277

\bibitem[{{Fabian} {et~al.}(2006){Fabian}, {Sanders}, {Taylor}, {Allen},
  {Crawford}, {Johnstone}, \& {Iwasawa}}]{Fabian_2006}
{Fabian}, A.~C., {Sanders}, J.~S., {Taylor}, G.~B., {Allen}, S.~W., {Crawford},
  C.~S., {Johnstone}, R.~M., \& {Iwasawa}, K. 2006, \mnras, 366, 417

\bibitem[{{Finoguenov} {et~al.}(2004){Finoguenov}, {Henriksen}, {Briel}, {de
  Plaa}, \& {Kaastra}}]{A3562_1}
{Finoguenov}, A., {Henriksen}, M.~J., {Briel}, U.~G., {de Plaa}, J., \&
  {Kaastra}, J.~S. 2004, \apj, 611, 811

\bibitem[{{Finoguenov} {et~al.}(2006){Finoguenov}, {Henriksen}, {Miniati},
  {Briel}, \& {Jones}}]{A3266_1}
{Finoguenov}, A., {Henriksen}, M.~J., {Miniati}, F., {Briel}, U.~G., \&
  {Jones}, C. 2006, \apj, 643, 790

\bibitem[{{Finoguenov} {et~al.}(2001){Finoguenov}, {Reiprich}, \&
  {B{\"o}hringer}}]{fino_2001}
{Finoguenov}, A., {Reiprich}, T.~H., \& {B{\"o}hringer}, H. 2001, \aap, 368,
  749

\bibitem[{{Frenk} {et~al.}(1999){Frenk}, {White}, {Bode}, {Bond}, {Bryan},
  {Cen}, {Couchman}, {Evrard}, {Gnedin}, {Jenkins}, {Khokhlov}, {Klypin},
  {Navarro}, {Norman}, {Ostriker}, {Owen}, {Pearce}, {Pen}, {Steinmetz},
  {Thomas}, {Villumsen}, {Wadsley}, {Warren}, {Xu}, \& {Yepes}}]{Frenk_1999}
{Frenk}, C.~S., {White}, S.~D.~M., {Bode}, P., {Bond}, J.~R., {Bryan}, G.~L.,
  {Cen}, R., {Couchman}, H.~M.~P., {Evrard}, A.~E., {Gnedin}, N., {Jenkins},
  A., {Khokhlov}, A.~M., {Klypin}, A., {Navarro}, J.~F., {Norman}, M.~L.,
  {Ostriker}, J.~P., {Owen}, J.~M., {Pearce}, F.~R., {Pen}, U.-L., {Steinmetz},
  M., {Thomas}, P.~A., {Villumsen}, J.~V., {Wadsley}, J.~W., {Warren}, M.~S.,
  {Xu}, G., \& {Yepes}, G. 1999, \apj, 525, 554

\bibitem[{{Fujita} {et~al.}(2008){Fujita}, {Tawa}, {Hayashida}, {Takizawa},
  {Matsumoto}, {Okabe}, \& {Reiprich}}]{A401_2}
{Fujita}, Y., {Tawa}, N., {Hayashida}, K., {Takizawa}, M., {Matsumoto}, H.,
  {Okabe}, N., \& {Reiprich}, T.~H. 2008, \pasj, 60, 343

\bibitem[{{Govoni} {et~al.}(2004){Govoni}, {Markevitch}, {Vikhlinin},
  {VanSpeybroeck}, {Feretti}, \& {Giovannini}}]{A2319_2}
{Govoni}, F., {Markevitch}, M., {Vikhlinin}, A., {VanSpeybroeck}, L.,
  {Feretti}, L., \& {Giovannini}, G. 2004, \apj, 605, 695

\bibitem[{{Govoni} {et~al.}(2006){Govoni}, {Murgia}, {Feretti}, {Giovannini},
  {Dolag}, \& {Taylor}}]{A2255_1}
{Govoni}, F., {Murgia}, M., {Feretti}, L., {Giovannini}, G., {Dolag}, K., \&
  {Taylor}, G.~B. 2006, \aap, 460, 425

\bibitem[{{Haiman} {et~al.}(2001){Haiman}, {Mohr}, \& {Holder}}]{haiman_2001}
{Haiman}, Z., {Mohr}, J.~J., \& {Holder}, G.~P. 2001, \apj, 553, 545

\bibitem[{{Hallman} {et~al.}(2006){Hallman}, {Motl}, {Burns}, \&
  {Norman}}]{hallman_2006}
{Hallman}, E.~J., {Motl}, P.~M., {Burns}, J.~O., \& {Norman}, M.~L. 2006, \apj,
  648, 852

\bibitem[{{Hayakawa} {et~al.}(2006){Hayakawa}, {Hoshino}, {Ishida}, {Furusho},
  {Yamasaki}, \& {Ohashi}}]{A1060_2}
{Hayakawa}, A., {Hoshino}, A., {Ishida}, M., {Furusho}, T., {Yamasaki}, N.~Y.,
  \& {Ohashi}, T. 2006, \pasj, 58, 695

\bibitem[{{Heinz} {et~al.}(2006){Heinz}, {Br{\"u}ggen}, {Young}, \&
  {Levesque}}]{Heinz_2006}
{Heinz}, S., {Br{\"u}ggen}, M., {Young}, A., \& {Levesque}, E. 2006, \mnras,
  373, L65

\bibitem[{{Hoeft} {et~al.}(2008){Hoeft}, {Brueggen}, {Yepes}, {Gottloeber}, \&
  {Schwope}}]{Hoeft_2008}
{Hoeft}, M., {Brueggen}, M., {Yepes}, G., {Gottloeber}, S., \& {Schwope}, A.
  2008, ArXiv e-prints

\bibitem[{{Hudaverdi} {et~al.}(2005){Hudaverdi}, {Yamashita}, \&
  {Furuzawa}}]{A3571_1}
{Hudaverdi}, M., {Yamashita}, K., \& {Furuzawa}, A. 2005, Advances in Space
  Research, 36, 643

\bibitem[{{Ikebe} {et~al.}(2002){Ikebe}, {Reiprich}, {B{\"o}hringer}, {Tanaka},
  \& {Kitayama}}]{Ikebe_2002}
{Ikebe}, Y., {Reiprich}, T.~H., {B{\"o}hringer}, H., {Tanaka}, Y., \&
  {Kitayama}, T. 2002, \aap, 383, 773

\bibitem[{{Jeltema} {et~al.}(2008){Jeltema}, {Hallman}, {Burns}, \&
  {Motl}}]{jeltema_2008}
{Jeltema}, T.~E., {Hallman}, E.~J., {Burns}, J.~O., \& {Motl}, P.~M. 2008,
  \apj, 681, 167

\bibitem[{{Jones} \& {Forman}(1984)}]{Jones_Forman_1984}
{Jones}, C., \& {Forman}, W. 1984, \apj, 276, 38

\bibitem[{{Jones} \& {Forman}(1999)}]{Jones_Forman_1999}
---. 1999, \apj, 511, 65

\bibitem[{{Kaiser}(1986)}]{Kaiser_1986}
{Kaiser}, N. 1986, \mnras, 222, 323

\bibitem[{{Kalberla} {et~al.}(2005){Kalberla}, {Burton}, {Hartmann}, {Arnal},
  {Bajaja}, {Morras}, \& {P{\"o}ppel}}]{kalberla_2005}
{Kalberla}, P.~M.~W., {Burton}, W.~B., {Hartmann}, D., {Arnal}, E.~M.,
  {Bajaja}, E., {Morras}, R., \& {P{\"o}ppel}, W.~G.~L. 2005, \aap, 440, 775

\bibitem[{{Kanov} {et~al.}(2006){Kanov}, {Sarazin}, \& {Hicks}}]{A2063_1}
{Kanov}, K.~N., {Sarazin}, C.~L., \& {Hicks}, A.~K. 2006, \apj, 653, 184

\bibitem[{{King}(1966)}]{king_1966}
{King}, I.~R. 1966, \aj, 71, 64

\bibitem[{{Kitzbichler} \& {Saurer}(2003)}]{A1656_1}
{Kitzbichler}, M.~G., \& {Saurer}, W. 2003, \apjl, 590, L9

\bibitem[{{LaRoque} {et~al.}(2006){LaRoque}, {Bonamente}, {Carlstrom}, {Joy},
  {Nagai}, {Reese}, \& {Dawson}}]{laroque_2006}
{LaRoque}, S.~J., {Bonamente}, M., {Carlstrom}, J.~E., {Joy}, M.~K., {Nagai},
  D., {Reese}, E.~D., \& {Dawson}, K.~S. 2006, \apj, 652, 917

\bibitem[{{Lin} {et~al.}(2003){Lin}, {Mohr}, \& {Stanford}}]{lin_2003}
{Lin}, Y.-T., {Mohr}, J.~J., \& {Stanford}, S.~A. 2003, \apj, 591, 749

\bibitem[{{{\L}okas} {et~al.}(2006){{\L}okas}, {Wojtak}, {Gottl{\"o}ber},
  {Mamon}, \& {Prada}}]{A2199_2}
{{\L}okas}, E.~L., {Wojtak}, R., {Gottl{\"o}ber}, S., {Mamon}, G.~A., \&
  {Prada}, F. 2006, \mnras, 367, 1463

\bibitem[{{Mahdavi} {et~al.}(2007){Mahdavi}, {Hoekstra}, {Babul}, {Sievers},
  {Myers}, \& {Henry}}]{A478_2}
{Mahdavi}, A., {Hoekstra}, H., {Babul}, A., {Sievers}, J., {Myers}, S.~T., \&
  {Henry}, J.~P. 2007, \apj, 664, 162

\bibitem[{{Markevitch} {et~al.}(2003){Markevitch}, {Bautz}, {Biller}, {Butt},
  {Edgar}, {Gaetz}, {Garmire}, {Grant}, {Green}, {Juda}, {Plucinsky},
  {Schwartz}, {Smith}, {Vikhlinin}, {Virani}, {Wargelin}, \&
  {Wolk}}]{mark_2003}
{Markevitch}, M., {Bautz}, M.~W., {Biller}, B., {Butt}, Y., {Edgar}, R.,
  {Gaetz}, T., {Garmire}, G., {Grant}, C.~E., {Green}, P., {Juda}, M.,
  {Plucinsky}, P.~P., {Schwartz}, D., {Smith}, R., {Vikhlinin}, A., {Virani},
  S., {Wargelin}, B.~J., \& {Wolk}, S. 2003, \apj, 583, 70

\bibitem[{{Markevitch} {et~al.}(1998){Markevitch}, {Forman}, {Sarazin}, \&
  {Vikhlinin}}]{Markevitch_1998}
{Markevitch}, M., {Forman}, W.~R., {Sarazin}, C.~L., \& {Vikhlinin}, A. 1998,
  \apj, 503, 77

\bibitem[{{Markevitch} {et~al.}(2000){Markevitch}, {Ponman}, {Nulsen}, {Bautz},
  {Burke}, {David}, {Davis}, {Donnelly}, {Forman}, {Jones}, {Kaastra},
  {Kellogg}, {Kim}, {Kolodziejczak}, {Mazzotta}, {Pagliaro}, {Patel}, {Van
  Speybroeck}, {Vikhlinin}, {Vrtilek}, {Wise}, \& {Zhao}}]{A2142_2}
{Markevitch}, M., {Ponman}, T.~J., {Nulsen}, P.~E.~J., {Bautz}, M.~W., {Burke},
  D.~J., {David}, L.~P., {Davis}, D., {Donnelly}, R.~H., {Forman}, W.~R.,
  {Jones}, C., {Kaastra}, J., {Kellogg}, E., {Kim}, D.-W., {Kolodziejczak}, J.,
  {Mazzotta}, P., {Pagliaro}, A., {Patel}, S., {Van Speybroeck}, L.,
  {Vikhlinin}, A., {Vrtilek}, J., {Wise}, M., \& {Zhao}, P. 2000, \apj, 541,
  542

\bibitem[{{Markwardt}(2009)}]{markwardt_2009}
{Markwardt}, C.~B. 2009, ArXiv e-prints

\bibitem[{{Mathews} {et~al.}(2006){Mathews}, {Faltenbacher}, \&
  {Brighenti}}]{Mathews_2006}
{Mathews}, W.~G., {Faltenbacher}, A., \& {Brighenti}, F. 2006, \apj, 638, 659

\bibitem[{{McNamara} \& {Nulsen}(2007)}]{McNamara_Nulsen_2007}
{McNamara}, B.~R., \& {Nulsen}, P.~E.~J. 2007, \araa, 45, 117

\bibitem[{{Miller} {et~al.}(2003){Miller}, {Owen}, \& {Hill}}]{A2256_2}
{Miller}, N.~A., {Owen}, F.~N., \& {Hill}, J.~M. 2003, \aj, 125, 2393

\bibitem[{{Miyoshi} {et~al.}(2005){Miyoshi}, {Tanaka}, {Yoshimura},
  {Yamashita}, {Furuzawa}, {Futamura}, \& {Hudaverdi}}]{A2029_2}
{Miyoshi}, S., {Tanaka}, N., {Yoshimura}, M., {Yamashita}, K., {Furuzawa}, A.,
  {Futamura}, T., \& {Hudaverdi}, M. 2005, Advances in Space Research, 36, 752

\bibitem[{{Mor\'e}(1977)}]{more_1977}
{Mor\'e}, J.~J. 1977, in Lecture Notes in Mathematics 630: Numerical Analysis,
  Vol. 630, 105--116

\bibitem[{{Morris} \& {Fabian}(2005)}]{A2597_1}
{Morris}, R.~G., \& {Fabian}, A.~C. 2005, \mnras, 358, 585

\bibitem[{{Morrison} \& {McCammon}(1983)}]{morrison_1983}
{Morrison}, R., \& {McCammon}, D. 1983, \apj, 270, 119

\bibitem[{{Motl} {et~al.}(2004){Motl}, {Burns}, {Loken}, {Norman}, \&
  {Bryan}}]{motl_2004}
{Motl}, P.~M., {Burns}, J.~O., {Loken}, C., {Norman}, M.~L., \& {Bryan}, G.
  2004, \apj, 606, 635

\bibitem[{{Mushotzky}(2004)}]{mushotzky_2004}
{Mushotzky}, R. 2004, {Clusters of galaxies: a cosmological probe} (Frontiers
  of X-ray astronomy), 149--164

\bibitem[{{Navarro} {et~al.}(1997){Navarro}, {Frenk}, \& {White}}]{nfw_1997}
{Navarro}, J.~F., {Frenk}, C.~S., \& {White}, S.~D.~M. 1997, \apj, 490, 493

\bibitem[{{Nusser} {et~al.}(2006){Nusser}, {Silk}, \& {Babul}}]{Nusser_2006}
{Nusser}, A., {Silk}, J., \& {Babul}, A. 2006, \mnras, 373, 739

\bibitem[{{O'Hara} {et~al.}(2006){O'Hara}, {Mohr}, {Bialek}, \&
  {Evrard}}]{ohara_2006}
{O'Hara}, T.~B., {Mohr}, J.~J., {Bialek}, J.~J., \& {Evrard}, A.~E. 2006, \apj,
  639, 64

\bibitem[{{O'Hara} {et~al.}(2004){O'Hara}, {Mohr}, \& {Guerrero}}]{A2319_1}
{O'Hara}, T.~B., {Mohr}, J.~J., \& {Guerrero}, M.~A. 2004, \apj, 604, 604

\bibitem[{{O'Shea} {et~al.}(2005){O'Shea}, {Bryan}, {Bordner}, {Norman},
  {Abel}, {Harkness}, \& {Kritsuk}}]{oshea_2004}
{O'Shea}, B., {Bryan}, G., {Bordner}, J., {Norman}, M., {Abel}, T., {Harkness},
  R., \& {Kritsuk}, A. 2005, Adaptive Mesh Refinement - Theory and
  Applications, ed. T.~Plewa, T.~Linde, \& G.~Weirs (Springer-Verlag)

\bibitem[{{Peacock}(1999)}]{peacock_1999}
{Peacock}, J.~A. 1999, {Cosmological Physics} (Cosmological Physics, by John
  A.~Peacock, pp.~704.~ISBN 052141072X.~Cambridge, UK: Cambridge University
  Press, January 1999.)

\bibitem[{{Peterson} {et~al.}(2003){Peterson}, {Kahn}, {Paerels}, {Kaastra},
  {Tamura}, {Bleeker}, {Ferrigno}, \& {Jernigan}}]{Peterson_2003}
{Peterson}, J.~R., {Kahn}, S.~M., {Paerels}, F.~B.~S., {Kaastra}, J.~S.,
  {Tamura}, T., {Bleeker}, J.~A.~M., {Ferrigno}, C., \& {Jernigan}, J.~G. 2003,
  \apj, 590, 207

\bibitem[{{Peterson} {et~al.}(2001){Peterson}, {Paerels}, {Kaastra}, {Arnaud},
  {Reiprich}, {Fabian}, {Mushotzky}, {Jernigan}, \&
  {Sakelliou}}]{Peterson_2001}
{Peterson}, J.~R., {Paerels}, F.~B.~S., {Kaastra}, J.~S., {Arnaud}, M.,
  {Reiprich}, T.~H., {Fabian}, A.~C., {Mushotzky}, R.~F., {Jernigan}, J.~G., \&
  {Sakelliou}, I. 2001, \aap, 365, L104

\bibitem[{{Pointecouteau} {et~al.}(2004){Pointecouteau}, {Arnaud}, {Kaastra},
  \& {de Plaa}}]{point_2004}
{Pointecouteau}, E., {Arnaud}, M., {Kaastra}, J., \& {de Plaa}, J. 2004, \aap,
  423, 33

\bibitem[{{Pracy} {et~al.}(2005){Pracy}, {Driver}, {De Propris}, {Couch}, \&
  {Nulsen}}]{A119_2}
{Pracy}, M.~B., {Driver}, S.~P., {De Propris}, R., {Couch}, W.~J., \& {Nulsen},
  P.~E.~J. 2005, \mnras, 364, 1147

\bibitem[{{Pratt} \& {Arnaud}(2002)}]{pratt_2002}
{Pratt}, G.~W., \& {Arnaud}, M. 2002, \aap, 394, 375

\bibitem[{{Puchwein} {et~al.}(2008){Puchwein}, {Sijacki}, \&
  {Springel}}]{Puchwein_2008}
{Puchwein}, E., {Sijacki}, D., \& {Springel}, V. 2008, ArXiv e-prints

\bibitem[{{Raymond} \& {Smith}(1977)}]{raymond_smith_1977}
{Raymond}, J.~C., \& {Smith}, B.~W. 1977, \apjs, 35, 419

\bibitem[{{Reiprich} {et~al.}(2007){Reiprich}, {Hudson}, {Nenestyan}, {Sato},
  {Ishisaki}, {Hoshino}, {Ohashi}, {Ota}, {Fujita}, \& {Hasinger}}]{A2204_1}
{Reiprich}, T.~H., {Hudson}, D.~S., {Nenestyan}, O., {Sato}, K., {Ishisaki},
  Y., {Hoshino}, A., {Ohashi}, T., {Ota}, N., {Fujita}, Y., \& {Hasinger}, G.
  2007, Progress of Theoretical Physics Supplement, 169, 33

\bibitem[{{Reiprich} {et~al.}(2008){Reiprich}, {Hudson}, {Zhang}, {Sato},
  {Ishisaki}, {Hoshino}, {Ohashi}, {Ota}, \& {Fujita}}]{Reiprich_2008}
{Reiprich}, T.~H., {Hudson}, D.~S., {Zhang}, Y.~., {Sato}, K., {Ishisaki}, Y.,
  {Hoshino}, A., {Ohashi}, T., {Ota}, N., \& {Fujita}, Y. 2008, ArXiv e-prints

\bibitem[{{Reynolds} {et~al.}(2008){Reynolds}, {Casper}, \&
  {Heinz}}]{Reynolds_08}
{Reynolds}, C.~S., {Casper}, E.~A., \& {Heinz}, S. 2008, \apj, 679, 1181

\bibitem[{{Rizza} {et~al.}(2000){Rizza}, {Loken}, {Bliton}, {Roettiger},
  {Burns}, \& {Owen}}]{A2052_2}
{Rizza}, E., {Loken}, C., {Bliton}, M., {Roettiger}, K., {Burns}, J.~O., \&
  {Owen}, F.~N. 2000, \aj, 119, 21

\bibitem[{{Rossetti} {et~al.}(2007){Rossetti}, {Ghizzardi}, {Molendi}, \&
  {Finoguenov}}]{A3558_1}
{Rossetti}, M., {Ghizzardi}, S., {Molendi}, S., \& {Finoguenov}, A. 2007, \aap,
  463, 839

\bibitem[{{Sakelliou} \& {Ponman}(2004)}]{A401_3}
{Sakelliou}, I., \& {Ponman}, T.~J. 2004, \mnras, 351, 1439

\bibitem[{{Sakelliou} \& {Ponman}(2006)}]{Sakelliou_2006}
---. 2006, \mnras, 367, 1409

\bibitem[{{Sanders} \& {Fabian}(2006)}]{A2199_1}
{Sanders}, J.~S., \& {Fabian}, A.~C. 2006, \mnras, 371, L65

\bibitem[{{Sanders} \& {Fabian}(2007)}]{A426_1}
---. 2007, \mnras, 381, 1381

\bibitem[{{Sanders} {et~al.}(2005){Sanders}, {Fabian}, \& {Taylor}}]{A2204_2}
{Sanders}, J.~S., {Fabian}, A.~C., \& {Taylor}, G.~B. 2005, \mnras, 356, 1022

\bibitem[{{Sanderson} \& {Ponman}(2003)}]{sanderson_2003}
{Sanderson}, A.~J.~R., \& {Ponman}, T.~J. 2003, \mnras, 345, 1241

\bibitem[{{Santos} {et~al.}(2008){Santos}, {Rosati}, {Tozzi}, {B{\"o}hringer},
  {Ettori}, \& {Bignamini}}]{santos_2008}
{Santos}, J.~S., {Rosati}, P., {Tozzi}, P., {B{\"o}hringer}, H., {Ettori}, S.,
  \& {Bignamini}, A. 2008, \aap, 483, 35

\bibitem[{{Sato} {et~al.}(2007){Sato}, {Yamasaki}, {Ishida}, {Ishisaki},
  {Ohashi}, {Kawahara}, {Kitaguchi}, {Kawaharada}, {Kokubun}, {Makishima},
  {Ota}, {Nakazawa}, {Tamura}, {Matsushita}, {Kawano}, {Fukazawa}, \&
  {Hughes}}]{A1060_1}
{Sato}, K., {Yamasaki}, N.~Y., {Ishida}, M., {Ishisaki}, Y., {Ohashi}, T.,
  {Kawahara}, H., {Kitaguchi}, T., {Kawaharada}, M., {Kokubun}, M.,
  {Makishima}, K., {Ota}, N., {Nakazawa}, K., {Tamura}, T., {Matsushita}, K.,
  {Kawano}, N., {Fukazawa}, Y., \& {Hughes}, J.~P. 2007, \pasj, 59, 299

\bibitem[{{Sauvageot} {et~al.}(2005){Sauvageot}, {Belsole}, \&
  {Pratt}}]{A3266_2}
{Sauvageot}, J.~L., {Belsole}, E., \& {Pratt}, G.~W. 2005, \aap, 444, 673

\bibitem[{{Sijacki} \& {Springel}(2006)}]{Sijacki_Springel_2006}
{Sijacki}, D., \& {Springel}, V. 2006, \mnras, 366, 397

\bibitem[{{Slee} \& {Roy}(1998)}]{A4038_1}
{Slee}, O.~B., \& {Roy}, A.~L. 1998, \mnras, 297, L86

\bibitem[{{Snowden} {et~al.}(1994){Snowden}, {McCammon}, {Burrows}, \&
  {Mendenhall}}]{snowden_1994}
{Snowden}, S.~L., {McCammon}, D., {Burrows}, D.~N., \& {Mendenhall}, J.~A.
  1994, \apj, 424, 714

\bibitem[{{Sun} \& {Vikhlinin}(2005)}]{A1367_2}
{Sun}, M., \& {Vikhlinin}, A. 2005, \apj, 621, 718

\bibitem[{{Tamura} {et~al.}(2001){Tamura}, {Kaastra}, {Peterson}, {Paerels},
  {Mittaz}, {Trudolyubov}, {Stewart}, {Fabian}, {Mushotzky}, {Lumb}, \&
  {Ikebe}}]{Tamura_2001}
{Tamura}, T., {Kaastra}, J.~S., {Peterson}, J.~R., {Paerels}, F.~B.~S.,
  {Mittaz}, J.~P.~D., {Trudolyubov}, S.~P., {Stewart}, G., {Fabian}, A.~C.,
  {Mushotzky}, R.~F., {Lumb}, D.~H., \& {Ikebe}, Y. 2001, \aap, 365, L87

\bibitem[{{Tittley} \& {Henriksen}(2005)}]{A2142_1}
{Tittley}, E.~R., \& {Henriksen}, M. 2005, \apj, 618, 227

\bibitem[{{Venturi} {et~al.}(2003){Venturi}, {Bardelli}, {Dallacasa},
  {Brunetti}, {Giacintucci}, {Hunstead}, \& {Morganti}}]{A3562_2}
{Venturi}, T., {Bardelli}, S., {Dallacasa}, D., {Brunetti}, G., {Giacintucci},
  S., {Hunstead}, R.~W., \& {Morganti}, R. 2003, \aap, 402, 913

\bibitem[{{Venturi} {et~al.}(2002){Venturi}, {Bardelli}, {Zagaria}, {Prandoni},
  \& {Morganti}}]{A3571_2}
{Venturi}, T., {Bardelli}, S., {Zagaria}, M., {Prandoni}, I., \& {Morganti}, R.
  2002, \aap, 385, 39

\bibitem[{{Vikhlinin} {et~al.}(1999){Vikhlinin}, {Forman}, \&
  {Jones}}]{vik_forman_jones_1999}
{Vikhlinin}, A., {Forman}, W., \& {Jones}, C. 1999, \apj, 525, 47

\bibitem[{{Vikhlinin} {et~al.}(2006){Vikhlinin}, {Kravtsov}, {Forman}, {Jones},
  {Markevitch}, {Murray}, \& {Van Speybroeck}}]{vik_doublebeta_2006}
{Vikhlinin}, A., {Kravtsov}, A., {Forman}, W., {Jones}, C., {Markevitch}, M.,
  {Murray}, S.~S., \& {Van Speybroeck}, L. 2006, \apj, 640, 691

\bibitem[{{Vikhlinin} {et~al.}(2009){Vikhlinin}, {Kravtsov}, {Burenin},
  {Ebeling}, {Forman}, {Hornstrup}, {Jones}, {Murray}, {Nagai}, {Quintana}, \&
  {Voevodkin}}]{vikhlinin_2009}
{Vikhlinin}, A., {Kravtsov}, A.~V., {Burenin}, R.~A., {Ebeling}, H., {Forman},
  W.~R., {Hornstrup}, A., {Jones}, C., {Murray}, S.~S., {Nagai}, D.,
  {Quintana}, H., \& {Voevodkin}, A. 2009, \apj, 692, 1060

\bibitem[{{Vikhlinin} {et~al.}(2005){Vikhlinin}, {Markevitch}, {Murray},
  {Jones}, {Forman}, \& {Van Speybroeck}}]{vik_2005}
{Vikhlinin}, A., {Markevitch}, M., {Murray}, S.~S., {Jones}, C., {Forman}, W.,
  \& {Van Speybroeck}, L. 2005, \apj, 628, 655

\bibitem[{{Voit}(2005)}]{voit_2005}
{Voit}, G.~M. 2005, Reviews of Modern Physics, 77, 207

\bibitem[{{Wall} \& {Jenkins}(2003)}]{wall_jenkins_2003}
{Wall}, J.~V., \& {Jenkins}, C.~R. 2003, {Practical Statistics for Astronomers}
  (Princeton Series in Astrophysics)

\bibitem[{{Wang} \& {Steinhardt}(1998)}]{wang_steinhardt_1998}
{Wang}, L., \& {Steinhardt}, P.~J. 1998, \apj, 508, 483

\bibitem[{{Whitaker} {et~al.}(2003){Whitaker}, {Kraft}, {Posson-Brown},
  {Jones}, \& {Donnelly}}]{A119_1}
{Whitaker}, K.~E., {Kraft}, R.~P., {Posson-Brown}, J., {Jones}, C., \&
  {Donnelly}, R.~H. 2003, in Bulletin of the American Astronomical Society,
  Vol.~35, Bulletin of the American Astronomical Society, 1282

\bibitem[{{White} {et~al.}(1993){White}, {Navarro}, {Evrard}, \&
  {Frenk}}]{White_1993}
{White}, S.~D.~M., {Navarro}, J.~F., {Evrard}, A.~E., \& {Frenk}, C.~S. 1993,
  \nat, 366, 429

\bibitem[{{Wise} {et~al.}(2007){Wise}, {McNamara}, {Nulsen}, {Houck}, \&
  {David}}]{Wise_2007}
{Wise}, M.~W., {McNamara}, B.~R., {Nulsen}, P.~E.~J., {Houck}, J.~C., \&
  {David}, L.~P. 2007, \apj, 659, 1153

\bibitem[{{Yuan} {et~al.}(2005){Yuan}, {Yan}, {Yang}, \& {Zhou}}]{A401_1}
{Yuan}, Q.-R., {Yan}, P.-F., {Yang}, Y.-B., \& {Zhou}, X. 2005, Chinese Journal
  of Astronomy and Astrophysics, 5, 126

\end{thebibliography}

\clearpage

\LongTables
\begin{deluxetable}{cccccccccc}
\tabletypesize{\scriptsize}
\tablewidth{0pt}
\tablecaption{Observed Cluster Properties and Pointing Data\label{clusters_table}}
\tablehead{
	\colhead{Cluster} & 
	  \colhead{\textit{(C)handra} Obs ID} &
	  \colhead{ACIS-I/S\tablenotemark{a}} &
	\colhead{Exp}&
	\colhead{z\tablenotemark{b}} & 
	\colhead{$T_{Chen}$\tablenotemark{d}} & 
	\colhead{M$_{200}$\tablenotemark{c,d}} & 
	\colhead{r$_{200}$\tablenotemark{c}} & 
	\colhead{Cool\tablenotemark{b}} &
	\colhead{Previous} \\
	\colhead{} & 
	  \colhead{(R)OSAT Obs ID} &
	  \colhead{PSPC-A/B\tablenotemark{a}} &
	\colhead{Time (ksec)} & 
	\colhead{} & 
	\colhead{(keV)} & 
	\colhead{(10$^{14}$ M$_{\odot}$)} & 
	\colhead{(Mpc)} & 
	\colhead{Core} &
	\colhead{Publications}	
	}
\startdata
A85 &(C) 904 & I & 39&0.0521 & 6.51 & 8.27 & 1.71 & Yes&1, 2\\
& (R) 174A & B & 2&&&&&&\\
& (R) 174A1 & B & 3&&&&&&\\
& (R) 250 & B & 10&&&&&&\\
\hline
A119 &(C) 4108 & S & 10& 0.0444 & 5.69 & 9.12 & 1.67 & No&3, 4\\
& (R) 251 & B & 15&&&&&&\\
\hline
A262&(C) 2215 & S & 29&  0.0163 & 2.25 & 0.93 & 1.00 & Yes&5, 6, 7\\
& (R) 254 & B & 9&&&&&&\\
\hline
A401 &(C) 518 & I & 18&0.0748 & 7.19 & 8.77  & 2.01 & No&8, 9, 10\\
&(C) 2309 & I & 12&&&&&&\\
& (R) 182 & B & 7&&&&&&\\
& (R) 235 & B & 7&&&&&&\\
\hline
A426\tablenotemark{e} &(C) 502 & I & 5 & 0.0179 & 5.28 & 13.7 & 1.59 & Yes&11, 12\\
&(C) 503 & S & 9&&&&&&\\
&(C) 1513 & S & 25&&&&&&\\
&(C) 3209 & S & 97&&&&&&\\
&(C) 4289 & S & 97&&&&&&\\
&(C) 4946 & S & 24&&&&&&\\
&(C) 4947 & S & 30&&&&&&\\
& (R) 32A & B & 1&&&&&&\\
& (R) 32A1 & B & 31&&&&&&\\
& (R) 33 & B & 15&&&&&&\\
& (R) 34A & B & 10&&&&&&\\
& (R) 34A1 & B & 24&&&&&&\\
& (R) 35A & B & 6&&&&&&\\
& (R) 35A1 & B & 26&&&&&&\\
& (R) 186 & B & 5&&&&&&\\
& (R) 521 & B & 3&&&&&&\\
\hline
A478 &(C) 1669 & S & 42&  0.0882 & 6.91  & 9.38 & 1.82 & Yes&13, 14\\
& (C) 6102 & I & 10&&&&&&\\
& (C) 7217 & I & 19&&&&&&\\
& (R) 193 & B & 22&&&&&&\\
\hline
A496 &(C) 931 & S & 19&  0.0331 & 3.91 & 4.83 & 1.35 & Yes&15, 16\\
&(C) 3361 & S & 10&&&&&&\\
&(C) 4976 & S & 76&&&&&&\\
& (R) 24 & B & 9&&&&&&\\
\hline
A1060 &(C) 2220 & I & 32&  0.0124 & 3.15 & 2.46 & 1.20 & Yes&17, 18\\
& (R) 135 & B & 20&&&&&&\\
& (R) 200 & B & 16&&&&&&\\
\hline
A1367 & (C) 514 & S & 41&  0.0214 & 3.55 & 7.36 & 1.29 & No&19, 20\\
& (R) 153 & B & 19&&&&&&\\
\hline
A1656\tablenotemark{f} &(C) 555 & I & 9& 0.0231 & 8.07 & 13.7 & 2.00 & No&21, 22\\
&(C) 556 & S & 10&&&&&&\\
&(C) 1086 & S & 10&&&&&&\\
&(C) 1112 & I & 10&&&&&&\\
&(C) 1113 & I & 10&&&&&&\\
&(C) 1114 & I & 10&&&&&&\\
& (R) 5 & B & 21&&&&&&\\
& (R) 6 & B & 22&&&&&&\\
& (R) 9 & B & 20&&&&&&\\
& (R) 13 & B & 21&&&&&&\\
\hline
A1795 & (C) 493 & S & 20&  0.0622 & 6.17 & 10.21 & 1.60 & Yes&6, 23\\
& (C) 494 & S & 20&&&&&&\\
& (C) 5289 & I & 15&&&&&&\\
& (R) 55 & B & 26&&&&&&\\
& (R) 105 & B & 36&&&&&&\\
& (R) 145 & B & 18&&&&&&\\
\hline
A2029 & (C) 891 & S & 20&  0.0766 & 7.93  & 10.43 & 2.08 & Yes&24, 25\\
& (C) 4977 & S & 79&&&&&&\\
& (C) 6101 & I & 10&&&&&&\\
& (R) 161 & B & 3&&&&&&\\
& (R) 249 & B & 13&&&&&&\\
\hline
A2052 &(C) 890 & S & 37&  0.0353 & 3.12 & 2.72 & 1.18 & Yes&26, 27\\
&(C) 5807 & S & 129&&&&&&\\
& (R) 275 & B & 6&&&&&&\\
\hline
A2063 &(C) 4187 & I & 9&  0.0355 & 3.56 & 2.38 & 1.31 & Yes&28, 29\\
&(C) 5795 & S & 10&&&&&&\\
&(C) 6262 & S & 14&&&&&&\\
&(C) 6263 & S & 17&&&&&&\\
& (R) 128 & B & 10&&&&&&\\
& (R) 184A1 & B & 10&&&&&&\\
& (R) 376 & B & 7&&&&&&\\
\hline
A2142 &(C) 1196 & S & 12&  0.0894 & 8.46 & 15.21 & 2.06 & Yes&30, 31\\
&(C) 1228 & S & 12&&&&&&\\
&(C) 5005 & I & 45&&&&&&\\
& (R) 96 & B & 6&&&&&&\\
& (R) 233 & B & 5&&&&&&\\
& (R) 415 & B & 19&&&&&&\\
& (R) 551 & B & 6&&&&&&\\
\hline
A2199 & (C) 497 & S & 20&  0.0299 & 4.28 & 4.29 & 1.40 & Yes&32, 33\\
& (C) 498 & S & 19&&&&&&\\
& (R) 644 & B & 41&&&&&&\\
\hline
A2204 &(C) 499 & S & 10&  0.1524 & 6.38 & 6.53 & 1.87 & Yes&34, 35\\
&(C) 6104 & I & 10&&&&&&\\
& (R) 281 & B & 5&&&&&&\\
\hline
A2255 &(C) 894 & I & 40&  0.0809 & 5.92 & 8.28 & 1.82 & No&36, 37\\
& (R) 512 & B & 15&&&&&&\\
\hline
A2256 &(C) 965 & S & 11&  0.0581 & 6.83 & 12.48 &  1.91& No&38, 39\\
&(C) 1386 & I & 13&&&&&&\\
&(C) 1521 & S & 3&&&&&&\\
&(C) 2419 & S & 12&&&&&&\\
& (R) 162A & B & 4&&&&&&\\
& (R) 162A1 & B & 5&&&&&&\\
& (R) 163 & B & 11&&&&&&\\
& (R) 339 & B & 5&&&&&&\\
& (R) 340 & B & 9&&&&&&\\
& (R) 341 & B & 10&&&&&&\\
\hline
A2319\tablenotemark{f} &(C) 3231 & I & 15&  0.0555 & 9.12 & 18.6& 2.12 & No&40, 41\\
& (R) 73 & B & 1&&&&&&\\
& (R) 73A1 & B & 3&&&&&&\\
\hline
A2597 &(C) 922 & S & 40&  0.0852 & 4.2 & 3.92 & 1.30 & Yes&42, 43\\
&(C) 6934 & S & 53&&&&&&\\
&(C) 7329 & S & 61&&&&&&\\
& (R) 112 & B & 7&&&&&&\\
\hline
A3158 & (C) 3201 & I & 25&  0.0590 & 5.41 & 5.93 & 1.68 & No&33\\
& (C) 3712 & I & 31&&&&&&\\
& (R) 310 & B & 3&&&&&&\\
\hline
A3266 &(C) 899 & I & 30&  0.0545 & 7.72 & 19.75 & 1.94 & No&44, 45\\
& (R) 211 & B & 7&&&&&&\\
& (R) 552 & B & 14&&&&&&\\
\hline
A3391 &(C) 4943 & I & 19&  0.0550 & 5.89 & 6.20 & 1.65 & No&29\\
& (R) 79 & B & 3&&&&&&\\
& (R) 80 & B & 7&&&&&&\\
\hline
A3558 & (C) 1646 & S & 15&  0.0477 & 5.37 & 6.84 & 1.66 & No&29, 46\\
& (R) 76 & B & 30&&&&&&\\
\hline
A3562 &(C) 899 & I & 30&  0.0502 & 5.16 & 3.59 & 1.57 & No&47, 48\\
& (R) 237 & B & 20&&&&&&\\
\hline
A3571 &(C) 4203 & S & 34&  0.0397 & 6.80 & 8.86 & 1.83 & Yes&49, 50\\
& (R) 287 & B & 6&&&&&&\\
\hline
A3667 & (C) 5751 & I & 130&  0.0530 & 6.28 & 5.41 & 1.85 & No&51\\
& (C) 5752 & I & 61&&&&&&\\
& (C) 5753 & I & 105&&&&&&\\
& (R) 234 & B & 13&&&&&&\\
\hline
A4038 &(C) 4188 & I & 6&  0.0281 & 3.22 & 2.58 & 1.21 & Yes&52\\
&(C) 4992 & I & 34&&&&&&\\
& (R) 354 & B & 3&&&&&&\\
\hline
A4059 &(C) 5785 & S & 93&  0.0456 & 3.94 & 4.49 & 1.39 & Yes&53, 54\\
& (R) 175 & B & 5&&&&&&\\
\enddata
\tablecomments{Observed X-ray properties and instrument pointing data for the selected Abell clusters.}
\tablenotetext{a}{\textit{Chandra} or \textit{ROSAT} detector used.}
\tablenotetext{b}{\citet{ohara_2006}}
\tablenotetext{c}{Calculated from scaling relations, see text.  \citet{ohara_2006}, \citet{fino_2001}, \citet{nfw_1997}, \citet{voit_2005}}
\tablenotetext{d}{$M_{500}$ and $T_{Chen}$ taken from \citet{chen_2007} and \citet{Ikebe_2002}}
\tablenotetext{e}{Mass and temperatures not from \citet{chen_2007}.}
\tablerefs{1:\citet{Durret_2005}, 2:\citet{A85_2}, 3:\citet{A119_1}, 4:\citet{A119_2}, 5:\citet{A262_1}, 6:\citet{vik_2005}, 7:\citet{A262_3}, 8:\citet{A401_1}, 9:\citet{A401_2}, 10:\citet{A401_3}, 11:\citet{A426_1}, 12:\citet{A426_2}, 13:\citet{A478_1}, 14:\citet{A478_2}, 15:\citet{A496_1}, 16:\citet{A496_2}, 17:\citet{A1060_1}, 18:\citet{A1060_2}, 19:\citet{A1367_1}, 20:\citet{A1367_2}, 21:\citet{A1656_1}, 22:\citet{A1656_2}, 23:\citet{A1795_1}, 24:\citet{A2029_1}, 25:\citet{A2029_2}, 26:\citet{A2052_1}, 27:\citet{A2052_2}, 28: \citet{A2063_1}, 29:\citet{A2063_2}, 30:\citet{A2142_1}, 31:\citet{A2142_2}, 32:\citet{A2199_1}, 33:\citet{A2199_2}, 34:\citet{A2204_1}, 35:\citet{A2204_2}, 36:\citet{A2255_1}, 37:\citet{Sakelliou_2006}, 38:\citet{A2256_1}, 39:\citet{A2256_2}, 40:\citet{A2319_1}, 41:\citet{A2319_2}, 42:\citet{A2597_1}, 43:\citet{A2597_2}, 44:\citet{A3266_1}, 45:\citet{A3266_2}, 46:\citet{A3558_1}, 47:\citet{A3562_1}, 48:\citet{A3562_2}, 49:\citet{A3571_1}, 50:\citet{A3571_2}, 51:\citet{A3667_1}, 52:\citet{A4038_1}, 53:\citet{Reynolds_08}, 54:\citet{A4059_2}}
\end{deluxetable}
\clearpage

\begin{deluxetable}{ccccc}
\tablecaption{Average $\beta$-Model Fit Results\label{beta_table}}
\tablewidth{0pt}
\tablehead{
	\colhead{} &
	\colhead{Average $\beta_1$} & 
	\colhead{Average $r_{core,1}$} & 
	\colhead{Average $\beta_2$} &
	  \colhead{Average $r_{core,2}$}}
\startdata
Observed CC clusters  &  0.54$\pm$0.07 &0.01$\pm$0.004  &0.55$\pm$0.04 &0.08$\pm$0.02\\
Observed NCC clusters  &... &... & 0.49$\pm$0.02&0.08$\pm$0.01\\
K-S Test Probabilities & ... & ... & 0.002 & 0.209 \\
\hline
Simulated CC clusters  & 0.98$\pm$0.41 &0.01$\pm$0.01  &0.70$\pm$0.10  &0.10$\pm$0.01\\
Simulated NCC clusters  &  ... & ... & 0.71$\pm$0.22&0.13$\pm$0.06\\
K-S Test Probabilities & ... & ... & 0.650 & 0.277 \\
\enddata
\tablecomments{The averaged values of individual $\beta$-models fit between 0 and $0.4\,r_{200}$ for observed and simulated clusters, along with the standard deviation of each population. (See \S \ref{sec_sim_surface_brightness} for more details on the simulated profiles).  Subscripts 1 and 2 denote the averages of the innermost and outermost $r_{core}$ and $\beta$ values, respectively.  The $r_{core}$ fits are in units of $r_{200}$.  K-S probabilities are probabilities that the fits for CC and NCC clusters come from the same parent distribution.}
\end{deluxetable}

\begin{deluxetable}{ccccccc}
\tabletypesize{\small}
\tablecaption{Observed Cluster $\beta$-Model Fit Parameters\label{all_beta_table}}
\tablewidth{0pt}
\tablehead{
	\colhead{Cluster}&
	\colhead{$\beta_1$}&
	\colhead{$r_{core,1}$}&
	\colhead{$\beta_2$}&
	\colhead{$r_{core,2}$}&
	\colhead{DOF}&
	\colhead{Reduced $\chi^2$}}
\startdata
A262 & 0.738 & 0.003 & 0.503 & 0.051 & 14 & 0.082 \\
A426 & 0.396 & 0.001 & 23.468 & 0.670 & 14 & 0.148 \\
A478 & 0.406 & 0.006 & 14.937 & 0.573 & 14 & 0.149 \\
A496 & 0.478 & 0.007 & 0.489 & 0.054 & 14 & 0.137 \\
A1060 & 0.494 & 0.008 & 0.463 & 0.049 & 14 & 0.130 \\
A1795 & 0.488 & 0.009 & 0.487 & 0.058 & 14 & 0.139 \\
A2029 & 0.504 & 0.012 & 0.515 & 0.070 & 14 & 0.138 \\
A2052 & 0.514 & 0.013 & 0.542 & 0.084 & 14 & 0.148 \\
A2063 & 0.513 & 0.014 & 0.557 & 0.088 & 14 & 0.135 \\
A2142 & 0.512 & 0.014 & 0.578 & 0.094 & 14 & 0.144 \\
A2199 & 0.520 & 0.014 & 0.579 & 0.090 & 14 & 0.130 \\
A2204 & 0.523 & 0.013 & 0.575 & 0.086 & 14 & 0.139 \\
A2597 & 0.531 & 0.014 & 0.575 & 0.087 & 14 & 0.135 \\
A3571 & 0.539 & 0.015 & 0.594 & 0.093 & 14 & 0.126 \\
A4038 & 0.536 & 0.016 & 0.589 & 0.089 & 14 & 0.107 \\
A4059 & 0.537 & 0.016 & 0.601 & 0.092 & 14 & 0.109 \\
\hline
A401 & ...&... & 0.498 & 0.082 & 17 & 0.065 \\
A1367 &...  &  ...& 0.448 & 0.078 & 17 & 0.065 \\
A1656 & ... &  ...& 0.488 & 0.074 & 17 & 0.062 \\
A2255 &  ...&  ...& 0.472 & 0.073 & 17 & 0.066 \\
A2256 &  ...&  ...& 0.499 & 0.084 & 17 & 0.052 \\
A3158 & ...&  ...& 0.520 & 0.090 & 17 & 0.057 \\
A2319 &  ...& ...& 0.503 & 0.082 & 17 & 0.052 \\
A3266 &  ...& ...& 0.505 & 0.082 & 17 & 0.054 \\
A3391 &  ...& ...& 0.495 & 0.081 & 17 & 0.064 \\
A3558 &  ...& ...& 0.491 & 0.078 & 17 & 0.144 \\
A3562 &  ...& ... & 0.487 & 0.074 & 17 & 0.205 \\
A3667 &  ...&...  & 0.487 & 0.075 & 17 & 0.197 \\
\enddata
\tablecomments{Individual fit values and statistics for first the observed CC and then NCC clusters.  ``1" represents the inner $\beta$-model fits (for CC clusters only) while ``2" represents the outer fits.  DOF show the degrees of freedom in the fits.  The fits for A426 and A478 are clear outliers and so are not included in the average values found in Table \ref{beta_table}.}
\end{deluxetable}
\clearpage

\begin{landscape}
\begin{deluxetable}{cccccccccccc}
\tabletypesize{\tiny}
\tablewidth{0pt}
\tablecaption{Observed Cluster Spectral Fitting Results}
\tablehead{
	\colhead{Cluster} & 
	  \colhead{CC/NCC} &
	  \colhead{Region ($r_{200}$)} &
	\colhead{$T_{Chen}$\tablenotemark{a}} & 
	\colhead{$T_{fit}$\tablenotemark{b}} & 
	\colhead{$T_{fit}/T_{Chen}$\tablenotemark{c}} & 
	\colhead{$N_H\,(10^{22} \mbox{cm}^{-2})$\tablenotemark{d}} & 
	\colhead{$z$\tablenotemark{e}} & 
	\colhead{$Z (Z_{\bigodot})$\tablenotemark{f}} &
	\colhead{Reduced $\chi^2$} &
	  \colhead{DOF\tablenotemark{g}} &
	  \colhead{Source \%\tablenotemark{h}}
		}
\startdata
A478 & CC & 0.14 - 0.18 & 6.91 & $7.63_{-0.37}^{+0.37}$ & 1.10 & 0.277 & 0.0882 & 0.211 & 1.011 & 52 & 67.5 \\
 &  & 0.18 - 0.22 & 6.91 & $7.04_{-0.42}^{+0.50}$ & 1.02 & 0.279 & 0.0882 & 0.189 & 1.356 & 52 & 57.4 \\
 &  & 0.22 - 0.26 & 6.91 & $5.91_{-0.56}^{+0.61}$ & 0.85 & 0.274 & 0.0882 & 0.212 & 1.530 & 52 & 47.6 \\
 &  & 0.26 - 0.30 & 6.91 & $5.09_{-0.85}^{+1.20}$ & 0.74 & 0.308 & 0.0882 & 0.347 & 0.558 & 52 & 38.7 \\
A401 & NCC  & 0.11 - 0.1575 & 7.19 & $6.95_{-0.00}^{+0.00}$ & 0.97 & 0.113 & 0.0748 & 0.355 & 2.017 & 52 & 90.5 \\
 &  & 0.1575 - 0.205 & 7.19 & $6.53_{-0.42}^{+0.49}$ & 0.91 & 0.123 & 0.0748 & 0.353 & 1.671 & 52 & 85.7 \\
 &  & 0.205 - 0.2525 & 7.19 & $6.65_{-0.49}^{+0.66}$ & 0.92 & 0.084 & 0.0748 & 0.353 & 1.237 & 52 & 80.5 \\
 &  & 0.2525 - 0.30 & 7.19 & $6.53_{-0.56}^{+0.73}$ & 0.91 & 0.121 & 0.0748 & 0.306 & 1.247 & 52 & 74.7 \\
\hline
A85 & CC & 0.12 - 0.165 & 6.51 & $6.30_{-0.25}^{+0.25}$ & 0.97 & 0.008 & 0.0521 & 0.351 & 0.967 & 52 & 88 \\
 &  & 0.165 - 0.21 & 6.51 & $6.32_{-0.42}^{+0.44}$ & 0.97 & 0.013 & 0.0521 & 0.355 & 1.574 & 52 & 81.3 \\
 &  & 0.21 - 0.255 & 6.51 & $6.20_{-0.43}^{+0.43}$ & 0.95 & 0.033 & 0.0521 & 0.148 & 1.341 & 52 & 73.4 \\
 &  & 0.255 - 0.30 & 6.51 & $5.86_{-0.47}^{+0.49}$ & 0.90 & 0.010 & 0.0521 & 0.240 & 0.997 & 52 & 69.9 \\
A2255 & NCC & 0.11 - 0.1575 & 5.92 & $5.64_{-0.40}^{+0.47}$ & 0.95 & 0.025 & 0.0809 & 0.300 & 1.208 & 52 & 83 \\
 &  & 0.1575 - 0.205 & 5.92 & $5.47_{-0.54}^{+0.62}$ & 0.92 & 0.028 & 0.0809 & 0.402 & 1.097 & 52 & 79.3 \\
 &  & 0.205 - 0.2525 & 5.92 & $5.34_{-0.35}^{+0.50}$ & 0.90 & 0.021 & 0.0809 & 0.342 & 1.331 & 52 & 74.9 \\
 &  & 0.2525 - 0.30 & 5.92 & $6.05_{-0.56}^{+0.57}$ & 1.02 & 0.000 & 0.0809 & 0.269 & 1.400 & 52 & 70.5 \\
\hline
A4059&CC&0.1 - 0.15&3.94&$4.12_{-0.44}^{+0.54}$&1.05&0.028&0.0456&0.605&1.019&52&87.3\\
&&0.15 - 0.20&3.94&$4.28_{-0.24}^{+0.35}$&1.09&0.021&0.0456&0.381&1.334&52&81.3\\
&&0.20 - 0.25&3.94&$3.86_{-0.34}^{+0.35}$&0.98&0.055&0.0456&0.142&1.457&52&68.3\\
&&0.25 - 0.30&3.94&$4.26_{-0.50}^{+0.74}$&1.08&0.061&0.0456&0.299&1.827&52&53.8\\
A3391&NCC&0.11 - 0.1575&5.89&$5.47_{-0.51}^{+0.71}$&0.93&0.059&0.0550&0.043&0.968&52&85.9\\
&&0.1575 - 0.205&5.89&$5.18_{-0.52}^{+0.62}$&0.88&0.039&0.0550&0.000&1.070&52&81.1\\
&&0.205 - 0.2525&5.89&$5.68_{-0.67}^{+0.85}$&0.96&0.046&0.0550&0.154&1.370&52&73.8\\
&&0.2525 - 0.30&5.89&$4.10_{-0.47}^{+0.54}$&0.70&0.104&0.0550&0.063&1.259&52&65.6\
\enddata
\tablecomments{Mass-matched CC/NCC pairs are separated by horizontal lines.}
\tablenotetext{a}{From \citet{chen_2007}.  All temperatures are in keV.\label{table_bulk_temperatures}}
\tablenotetext{b}{Fitted temperature errors are given at the 90\% confidence level.}
\tablenotetext{c}{Normalized temperatures plotted in Figure \ref{fig_HR_profiles} are expressed here as $T_{fit}/T_{Chen}$, where $T_{Chen}$ have been corrected for cool cores.}
\tablenotetext{d}{Fit results for galactic hydrogen column density.}
\tablenotetext{e}{Redshifts from \citet{ohara_2006}.}
\tablenotetext{f}{Fit results for cluster metallicity.}
\tablenotetext{g}{Degrees of freedom in the fit.}
\tablenotetext{h}{Percent of extracted spectra counts attributable to the cluster in the corresponding spectral region.}
\end{deluxetable}
\clearpage
\end{landscape}

\end{document}